\documentclass[a4paper,onecolumn,12pt]{elsarticle}
\usepackage[labelfont=,labelsep=period]{caption}
\usepackage{setspace}
\usepackage{ragged2e}
\usepackage{graphicx}
\usepackage{subfigure}   
\usepackage{float}
\usepackage[T1]{fontenc}
\usepackage{tgbonum}
\usepackage{booktabs}
\usepackage{subcaption}
\usepackage[english]{babel}
\usepackage{blindtext}
\usepackage{layout}
\usepackage{subcaption}
\usepackage{rotating}
\usepackage{multirow}
\usepackage[legalpaper, lmargin=0.7in, rmargin=.7in, tmargin=0.8in, bmargin=0.8in]{geometry}
\usepackage{enumerate,dsfont}
\usepackage{geometry}
\usepackage{lipsum} 
\usepackage[citecolor=blue]{hyperref}
\usepackage{xcolor}
\usepackage{float}
\usepackage{graphicx}
\usepackage{eucal}
\usepackage{amssymb}
\usepackage{amsmath}
\usepackage{mathtools}
\usepackage{enumitem}
\usepackage{amsthm}
\usepackage{mathrsfs,bm}
\numberwithin{equation}{section}
\newcommand\highlightReference[1]{%
	\expandafter\newcommand\csname highlightReference-#1\endcsname{}%
}
\let\oldbibitem\bibitem
\def\bibitem#1 #2\par{%
	\expandafter\ifx\csname highlightReference-#1\endcsname\relax
	\oldbibitem{#1}#2\par
	\else
	\oldbibitem{#1}\highlight{#2}\par
	\fi
}
\usepackage{color}
\newcommand\highlight[1]{\textcolor{blue}{#1}}
\newtheorem{theorem}{Theorem}[section]

\newtheorem{definition}{Definition}[section]

\allowdisplaybreaks
\allowdisplaybreaks[1]
\makeatletter
\def\ps@pprintTitle{} % Remove Elsevier preprint watermark
\makeatother
\begin{document}
	\begin{frontmatter}
%		\title{GeoDiv-KAN: Geometry-Gated Divergence-Free KAN Solver for Time-Fractional Navier–Stokes Equations on Complex Domains}
\title{GeoDiv-KAN Solver for Time-Fractional Navier–Stokes Equations on Complex Domains to Realistic CFD Reconstruction}
		
	\author[1]{Yatin Kumar\corref{cor1}}
	\ead{yatindhiman2@gmail.com}
	
	\author[1]{Shivani Saini}
	\ead{shivanis.nith@gmail.com}
	
	\author[1]{Ramesh Kumar Vats}
	\ead{rkvats@nith.ac.in}
	
	\cortext[cor1]{Corresponding author}
		
%	\author[2]{Takaharu Yaguchi}
%	\ead{yaguchi@pearl.kobe-u.ac.jp}

	\affiliation[1]{
		organization={Department of Mathematics and Scientific Computing, 
			National Institute of Technology Hamirpur},
		city={Hamirpur},
		state={Himachal Pradesh},
		postcode={177005},
		country={India}
	}
	
%	\affiliation[2]{
%		organization={IMI, Kyushu University 
%			and RIKEN Center for Advanced Intelligence Project},
%		addressline={744 Motooka, Nishi-ku},
%		city={Fukuoka},
%		country={Japan}
%	}
\begin{abstract}
	Scientific machine learning provides a promising framework for solving complex problems in science and engineering. Notably, physics-informed neural networks (PINNs) have shown considerable potential for solving differential equations and modelling engineering systems. Solving time-fractional Navier--Stokes equations (TFNSEs) on complex domains is challenging due to temporal nonlocality, nonlinear flow dynamics, geometric complexity, and incompressibility constraints.
	To address these challenges, we propose  a geometry-gated, divergence-free Gaussian radial basis function Kolmogorov--Arnold network (GeoDiv-KAN). The proposed architecture uses geometry-dependent feature gating and a streamfunction formulation to preserve incompressibility by construction. GeoDiv-KAN is evaluated for the time-fractional incompressible Navier--Stokes equations on square, circular, and star-shaped domains. Numerical experiments show that GeoDiv-KAN achieves higher predictive accuracy than PINN, PIKAN, RBF-KAN, and DF-RBFKAN across the considered benchmark problems. To establish practical relevance, GeoDiv-KAN is applied to two
	data-driven CFD cases: unsteady flow past a circular cylinder using the
	\href{https://doi.org/10.5281/zenodo.5039610}{Zenodo dataset} and
	turbulent RANS flow through a complex U-bend using the
	\href{https://huggingface.co/datasets/JDecke/ubend-cfd}{U-bend dataset}.
\end{abstract}

\begin{keyword}\justifying
	Scientific Machine Learning, Kolmogorov--Arnold Networks, Fractional Calculus, Navier--Stokes Equations, Complex Geometries.
\end{keyword}

	\end{frontmatter}
	
	\section{Introduction}\justifying
	Deep learning has become an important computational tool for function
	approximation, pattern recognition, and data-driven modelling
	\cite{goodfellow2016deep,raschka2019python,kumar2026deep}. Its ability to
	learn nonlinear relationships directly from data has led to successful
	applications in computer vision, natural language processing, and
	scientific computing
	\cite{voulodimos2018deep,cornelio2025neutrosophic}. In computational
	science, particular interest arises from the expressive power of neural
	networks and their ability to incorporate physical constraints into the
	learning process. This has encouraged the development of neural solvers that can complement conventional numerical
	methods and, in suitable settings, provide flexible mesh-free
	approximations for complex and high-dimensional systems
	\cite{zhang2022application,Rao2024TFColdPlasma, saini2026pinn}.
	
Many physical systems retain the influence of their previous states,
whereas classical integer-order models are local in time. Fractional
calculus provides a mathematical framework for describing such memory
through derivatives and integrals of non-integer order
\cite{podlubny1999introduction}. This framework has been successfully
used to model viscoelastic materials, anomalous diffusion, porous-media
transport, biological processes, financial dynamics and fluid flow
\cite{pandey2022numerical,jin2021monotonicity,han2024application,
	chatterjee2021fractional,chen2022fractional}. 
	Among these applications, fluid-flow modelling is particularly
	important. The motion of an incompressible fluid is classically
	described by the Navier--Stokes equations \cite{lukaszewicz2016navier}, which are widely used in
	aerodynamics, marine systems, biomedical transport, energy systems, and
	industrial processes \cite{wang2025aerodit, avallone2026metamaterials, hrizi2026inverse}. They provide the mathematical basis for analysing
	flow separation, wake formation, mixing, and pressure loss in engineering
	configurations.
	
	Some complex fluids, however, exhibit history-dependent behaviour that
	is not explicitly represented by the classical local-in-time
	formulation. To account for this temporal memory, the first-order time
	derivative in the momentum equations can be generalized using a
	fractional derivative, leading to the TFNSEs \cite{gao2024energy}. These equations retain the nonlinear convection, viscous
	diffusion, pressure, and incompressibility characteristics of the
	classical model while incorporating the influence of the flow history. The two-dimensional TFNSEs considered in this study are given by \cite{liu2024meshfree}
	\begin{equation}
		\label{eq:intro_tfnse}
		\begin{cases}
			\prescript{C}{}{D}_{t}^{\alpha}u
			+u\dfrac{\partial u}{\partial x}
			+v\dfrac{\partial u}{\partial y}
			-\dfrac{1}{Re}
			\left(
			\dfrac{\partial^{2}u}{\partial x^{2}}
			+\dfrac{\partial^{2}u}{\partial y^{2}}
			\right)
			+\dfrac{\partial p}{\partial x}
			=f_{1}(x,y,t,u,v),
			\\[1.2ex]
			\prescript{C}{}{D}_{t}^{\alpha}v
			+u\dfrac{\partial v}{\partial x}
			+v\dfrac{\partial v}{\partial y}
			-\dfrac{1}{Re}
			\left(
			\dfrac{\partial^{2}v}{\partial x^{2}}
			+\dfrac{\partial^{2}v}{\partial y^{2}}
			\right)
			+\dfrac{\partial p}{\partial y}
			=f_{2}(x,y,t,u,v),
			\\[1.2ex]
			\dfrac{\partial u}{\partial x}
			+\dfrac{\partial v}{\partial y}=0,
		\end{cases}
		\qquad
		(x,y,t)\in\Omega\times[0,T].
	\end{equation}
	The corresponding initial conditions are
	\begin{equation}
		\label{eq:intro_initial}
		u(x,y,0)=u_{0}(x,y),
		\qquad
		v(x,y,0)=v_{0}(x,y),
		\qquad (x,y)\in\Omega,
	\end{equation}
	while the boundary conditions are prescribed as
	\begin{equation}
		\label{eq:intro_boundary}
		u(x,y,t)=g_{u}(x,y,t),
		\qquad
		v(x,y,t)=g_{v}(x,y,t),
		\qquad
		(x,y,t)\in\partial\Omega\times[0,T],
	\end{equation}
	where, \(u(x,y,t)\) and \(v(x,y,t)\) denote the velocity components,
	\(p(x,y,t)\) is the pressure, \(Re\) represents the Reynolds number, and
	\(\prescript{C}{}{D}_{t}^{\alpha}\) denotes the Caputo derivative of
	order \(0<\alpha\leq1\). The functions \(f_{1}\) and \(f_{2}\) are the
	source terms. For \(\alpha=1\), the system reduces to the classical
	incompressible Navier--Stokes equations. For \(0<\alpha<1\), however,
	the Caputo derivative introduces a dependence on the previous states of
	the flow, which makes the numerical treatment more demanding.
	
	Several established numerical methods, including finite difference \cite{li2012finite}, Homotopy analysis technique \cite{rao2024numerical}
	finite element \cite{shang2022stabilized}, finite volume \cite{rawat2026finite}, and spectral schemes \cite{zheng2019spectral}, have been employed
	to solve TFNSEs. These methods discretize the spatial domain and
	approximate the Caputo derivative using techniques such as the \(L1\)
	scheme or convolution quadrature. Because the fractional derivative
	depends on the solution history, the numerical procedure must retain
	and process information from previous time levels. This history dependence makes long-time simulations computationally
	expensive and memory intensive.
	
	The computational requirements of conventional solvers have encouraged
	the use of neural networks for approximating differential equation
	solutions \cite{lagaris1998}. In this
	approach, the spatial and temporal coordinates are supplied directly to
	the network, which represents the unknown solution fields as continuous
	functions. PINNs  \cite{raissi2019} extend this idea by
	incorporating the governing equations, initial conditions, and boundary
	conditions into the loss function. Consequently,
	PINNs can approximate the solution without constructing a conventional
	computational mesh or generating labelled data throughout the domain. Pang et al.~\cite{pang2019} proposed the fractional PINN (fPINN) for solving fractional differential equations, in which the fractional derivatives are approximated using the Grünwald--Letnikov formula. Oldenburg et al. \cite{oldenburg2022gapinn} proposed a geometry-aware PINN for Navier–Stokes equations on irregular non-parameterized geometries, combining variational autoencoder-based geometry encoding with physics-driven training and hard boundary enforcement. Shi et al. \cite{shi2024novel} applied fPINNs to the time-fractional
	Huxley equation, demonstrating their capability for nonlinear reaction–diffusion systems. Kumar et al. \cite{kumar2026robust} proposed a robust Hermite interpolation neural network, in which the neural network incorporates
	additional connections, known as lateral connections. Wang and Zhang \cite{wang2026deep} proposed Sumudu transform
	PINNs for solving time fractional equations for special stochastic processes in foreign exchange market.
	Liu et al. \cite{liu2025kan}  introduced KANs with learnable edge-based activation functions, achieving improved accuracy, efficiency, and interpretability compared with conventional multilayer perceptrons. Alireza Afzal Aghaei \cite{aghaei2025kantrol} proposed KANtrol for continuous-time optimal control, combining KANs with efficient fractional derivative approximation and achieving higher accuracy. 	Kumar
	and Vats \cite{Kumar2025KAN} introduced a deep KAN framework for solving fractional partial differential equations. 
	Cui et al. \cite{cui2025physics} proposed a physics-informed KAN (PIKAN) for solving Navier–Stokes equations and showed that suitable network architectures can achieve high accuracy while addressing multi-output information bottlenecks and training stability. Zhao and  Wang \cite{zhao2025fpinns} developed fPINN and extended fractional PINNs for fractional Navier–Stokes equations, with rigorous convergence analysis and numerical validation for time- and space-fractional cases.  Zhang et al. \cite{zhang2025p2pinn} introduced a pseudo-time stepping and parameterized PINN for Navier–Stokes equations, enabling efficient solutions across varying parameters. Aghaei et al. \cite{aghaei2026rkan} proposed rational KANs using Padé and rational Jacobi basis functions, demonstrating strong performance in deep learning and physics-informed learning tasks. Zhang
	 et al. \cite{zhang2026turbulence} developed a PINN-based data-driven Reynolds-averaged Navier–Stokes closure for flow past a circular cylinder, improving predictions of mean flow and turbulence statistics over conventional turbulence models. Hu et al. \cite{hu2026adaptive} introduced an adaptive error estimator for variational PINNs applied to stationary Navier–Stokes equations, enabling reliable detection of sharp gradients and singularities while improving solution accuracy.
	
Despite these advances, the combined treatment of temporal memory,
complex geometry, and incompressibility in TFNSEs remains insufficiently
explored. Many coordinate-based PINNs represent the domain geometry
implicitly through interior and boundary collocation points, without
directly encoding a point's relation to the boundary. Although
geometry-aware variants improve this representation, TFNSEs additionally
require the evaluation of the nonlocal Caputo derivative, the resolution
of nonlinear momentum coupling, and the enforcement of continuity.
Moreover, imposing the continuity equation as a soft loss term does not
guarantee an exactly divergence-free velocity field. These limitations
motivate a unified geometry-aware and divergence-free architecture for
solving TFNSEs on complex domains.
	
In this work, we develop GeoDiv-KAN, a geometry-gated, divergence-free Gaussian RBF based KAN. The model enriches the spatial and temporal coordinates with geometry-dependent features and uses a learnable gate to control their
contribution to the network representation. Gaussian radial basis functions parameterize the KAN edge functions, providing smooth and
flexible approximations of the solution fields. Instead of predicting
 the two velocity components independently, GeoDiv-KAN predicts a streamfunction and obtains the velocity as
	\(u=\partial\psi/\partial y\) and
	\(v=-\partial\psi/\partial x\). This construction satisfies the
	incompressibility condition by design, while pressure is predicted
	through a separate network output.\\	
	The main contributions of this study are summarized as follows:
	\begin{itemize}
		\item A geometry-aware KAN architecture is developed by integrating
		Gaussian RBF edge functions with a learnable feature gate that explicitly
		incorporates domain-shape information into the network representation.
		
		\item A streamfunction-based output formulation is incorporated to
		satisfy the incompressibility condition exactly, without introducing
		an additional divergence penalty in the loss function.
		
		\item GeoDiv-KAN is applied to the TFNSEs on square, circular , and star-shaped domains and is compared
		with PINN, PIKAN, RBF-KAN, and DF-RBFKAN using standard error measures.
		
		\item The practical applicability of the model is examined through
		two data-driven CFD reconstruction problems: unsteady flow past a
		fixed circular cylinder at \(Re=100\) and turbulent RANS flow through
		a U-bend at \(Re=40{,}000\).
	\end{itemize}
The remainder of this paper is organized as follows. Section~\ref{sec3} introduces the mathematical
preliminaries and essential definitions. Section~\ref{sec4} presents
the methodology of the proposed GeoDiv-KAN solver.
Section~\ref{sec5} reports the numerical results and comparisons with
the baseline methods. Section~\ref{sec6} implement GeoDiv-KAN on two
CFD datasets. Finally, Section~\ref{sec7} summarizes the main findings,
discusses the limitations, and outlines directions for future research.

\section{Preliminary}\label{sec3}
\begin{definition}\cite{podlubny1999introduction}
	Let $f\in L^{1}(0,t)$ and $\alpha>0$, the Riemann--Liouville fractional integral of order $\alpha$ is given by
	\begin{equation}
		\mathcal{I}_{0}^{\alpha} f(t) 
		= \frac{1}{\Gamma(\alpha)} 
		\int_{0}^{t} (t-\tau)^{\alpha-1} f(\tau)\,d\tau,
	\end{equation}
	with the limiting case
	\begin{equation}
		\mathcal{I}_{0}^{0} f(t) = f(t).
	\end{equation}
\end{definition}

\begin{definition}\cite{podlubny1999introduction}
	Let $f \in AC^{k}[0,\infty)$ be a $k$-times continuously differentiable function.  
	The Caputo fractional derivative of order $\alpha>0$ is defined as
	\begin{equation}
		{}^{C}D_{0}^{\alpha} f(t) 
		= \frac{1}{\Gamma(k-\alpha)} 
		\int_{0}^{t} (t-\tau)^{k-\alpha-1} f^{(k)}(\tau)\,d\tau,
	\end{equation}
	where $(k-1)\leq\alpha < k$, with $k \in \mathbb{N}$.
\end{definition}

\subsection{Kolmogorov--Arnold Representation}
The theoretical motivation for KAN follows from the
Kolmogorov--Arnold representation theorem.
\begin{theorem}[Kolmogorov--Arnold representation theorem \cite{kolmogorov1957representations}]
	\label{thm:kolmogorov_arnold}
	Let \(G:[0,1]^n\rightarrow\mathbb{R}\) be a continuous multivariate
	function. Then \(G\) admits a representation of the form
	\begin{equation}
		G(\boldsymbol{z})
		=
		\sum_{q=0}^{2n}
		\Phi_q
		\left(
		\sum_{j=1}^{n}\phi_{qj}(z_j)
		\right),
		\qquad
		\boldsymbol{z}
		=
		(z_1,\ldots,z_n)^{\mathsf T}\in[0,1]^n,
		\label{eq:kar_theorem}
	\end{equation}
	where \(\phi_{qj}:[0,1]\rightarrow\mathbb{R}\) are continuous univariate
	inner functions and
	\(\Phi_q:\mathbb{R}\rightarrow\mathbb{R}\) are continuous univariate
	outer functions.
\end{theorem}
Theorem~\ref{thm:kolmogorov_arnold} states that a continuous multivariate
mapping can be represented through sums and compositions of univariate
functions. Motivated by this result, KAN replaces the fixed scalar
activations used in conventional neural networks with learnable univariate
functions defined on network edges \cite{liu2025kan}. 
\subsection{Kolmogorov--Arnold Network Layer}
Let
\(\boldsymbol{h}^{(\ell)}
\in\mathbb{R}^{n_\ell}\)
denote the input to layer \(\ell\). A KAN layer maps
\(\boldsymbol{h}^{(\ell)}\) to
\(\boldsymbol{h}^{(\ell+1)}
\in\mathbb{R}^{n_{\ell+1}}\) according to
\begin{equation}
	h_i^{(\ell+1)}
	=
	\sum_{j=1}^{n_\ell}
	\phi_{ij}^{(\ell)}
	\left(h_j^{(\ell)}\right),
	\qquad
	i=1,\ldots,n_{\ell+1},
	\label{eq:kan_component}
\end{equation}
where
\(\phi_{ij}^{(\ell)}:\mathbb{R}\rightarrow\mathbb{R}\)
is a learnable univariate edge function.\\
In matrix form, define the functional matrix
\begin{equation}
	\boldsymbol{\Phi}^{(\ell)}
	=
	\begin{bmatrix}
		\phi_{11}^{(\ell)} & \cdots & \phi_{1n_\ell}^{(\ell)}\\
		\vdots             & \ddots & \vdots\\
		\phi_{n_{\ell+1}1}^{(\ell)}
		& \cdots
		& \phi_{n_{\ell+1}n_\ell}^{(\ell)}
	\end{bmatrix}.
\end{equation}
The corresponding layer operation is written as
\begin{equation}
	\boldsymbol{h}^{(\ell+1)}
	=
	\boldsymbol{\Phi}^{(\ell)}
	\odot_{\mathrm{KAN}}
	\boldsymbol{h}^{(\ell)},
	\label{eq:kan_matrix}
\end{equation}
where \(\odot_{\mathrm{KAN}}\) denotes functional matrix application,
defined componentwise by Eq.~\eqref{eq:kan_component}. This notation
distinguishes a KAN layer from conventional matrix multiplication because
each matrix entry is a univariate function rather than a scalar weight.\\
For an \(L\)-layer KAN, the complete mapping is
\begin{equation}
	\mathcal{K}_{\boldsymbol{\Theta}}
	=
	\boldsymbol{\Phi}^{(L-1)}
	\circ
	\boldsymbol{\Phi}^{(L-2)}
	\circ\cdots\circ
	\boldsymbol{\Phi}^{(0)},
\end{equation}
where \(\boldsymbol{\Theta}\) denotes the collection of all trainable
parameters.

\subsection{Gaussian Radial Basis Edge Functions}

In the proposed model, the learnable KAN edge functions are parameterized
using Gaussian radial basis functions \cite{chao2026rbfkan}. For an input \(z\), the edge
function connecting node \(j\) in layer \(\ell\) to node \(i\) in layer
\(\ell+1\) is defined as
\begin{equation}
	\phi_{ij}^{(\ell)}(z)
	=
	b_{ij}^{(\ell)}
	+
	\sum_{k=1}^{K}
	c_{ijk}^{(\ell)}
	\exp
	\left[
	-\frac{
		\left(z-\mu_{ijk}^{(\ell)}\right)^2
	}{
		2\left(\sigma_{ijk}^{(\ell)}\right)^2
	}
	\right],
	\label{eq:gaussian_rbf_edge}
\end{equation}
where \(K\) is the number of Gaussian basis functions,
\(c_{ijk}^{(\ell)}\) is the expansion coefficient,
\(\mu_{ijk}^{(\ell)}\) is the centre,
\(\sigma_{ijk}^{(\ell)}>0\) is the width, and
\(b_{ij}^{(\ell)}\) is the edge bias.\\
The corresponding Gaussian basis vector is defined as
\begin{equation}
	\boldsymbol{\rho}_{ij}^{(\ell)}(z)
	=
	\begin{bmatrix}
		\exp\!\left(
		-\dfrac{(z-\mu_{ij1}^{(\ell)})^2}
		{2(\sigma_{ij1}^{(\ell)})^2}
		\right)\\
		\vdots\\
		\exp\!\left(
		-\dfrac{(z-\mu_{ijK}^{(\ell)})^2}
		{2(\sigma_{ijK}^{(\ell)})^2}
		\right)
	\end{bmatrix}.
\end{equation}
Consequently, Eq.~\eqref{eq:gaussian_rbf_edge} can be written compactly as
\begin{equation}
	\phi_{ij}^{(\ell)}(z)
	=
	b_{ij}^{(\ell)}
	+
	\left(
	\boldsymbol{c}_{ij}^{(\ell)}
	\right)^{\mathsf T}
	\boldsymbol{\rho}_{ij}^{(\ell)}(z).
\end{equation}
%This parameterization yields smooth and differentiable edge functions,
%enabling the first- and second-order derivatives required to evaluate the
%governing-equation residuals.
\section{Methodology}\label{sec4}
\label{sec:proposed_tfnse_gg_dfrbfkan}
\begin{figure}[H]
	\centering	\includegraphics[width=1.0\textwidth]{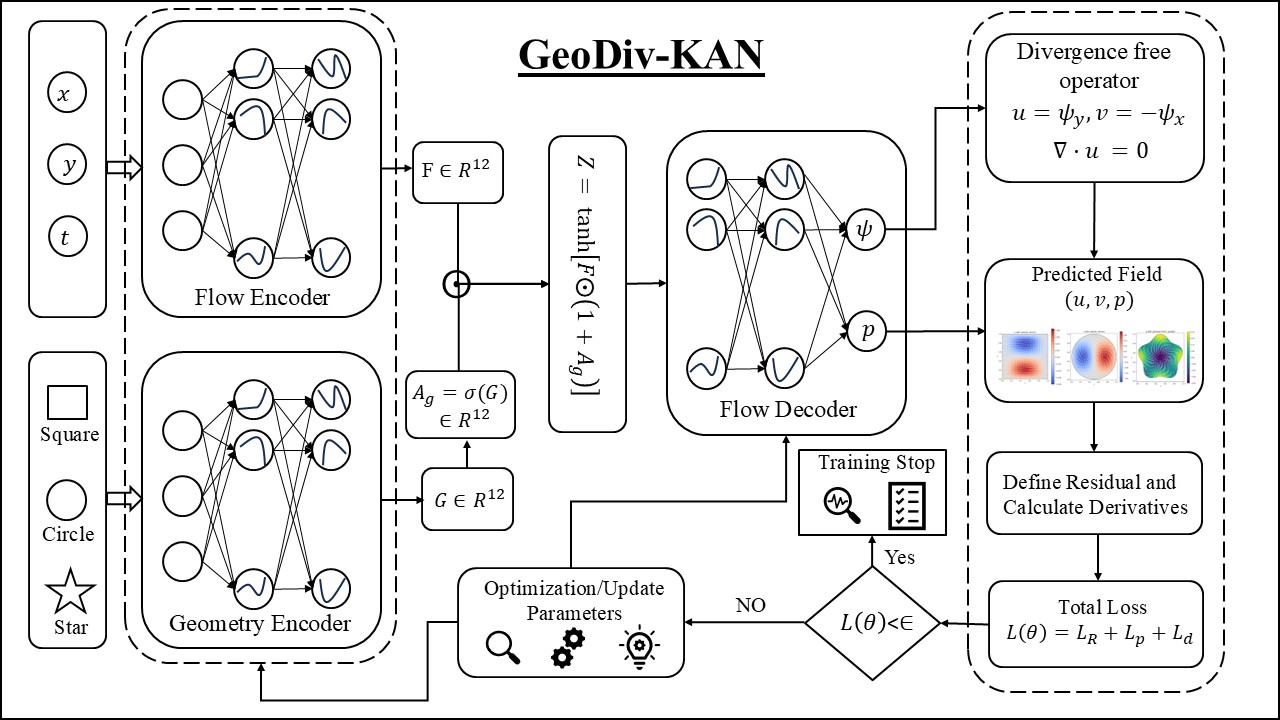}
	\caption{Architecture and training workflow of the proposed GeoDiv-KAN solver.}
	\label{M-D_KAN}
\end{figure}

This section presents the mathematical formulation and training
procedure of GeoDiv-KAN for solving the two-dimensional incompressible
TFNSEs. As illustrated in Fig.~\ref{M-D_KAN}, the architecture combines
Gaussian RBF-based KAN layers, geometry-dependent feature gating, and a
streamfunction representation that preserves incompressibility by
construction. Let \(\Omega\subset\mathbb{R}^{2}\) be a bounded spatial
domain and \(t\in[0,T]\) denote time. The governing TFNSE system is
written as
\begin{equation}
	\begin{cases}
		{}^{C}D_{t}^{\alpha}u
		+u\displaystyle\frac{\partial u}{\partial x}
		+v\displaystyle\frac{\partial u}{\partial y}
		-\displaystyle\frac{1}{Re}
		\left(
		\frac{\partial^{2}u}{\partial x^{2}}
		+\frac{\partial^{2}u}{\partial y^{2}}
		\right)
		+\displaystyle\frac{\partial p}{\partial x}
		=f_{1}(x,y,t,u,v),
		\\[1.2ex]
		{}^{C}D_{t}^{\alpha}v
		+u\displaystyle\frac{\partial v}{\partial x}
		+v\displaystyle\frac{\partial v}{\partial y}
		-\displaystyle\frac{1}{Re}
		\left(
		\frac{\partial^{2}v}{\partial x^{2}}
		+\frac{\partial^{2}v}{\partial y^{2}}
		\right)
		+\displaystyle\frac{\partial p}{\partial y}
		=f_{2}(x,y,t,u,v),
		\\[1.2ex]
		\displaystyle\frac{\partial u}{\partial x}
		+\displaystyle\frac{\partial v}{\partial y}=0,
	\end{cases}
	\qquad
	(x,y,t)\in\Omega\times[0,T].
\end{equation}
For $0<\alpha<1$,
\begin{equation}
	{}^{C}_{0}D_t^\alpha q=
	\frac{1}{\Gamma(1-\alpha)}\int_0^t(t-s)^{-\alpha}q_s(s)\,\mathrm{d}s,
\end{equation}
whereas the integer-order limit is \( {}^{C}_{0}D_t^1q=q_t \). The normalized coordinate is
\begin{equation}
	\widehat{\mathbf{x}}=
	\begin{bmatrix}
			2/(x_{\max}-x_{\min})&0&0\\
			0&2/(y_{\max}-y_{\min})&0\\
			0&0&2/T
	\end{bmatrix}
	\left(\begin{bmatrix}x\\y\\t\end{bmatrix}-
	\begin{bmatrix}x_{\min}\\y_{\min}\\0\end{bmatrix}\right)
	-\begin{bmatrix}1\\1\\1\end{bmatrix}
	=\begin{bmatrix}\xi\\\eta\\\tau\end{bmatrix}.
\end{equation}
Thus, a batch is $\widehat{\mathbf{X}}=[\boldsymbol\xi,\boldsymbol\eta,\boldsymbol\tau]\in\mathbb{R}^{N\times3}$.\\
For an RBF-KAN layer with $m$ inputs, $n$ outputs, and $K$ bases per edge,
\begin{equation}
	c_k=-1+\frac{2(k-1)}{K-1},\qquad h=s_h\frac{2}{K-1},\qquad
	\rho_{ik}(z_i)=\exp\!\left[-\frac12\left(\frac{z_i-c_k}{h}\right)^2\right].
\end{equation}
The complete basis matrix is
\begin{equation}
	\boldsymbol{\Phi}(\mathbf{z})=
	\begin{bmatrix}
		\rho_{11}(z_1)&\cdots&\rho_{1K}(z_1)\\
		\vdots&\ddots&\vdots\\
		\rho_{m1}(z_m)&\cdots&\rho_{mK}(z_m)
	\end{bmatrix}\in\mathbb{R}^{m\times K}.
\end{equation}
The edge function is $\varphi_{ji}(z_i)=b_{ji}\operatorname{SiLU}(z_i)+s_{ji}\sum_{k=1}^K a_{jik}\rho_{ik}(z_i)$. Defining
\begin{equation}
	\mathbf{B}=[b_{ji}]\in\mathbb{R}^{n\times m},\quad
	\mathbf{S}=[s_{ji}]\in\mathbb{R}^{n\times m},\quad
	\mathbf{A}^{(k)}=[a_{jik}]\in\mathbb{R}^{n\times m},
\end{equation}
and $\boldsymbol\rho^{(k)}=[\rho_{1k},\ldots,\rho_{mk}]^{\mathsf T}$ gives
\begin{equation}
	\boxed{\mathcal{K}_{m,n}(\mathbf{z})=
		\mathbf{B}\operatorname{SiLU}(\mathbf{z})
		+\sum_{k=1}^K(\mathbf{A}^{(k)}\odot\mathbf{S})\boldsymbol\rho^{(k)}}.
	\label{eq:main_rbfkan}
\end{equation}
For $\mathbf{Z}\in\mathbb{R}^{N\times m}$,
\begin{equation}
	\mathbf{Y}=\operatorname{SiLU}(\mathbf{Z})\mathbf{B}^{\mathsf T}
	+\mathbf{\Phi}_{\rm flat}\mathbf{W}_{\rm RBF}^{\mathsf T},
	\quad \mathbf{\Phi}_{\rm flat}\in\mathbb{R}^{N\times mK},\quad
	\mathbf{W}_{\rm RBF}\in\mathbb{R}^{n\times mK}.
\end{equation}
GeoDiv-KAN uses a flow encoder $3\rightarrow12\rightarrow12$, a geometry encoder $4\rightarrow12\rightarrow12$, and a decoder $12\rightarrow12\rightarrow2$, with $K=8$ and $s_h=1$. On the square domain, the geometry vector is
\begin{equation}
	\mathbf{g}_{\square}=
	\begin{bmatrix}
		(1-\xi^2)(1-\eta^2)&-2\xi(1-\eta^2)&-2\eta(1-\xi^2)&0
	\end{bmatrix}^{\mathsf T},
\end{equation}
whereas the circular domain centred at $(0.5,0.5)$ is represented by
\begin{equation}
	\mathbf{g}_{\circ}=
	\begin{bmatrix}1-\xi^2-\eta^2&-\xi&-\eta&1\end{bmatrix}^{\mathsf T}.
\end{equation}
For the normalized star boundary $r_b(\vartheta)=0.9+0.1\sin(5\vartheta)$, let $r^2=\xi^2+\eta^2$ and
\begin{equation}
	H_5=5\xi^4\eta-10\xi^2\eta^3+\eta^5=r^5\sin(5\vartheta),\qquad
	\phi_\star=0.9r^5+0.1H_5-r^6.
\end{equation}
The nonsingular Cartesian descriptors are
\begin{align}
	\phi_{\star,\xi}&=4.5\xi r^3+0.1(20\xi^3\eta-20\xi\eta^3)-6\xi r^4,\\
	\phi_{\star,\eta}&=4.5\eta r^3+0.1(5\xi^4-30\xi^2\eta^2+5\eta^4)-6\eta r^4,\\
	\Delta\phi_\star&=22.5r^3-36r^4,
\end{align}
and
\begin{equation}
	\mathbf{g}_{\star}=
	\begin{bmatrix}
		\tanh(4\phi_\star)&\tanh(\phi_{\star,\xi})&
		\tanh(\phi_{\star,\eta})&\tanh(\Delta\phi_\star)
	\end{bmatrix}^{\mathsf T}.
\end{equation}
For $N$ collocation points, $\mathbf{G}_{\Omega}\in\mathbb{R}^{N\times4}$ collects the relevant geometry vectors. The encoded matrices are
\begin{equation}
	\mathbf{F}=\mathcal{K}^{f}_{12,12}(\mathcal{K}^{f}_{3,12}(\widehat{\mathbf{X}}))
	\in\mathbb{R}^{N\times12},\qquad
	\mathbf{G}=\mathcal{K}^{g}_{12,12}(\mathcal{K}^{g}_{4,12}(\mathbf{G}_{\Omega}))
	\in\mathbb{R}^{N\times12}.
\end{equation}
The complete gate and fused representation are
\begin{equation}
	\mathbf{A}_g=\sigma(\mathbf{G})=
	\begin{bmatrix}
		a_{11}&\cdots&a_{1,12}\\
		\vdots&\ddots&\vdots\\
		a_{N1}&\cdots&a_{N,12}
	\end{bmatrix}\in(0,1)^{N\times12},
\end{equation}
\begin{equation}
	\boxed{\mathbf{Z}=\tanh[\mathbf{F}\odot(\mathbf{1}_{N\times12}+\mathbf{A}_g)]},
	\qquad Z_{ij}=\tanh[F_{ij}(1+A_{g,ij})].
\end{equation}
%The decoder returns $\mathbf{N}_\theta=[N_\psi,N_p]^{\mathsf T}$. Since an $m\rightarrow n$ RBF-KAN layer has $nm(K+2)$ trainable coefficients,
%\begin{align}
%	P_f&=10(12\cdot3+12\cdot12)=1800,\\
%	P_g&=10(12\cdot4+12\cdot12)=1920,\\
%	P_d&=10(12\cdot12+2\cdot12)=1680,
%\end{align}
%and $\boxed{P_{\mathrm{total}}=5400}$.
Initial conditions are imposed strongly through
\begin{equation}
	\boxed{
		\begin{bmatrix}\widehat\psi_\Theta(x,y,t)\\\widehat p_\Theta(x,y,t)\end{bmatrix}
		=
		\begin{bmatrix}\psi_0(x,y)\\p_0(x,y)\end{bmatrix}
		+t\begin{bmatrix}N_\psi(\xi,\eta,\tau)\\N_p(\xi,\eta,\tau)\end{bmatrix}}.
	\label{eq:main_trial_solution}
\end{equation}
Since the network correction in Eq.~\eqref{eq:main_trial_solution} is
multiplied by \(t\), it vanishes at \(t=0\). Therefore,
\[
\widehat{\psi}_{\Theta}(x,y,0)=\psi_0(x,y),
\qquad
\widehat{p}_{\Theta}(x,y,0)=p_0(x,y).
\]
Consequently, the velocity components derived from the streamfunction
also satisfy the prescribed initial conditions,
$
\widehat{u}(x,y,0)=u_0(x,y),
\quad
\widehat{v}(x,y,0)=v_0(x,y).
$
The velocity components and the diagnostic vorticity are subsequently
computed as
\begin{equation}
	\boxed{\widehat u=\widehat\psi_{\Theta,y},\qquad
		\widehat v=-\widehat\psi_{\Theta,x},\qquad
		\widehat\omega=-\Delta\widehat\psi_\Theta}.
\end{equation}
Equivalently,
\begin{equation*}
	\begin{bmatrix}\widehat u\\\widehat v\end{bmatrix}
	=\underbrace{\begin{bmatrix}0&1\\-1&0\end{bmatrix}}_{\mathbf{J}}
	\begin{bmatrix}\widehat\psi_{\Theta,x}\\\widehat\psi_{\Theta,y}\end{bmatrix},
\end{equation*}
which gives $\widehat u_x+\widehat v_y=\widehat\psi_{\Theta,yx}-\widehat\psi_{\Theta,xy}=0$. Hence, incompressibility is exact and does not require a penalty. Vorticity is not an independent network output and no vorticity-transport or compatibility equation is trained.\\
On $t_n=n\Delta t$, $n=0,\ldots,N_t$, the L1 formula is
\begin{equation}
	\left.{}^{C}_{0}D_t^\alpha q\right|_{t_n}
	\approx\frac{\Delta t^{-\alpha}}{\Gamma(2-\alpha)}
	\sum_{k=0}^{n-1}b_{n-1-k}(q^{k+1}-q^k),\qquad
	b_j=(j+1)^{1-\alpha}-j^{1-\alpha}.
\end{equation}
For $\mathbf{q}=[q^0,\ldots,q^{N_t}]^{\mathsf T}$, $\mathbf{d}_\alpha=\mathbf{L}_\alpha\mathbf{q}$, with the complete matrix
\begin{equation}
	\mathbf{L}_\alpha=\frac{\Delta t^{-\alpha}}{\Gamma(2-\alpha)}
	\begin{bmatrix}
		-b_0&b_0&0&\cdots&0\\
		-b_1&b_1-b_0&b_0&\cdots&0\\
		-b_2&b_2-b_1&b_1-b_0&\ddots&\vdots\\
		\vdots&\vdots&\ddots&\ddots&b_0\\
		-b_{N_t-1}&b_{N_t-1}-b_{N_t-2}&\cdots&b_1-b_0&b_0
	\end{bmatrix}
	\in\mathbb{R}^{N_t\times(N_t+1)}.
	\label{eq:main_l1_matrix}
\end{equation}
If $\mathbf{Q}=[q_i^n]\in\mathbb{R}^{N_s\times(N_t+1)}$ contains all spatial points and times, then the code evaluates
\begin{equation}
	\boxed{\mathbf{D}_t^\alpha\mathbf{Q}=\mathbf{Q}\mathbf{L}_\alpha^{\mathsf T}}
	\in\mathbb{R}^{N_s\times N_t}.
\end{equation}
Automatic differentiation supplies all spatial derivatives. At every non-initial interior point, the two trained residuals are
\begin{align}
	\label{L_1}
	\mathcal{R}_u&=\mathbf{D}_t^\alpha\widehat u+
	\widehat u\,\widehat u_x+\widehat v\,\widehat u_y
	-Re^{-1}(\widehat u_{xx}+\widehat u_{yy})+\widehat p_x-f_1,\\ \nonumber
	\mathcal{R}_v&=\mathbf{D}_t^\alpha\widehat v+
	\widehat u\,\widehat v_x+\widehat v\,\widehat v_y
	-Re^{-1}(\widehat v_{xx}+\widehat v_{yy})+\widehat p_y-f_2.
\end{align}
Exact fields are used only to construct initial values, boundary values, and evaluation references; they are not supplied as interior observations. For the square,
\begin{align}
	u^*&=2(x^2-2x^3+x^4)(y-3y^2+2y^3)(1+t)^2,\\ \nonumber
	v^*&=-2(y^2-2y^3+y^4)(x-3x^2+2x^3)(1+t)^2,\\
	p^*&=[x^2-2x^3+x^4-(y^2-2y^3+y^4)](1+t)^2, \nonumber
\end{align}
with
\begin{equation}
	\psi^*=(x^2-2x^3+x^4)(y^2-2y^3+y^4)(1+t)^2.
\end{equation}
For the circle and star,
\begin{align}
	u^*&=\tfrac12\sin^2(\pi x)\sin(2\pi y)(1+t)^2,\\ \nonumber
	v^*&=-\tfrac12\sin^2(\pi y)\sin(2\pi x)(1+t)^2,\\
	p^*&=-\tfrac1{16}\cos(2\pi x)\cos(2\pi y)(1+t)^2, \nonumber
\end{align}
and
\begin{equation}
	\psi^*=\left[-\frac{\sin^2(\pi x)\cos(2\pi y)}{4\pi}
	-\frac{\cos(2\pi x)}{8\pi}\right](1+t)^2.
\end{equation}
For $N_r$ residual points, define
\begin{equation}
	\mathbf{r}_u=
	\begin{bmatrix}\mathcal{R}_{u,1}&\cdots&\mathcal{R}_{u,N_r}\end{bmatrix}^{\mathsf T},
	\qquad
	\mathbf{r}_v=
	\begin{bmatrix}\mathcal{R}_{v,1}&\cdots&\mathcal{R}_{v,N_r}\end{bmatrix}^{\mathsf T}.
\end{equation}
Then
\begin{equation}
	\mathcal{L}_u^{\mathrm{raw}}=\frac{\mathbf{r}_u^{\mathsf T}\mathbf{r}_u}{N_r},
	\qquad
	\mathcal{L}_v^{\mathrm{raw}}=\frac{\mathbf{r}_v^{\mathsf T}\mathbf{r}_v}{N_r}.
\end{equation}
For the circle and star, fixed scales are computed once from the initial network,
\begin{equation}
	s_u=\max\!\left(\sqrt{\mathcal{L}_u^{\mathrm{raw}}(\boldsymbol\Theta_0)},0.1\right),
	\qquad
	s_v=\max\!\left(\sqrt{\mathcal{L}_v^{\mathrm{raw}}(\boldsymbol\Theta_0)},0.1\right),
\end{equation}
and
\begin{equation}\label{L_u_v}
	\mathcal{L}_u=\mathcal{L}_u^{\mathrm{raw}}/s_u^2,\qquad
	\mathcal{L}_v=\mathcal{L}_v^{\mathrm{raw}}/s_v^2.
\end{equation}
For the square, the raw residuals are used, which is equivalent to $s_u=s_v=1$. For $N_b$ boundary space--time points,
\begin{equation}
	\widehat{\mathbf{U}}_b=
	\begin{bmatrix}
		\widehat u_1&\widehat v_1\\
		\vdots&\vdots\\
		\widehat u_{N_b}&\widehat v_{N_b}
	\end{bmatrix},\qquad
	\mathbf{U}_b=
	\begin{bmatrix}
		u_{b,1}&v_{b,1}\\
		\vdots&\vdots\\
		u_{b,N_b}&v_{b,N_b}
	\end{bmatrix}\in\mathbb{R}^{N_b\times2},
\end{equation}
and
\begin{equation}\label{L_b}
	\mathcal{L}_b=\frac{1}{2N_b}
	\|\widehat{\mathbf{U}}_b-\mathbf{U}_b\|_F^2.
\end{equation}
Pressure uniqueness is imposed at one anchor over all times. With
\begin{equation}
	\widehat{\mathbf{p}}_a=
	\begin{bmatrix}
		\widehat p(x_a,y_a,t_0)&\cdots&\widehat p(x_a,y_a,t_{N_t})
	\end{bmatrix}^{\mathsf T},
\end{equation}
the pressure loss is
\begin{equation}\label{L_p}
	\mathcal{L}_p=\frac{1}{N_t+1}
	\|\widehat{\mathbf{p}}_a-\mathbf{p}_a^*\|_2^2.
\end{equation}
The anchor is $(0.5,0.5)$ for the square and circle and $(0.5,0.6)$ for the star. For the square, smoothness of the effective RBF coefficients $q_{jik}=s_{ji}a_{jik}$ is additionally controlled by
\begin{equation}
	\label{L_k}
	\mathcal{L}_{\kappa}=
	\sum_{\ell}\frac{1}{n_\ell m_\ell(K-2)}
	\sum_{j=1}^{n_\ell}\sum_{i=1}^{m_\ell}\sum_{k=1}^{K-2}
	(q_{ji,k+2}^{(\ell)}-2q_{ji,k+1}^{(\ell)}+q_{jik}^{(\ell)})^2,
\end{equation}
where $\ell$ covers all six RBF-KAN layers.Using Eqs.~\eqref{L_u_v}, \eqref{L_b}, \eqref{L_p}, and \eqref{L_k},
the overall training objective is defined as
\begin{equation}
	\boxed{\mathcal{L}(\boldsymbol\Theta)=
		\lambda_m(\mathcal{L}_u+\mathcal{L}_v)
		+\lambda_b\mathcal{L}_b+\lambda_p\mathcal{L}_p
		+\lambda_\kappa\mathcal{L}_\kappa}.
	\label{eq:main_total_loss}
\end{equation}
The trainable vector $\boldsymbol\Theta\in\mathbb{R}^{5400}$ is optimized by limited-memory BFGS. With $\mathbf{s}_k=\boldsymbol\Theta_{k+1}-\boldsymbol\Theta_k$, $\mathbf{y}_k=\nabla\mathcal{L}_{k+1}-\nabla\mathcal{L}_k$, and $\rho_k=(\mathbf{y}_k^{\mathsf T}\mathbf{s}_k)^{-1}$, the inverse-Hessian update is
\begin{equation}
	\mathbf{H}_{k+1}=
	(\mathbf{I}-\rho_k\mathbf{s}_k\mathbf{y}_k^{\mathsf T})
	\mathbf{H}_k
	(\mathbf{I}-\rho_k\mathbf{y}_k\mathbf{s}_k^{\mathsf T})
	+\rho_k\mathbf{s}_k\mathbf{s}_k^{\mathsf T},
\end{equation}
followed by
\begin{equation}
	\boldsymbol\Theta_{k+1}=\boldsymbol\Theta_k-
	\gamma_k\mathbf{H}_k\nabla\mathcal{L}(\boldsymbol\Theta_k).
\end{equation}
In L-BFGS, the action of \(\mathbf H_k\) is approximated
using a limited history of the correction pairs
\((\mathbf s_k,\mathbf y_k)\), thereby avoiding storage of the full
inverse-Hessian matrix. The optimization jointly updates the parameters
of the flow encoder, geometry encoder, feature gate, and decoder until
the prescribed convergence criterion is satisfied.

The main architectural differences between the proposed GeoDiv-KAN and
the baseline models are summarized in Table~\ref{model_comparison}.

\begin{table}[H]
	\centering
	\caption{Architectural features of the models considered in this study.}
	\label{model_comparison}
	\small
	\renewcommand{\arraystretch}{1.15}
	\setlength{\tabcolsep}{7pt}
	\begin{tabular}{lcccc}
		\toprule
		\textbf{Model} & \textbf{KAN edges} & \textbf{Gaussian RBFs} &
		\textbf{Divergence-free} & \textbf{Geometry gate} \\
		
		\midrule
		PINN       & No  & No  & No  & No  \\
		Base PIKAN & Yes & No  & No  & No  \\
		RBF-KAN    & Yes & Yes & No  & No  \\
		DF-RBFKAN  & Yes & Yes & Yes & No  \\
		GeoDiv-KAN & Yes & Yes & Yes & Yes \\
		\bottomrule
	\end{tabular}
\end{table}
\section{Numerical Experiments}\label{sec5}
To evaluate GeoDiv-KAN across complex geometries, we solve the two-dimensional time-fractional incompressible Navier--Stokes equations on square, circular, and star-shaped domains with initial and boundary conditions. The considered TFNSE system is expressed as follows \cite{liu2024meshfree}:
\begin{equation}
	\label{eq:tfnse_system_M}
	\begin{cases}
		{}^{C}D_{t}^{\alpha}u
		+u\displaystyle\frac{\partial u}{\partial x}
		+v\displaystyle\frac{\partial u}{\partial y}
		-\displaystyle\frac{1}{Re}
		\left(
		\frac{\partial^{2}u}{\partial x^{2}}
		+\frac{\partial^{2}u}{\partial y^{2}}
		\right)
		+\displaystyle\frac{\partial p}{\partial x}
		=f_{1}(x,y,t,u,v),
		\\[1.2ex]
		{}^{C}D_{t}^{\alpha}v
		+u\displaystyle\frac{\partial v}{\partial x}
		+v\displaystyle\frac{\partial v}{\partial y}
		-\displaystyle\frac{1}{Re}
		\left(
		\frac{\partial^{2}v}{\partial x^{2}}
		+\frac{\partial^{2}v}{\partial y^{2}}
		\right)
		+\displaystyle\frac{\partial p}{\partial y}
		=f_{2}(x,y,t,u,v),
		\\[1.2ex]
		\displaystyle\frac{\partial u}{\partial x}
		+\displaystyle\frac{\partial v}{\partial y}=0,
	\end{cases}
	\qquad
	(x,y,t)\in\Omega\times[0,T].
\end{equation}
where, \(u(x,y,t)\) and \(v(x,y,t)\) denote the velocity components,
\(p(x,y,t)\) represents the pressure,  \(Re\) is the Reynolds number, \({}^{C}D_t^\alpha\) denotes the Caputo fractional derivative
of order \(0<\alpha\leq 1\), and  \(f_1\) and \(f_2\) are source terms.

The prediction accuracy is evaluated using the absolute error, root-mean-square
error (RMSE), maximum absolute error, and relative \(L_2\) error:
\begin{equation}
	\begin{aligned}
		E_{q,i}^{\mathrm{abs}}
		&=\left|\widehat{q}_{i}-q_{i}\right|, \qquad
		\operatorname{RMSE}_{q}
		=\sqrt{\frac{1}{N}\sum_{i=1}^{N}
			\left(\widehat{q}_{i}-q_{i}\right)^{2}},\\
		E_{q}^{\max}
		&=\max_{1\leq i\leq N}
		\left|\widehat{q}_{i}-q_{i}\right|,\qquad
		E_{q}^{\mathrm{rel}}
		=\frac{\left\|\widehat{ q}- q\right\|_{2}}
		{\left\|\ q\right\|_{2}}
		=\sqrt{
			\frac{\sum_{i=1}^{N}\left(\widehat{q}_{i}-q_{i}\right)^{2}}
			{\sum_{i=1}^{N}q_{i}^{2}}
		},
		\qquad q\in\{u,v,p\}.
	\end{aligned}
	\label{eq:error_metrics}
\end{equation}
Here, \(q_i\) and \(\widehat q_i\) denote the reference and predicted values,
respectively, and \(N\) is the total number of evaluation points. All experiments were implemented in Python~3.11.6 using
PyTorch~2.12.0 and executed without GPU acceleration on a
13$^{th}$ generation Intel Core i7-1355U processor with 10 cores, and 16~GB RAM at Windows~11.
\begin{table}[H]
	\centering
	\caption{Final architectures and parameter counts of the comparison models for Square Case 1.}
	\label{tab:square_case1_architecture}
	\renewcommand{\arraystretch}{1.25}
	\setlength{\tabcolsep}{5pt}
	
	\begin{tabular}{lcccc}
		\toprule
		\textbf{Model} &
		\textbf{Layers / Width} &
		\textbf{Basis / Activation} &
		\textbf{Parameters} &
		\textbf{Outputs} \\
		\midrule
		
		PINN
		& $5 \times 35$
		& Tanh
		& 5,288
		& $(u,v,p)$ \\
		
		Base PIKAN
		& $4 \times 10$
		& Spline grid = 10
		& 5,400
		& $(u,v,p)$ \\
		
		RBF-KAN
		& $3 \times 15$
		& 8 RBF bases
		& 5,400
		& $(u,v,p)$ \\
		
		DF-RBFKAN
		& $3 \times 15$
		& 8 RBF bases
		& 5,250
		& $(\psi,p)$ \\
		
		GeoDiv-KAN
		& Branched width = 12
		& 8 RBF bases
		& 5,400
		& $(\psi,p)$ \\
		
		\bottomrule
	\end{tabular}
\end{table}

\subsection{\textbf{Case 1: Square Domain}}

\begin{figure}[H]
	\centering	\includegraphics[width=0.5\textwidth]{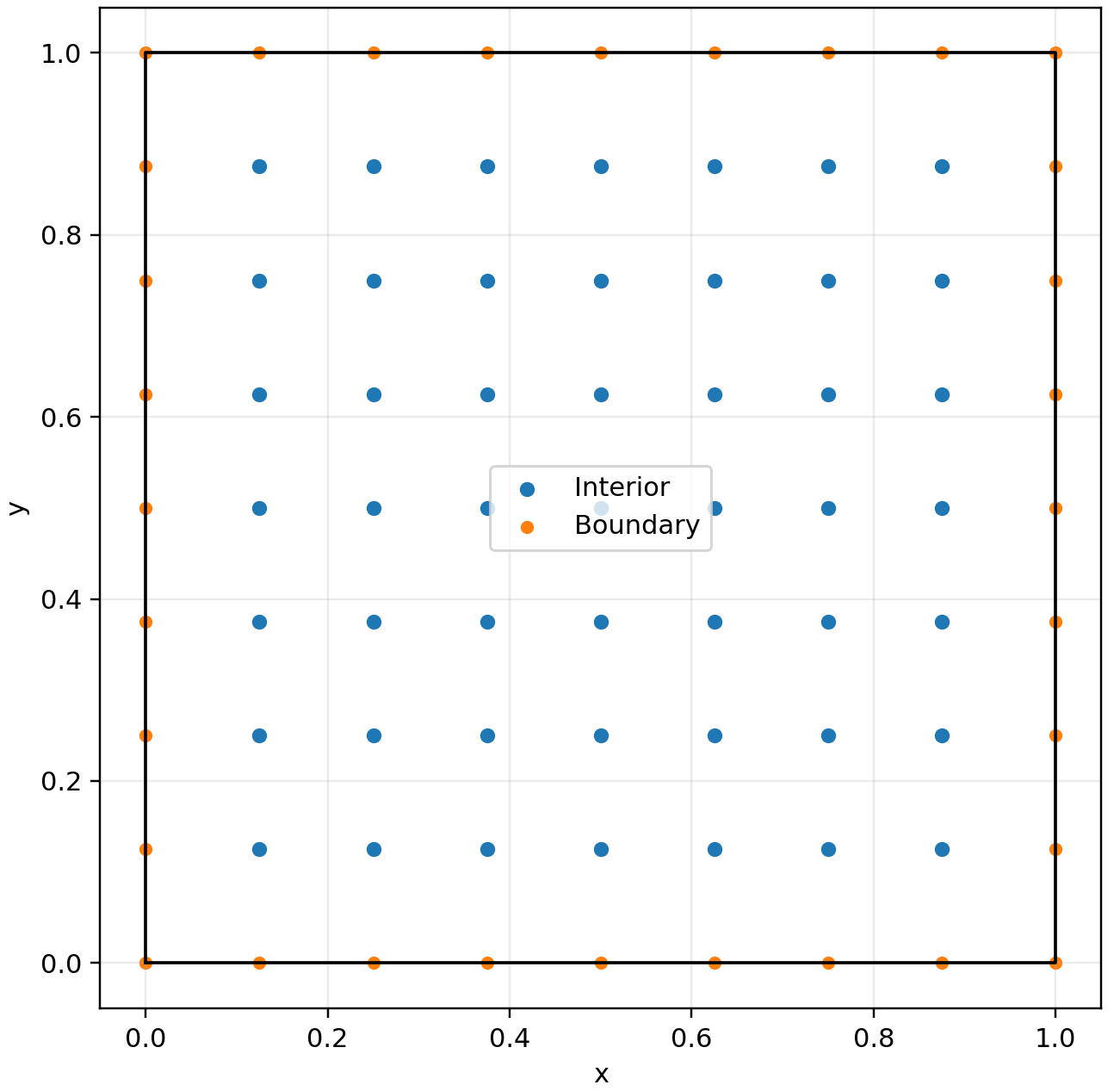}
	\caption{Distribution of interior and boundary collocation points on
		the square domain.}
	\label{Sq_D_P_1}
\end{figure}

In the first case, the TFNSE system  is considered in Eq. \eqref{eq:tfnse_system_M} on the unit-square domain
shown in Fig.~\ref{Sq_D_P_1}, defined as
\begin{equation}
	\Omega_{\mathrm{sq}}
	=
	\left\{(x,y)\in\mathbb{R}^{2}
	\,\middle|\,
	0\leq x\leq1,\;
	0\leq y\leq1
	\right\}.
\end{equation}
The source terms are given by
\begin{equation}
	\begin{aligned}
		&f_{1}(x,y,t,u,v)
		={}
		4(x^{2}-2x^{3}+x^{4})(y-3y^{2}+2y^{3})
		\left[
		\frac{t^{1-\alpha}}{\Gamma(2-\alpha)}
		+
		\frac{t^{2-\alpha}}{\Gamma(3-\alpha)}
		\right]+2(1+t)^{2}
		\\
		&\times\left[
		2u(x-3x^{2}+2x^{3})(y-3y^{2}+2y^{3})
		\right.
		\left.
		+v(x^{2}-2x^{3}+x^{4})(1-6y+6y^{2})
		\right]-\frac{(1+t)^{2}}{Re}
		\\
		&\times\left[
		4(y-3y^{2}+2y^{3})(1-6x+6x^{2})
		\right.
		\left.
		+2(x^{2}-2x^{3}+x^{4})(12y-6)
		\right]
		+2(1+t)^{2}(x-3x^{2}+2x^{3}),
	\end{aligned}
	\label{eq:case1_f1}
\end{equation}
and
\begin{equation}
	\begin{aligned}
		&f_{2}(x,y,t,u,v)
		={}
		-4(y^{2}-2y^{3}+y^{4})(x-3x^{2}+2x^{3})
		\left[
		\frac{t^{1-\alpha}}{\Gamma(2-\alpha)}
		+
		\frac{t^{2-\alpha}}{\Gamma(3-\alpha)}
		\right]
		-2(1+t)^{2}\\
		&
		\times \left[
		u(1-6x+6x^{2})(y^{2}-2y^{3}+y^{4})
		\right.\left.
		+2v(x-3x^{2}+2x^{3})(y-3y^{2}+2y^{3})
		\right]+\frac{(1+t)^{2}}{Re}
		\\
		&
		\times\left[
		2(y^{2}-2y^{3}+y^{4})(12x-6)
		\right.\left.
		+4(x-3x^{2}+2x^{3})(1-6y+6y^{2})
		\right]-2(1+t)^{2}(y-3y^{2}+2y^{3}).
	\end{aligned}
	\label{eq:case1_f2}
\end{equation}
The corresponding exact  fields are \
\begin{equation}
	\begin{cases}
		u(x,y,t)
		=
		2(x^{2}-2x^{3}+x^{4})
		(y-3y^{2}+2y^{3})(1+t)^{2},
		\\[0.6ex]
		v(x,y,t)
		=
		-2(y^{2}-2y^{3}+y^{4})
		(x-3x^{2}+2x^{3})(1+t)^{2},
		\\[0.6ex]
		p(x,y,t)
		=
		\left(
		x^{2}-2x^{3}+x^{4}
		-y^{2}+2y^{3}-y^{4}
		\right)(1+t)^{2}+c,
	\end{cases}
	\label{eq:case1_exact}
\end{equation}
To solve square case, a uniform \(7\times7\) grid, comprising 49 interior spatial collocation points, and 32 boundary points was employed. The temporal interval \(t\in[0,1]\) was discretized into 41 levels with \(\Delta t=0.025\). Consequently, the governing-equation residuals were evaluated at \(49\times40=1960\) interior space–time points, while the boundary conditions were imposed at \(32\times41=1312\) points. The initial conditions were exactly embedded in the trial solution. The computations were performed using \(\alpha=0.8\) and \(Re=400\), and the predicted fields were visualized on an \(81\times81\) evaluation grid.
\begin{figure}[H]
	\centering	\includegraphics[width=1.0\textwidth]{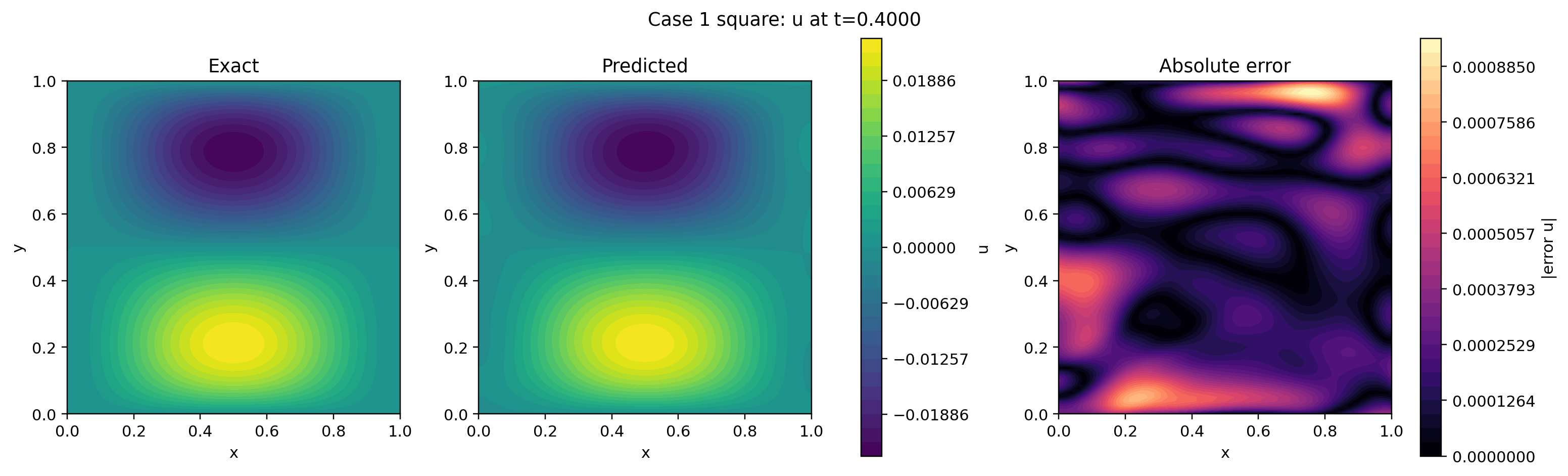}
	\caption{Comparison of the exact and GeoDiv-KAN predicted fields $u$,
		together with absolute errors, for square domain at \(t=0.4\).}
	\label{U_1_0.4}
\end{figure}

\begin{figure}[H]
	\centering	\includegraphics[width=1.0\textwidth]{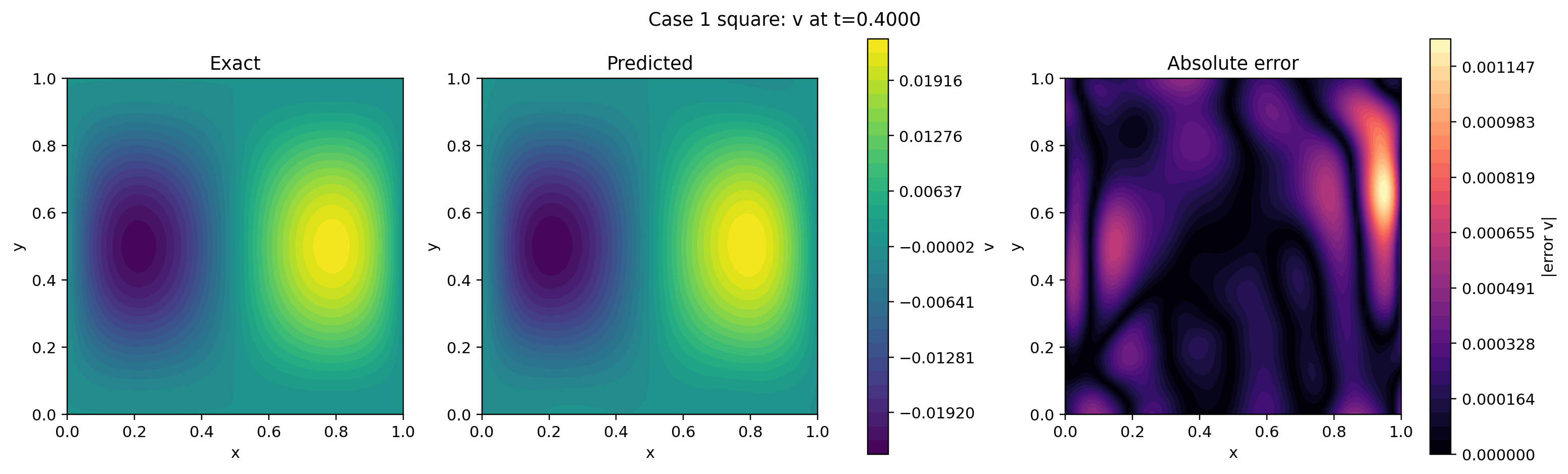}
	\caption{Comparison of the exact and GeoDiv-KAN predicted fields $v$,
		together with absolute errors, for square domain at \(t=0.4\).}
	\label{V_1_0.4}
\end{figure}

\begin{figure}[H]
	\centering	\includegraphics[width=1.0\textwidth]{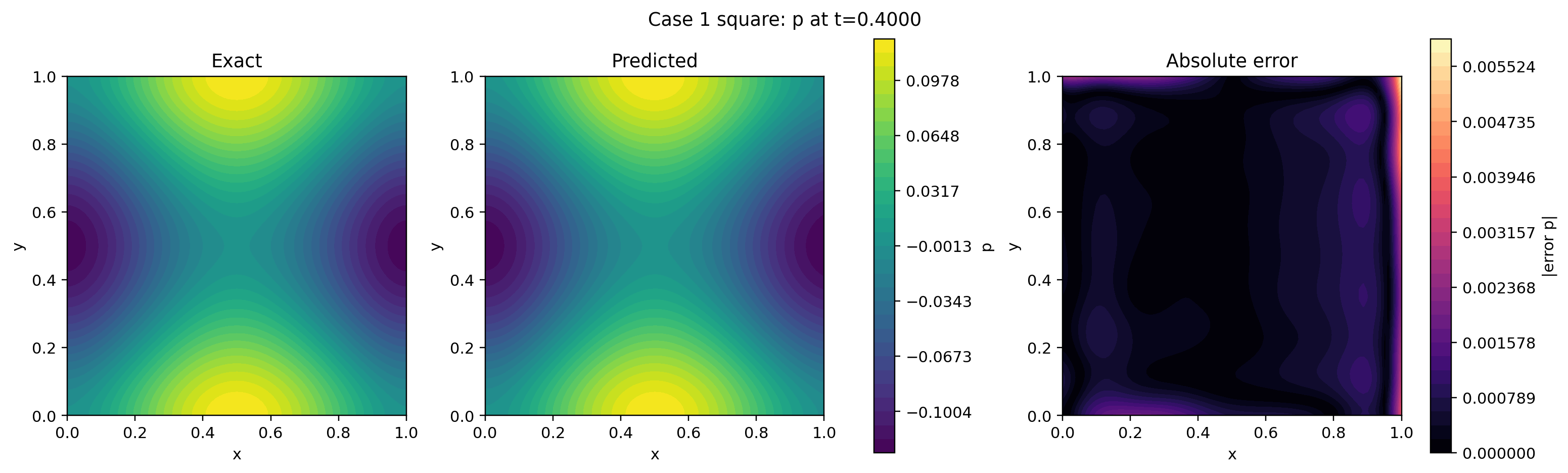}
	\caption{Comparison of the exact and GeoDiv-KAN predicted fields $p$,
		together with absolute errors, for square domain at \(t=0.4\).}
	\label{P_1_0.4}
\end{figure}

\begin{figure}[H]
	\centering	\includegraphics[width=1.0\textwidth]{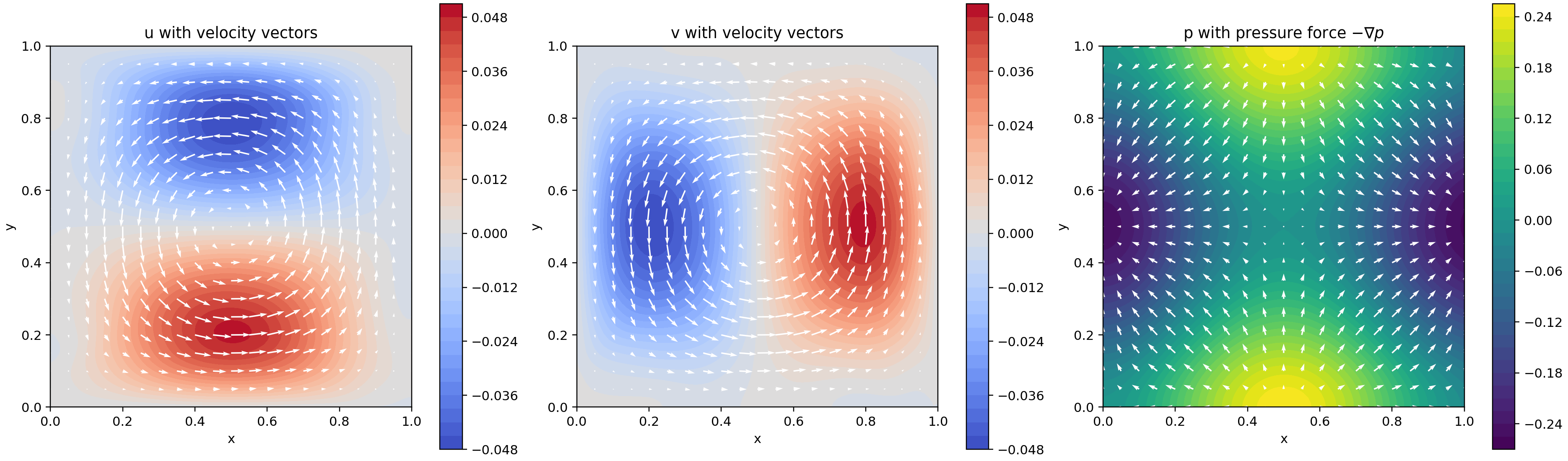}
	\caption{GeoDiv-KAN predicted flow field diagnostics for
		square domain at \(t=1\), first two panels show \(u\) and
		\(v\) with the velocity vectors, and third presents
		\(p\) with the pressure-force vectors.}
	\label{flow_field_diagnostics_1}
\end{figure}
In Figs.~ \ref{U_1_0.4}, \ref{V_1_0.4}, and \ref{P_1_0.4}, compares the exact and
GeoDiv-KAN predicted solutions fields for \(p\), \(u\), and \(v\) respectively on the square-domain at \(t=0.4\). The predicted contours closely reproduce the spatial
patterns, amplitudes, and symmetries of the corresponding exact solutions. The
absolute errors are approximately \(8.85\times10^{-4}\) for \(u\),
\(1.15\times10^{-3}\) for \(v\), and
\(5.52\times10^{-3}\) for \(p\). The errors remain small compared with the corresponding field
magnitudes and are primarily localized near the domain boundaries.

\begin{figure}[H]
	\centering	\includegraphics[width=1.0\textwidth]{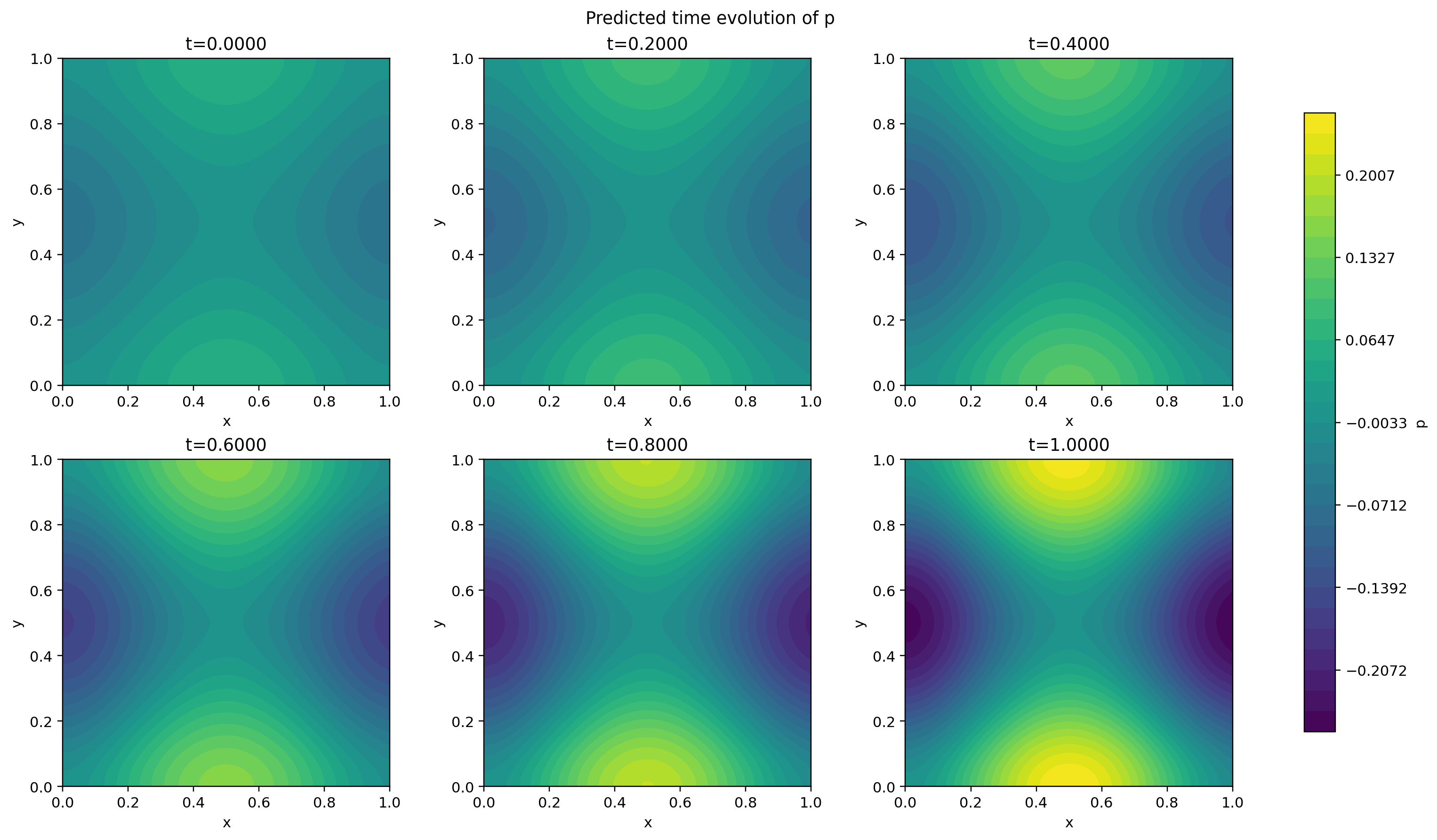}
	\caption{GeoDiv-KAN predicted temporal evolution of  the solution fields $p$ on the square domain at
		\(t=0,\,0.2,\,0.4,\,0.6,\,0.8,\) and \(1.0\).}
	\label{time_evolution_predicted_p_1}
\end{figure}

Fig.~\ref{flow_field_diagnostics_1} presents the predicted 
flow-field diagnostics. The first two panels show 
the horizontal and vertical velocity components, \(u\) and \(v\), 
respectively, superimposed with the velocity vector field \((u,v)\). 
%The \(u\)-field changes sign across the horizontal centreline, whereas 
%the \(v\)-field exhibits antisymmetric behaviour about the vertical 
%centreline. Together, these components form two oppositely rotating 
%circulation regions within the square domain. The velocity magnitude 
%gradually approaches zero near the boundary, consistently reproducing 
%the prescribed boundary behaviour.
The third panel shows the predicted pressure distribution together with 
the pressure-force vectors, \(-\nabla p\). High-pressure regions occur 
near the upper and lower central boundaries, whereas comparatively low 
pressure is observed near the lateral midpoints. 
%The pressure-force 
%vectors point from high- toward low-pressure regions and exhibit the 
%expected symmetric structure. Overall, the smooth fields, coherent 
%velocity vectors, and physically consistent pressure-gradient directions 
%demonstrate that GeoDiv-KAN accurately captures the coupled kinematic and 
%pressure characteristics of the manufactured TFNSE solution.
\begin{figure}[H]
	\centering	\includegraphics[width=1.0\textwidth]{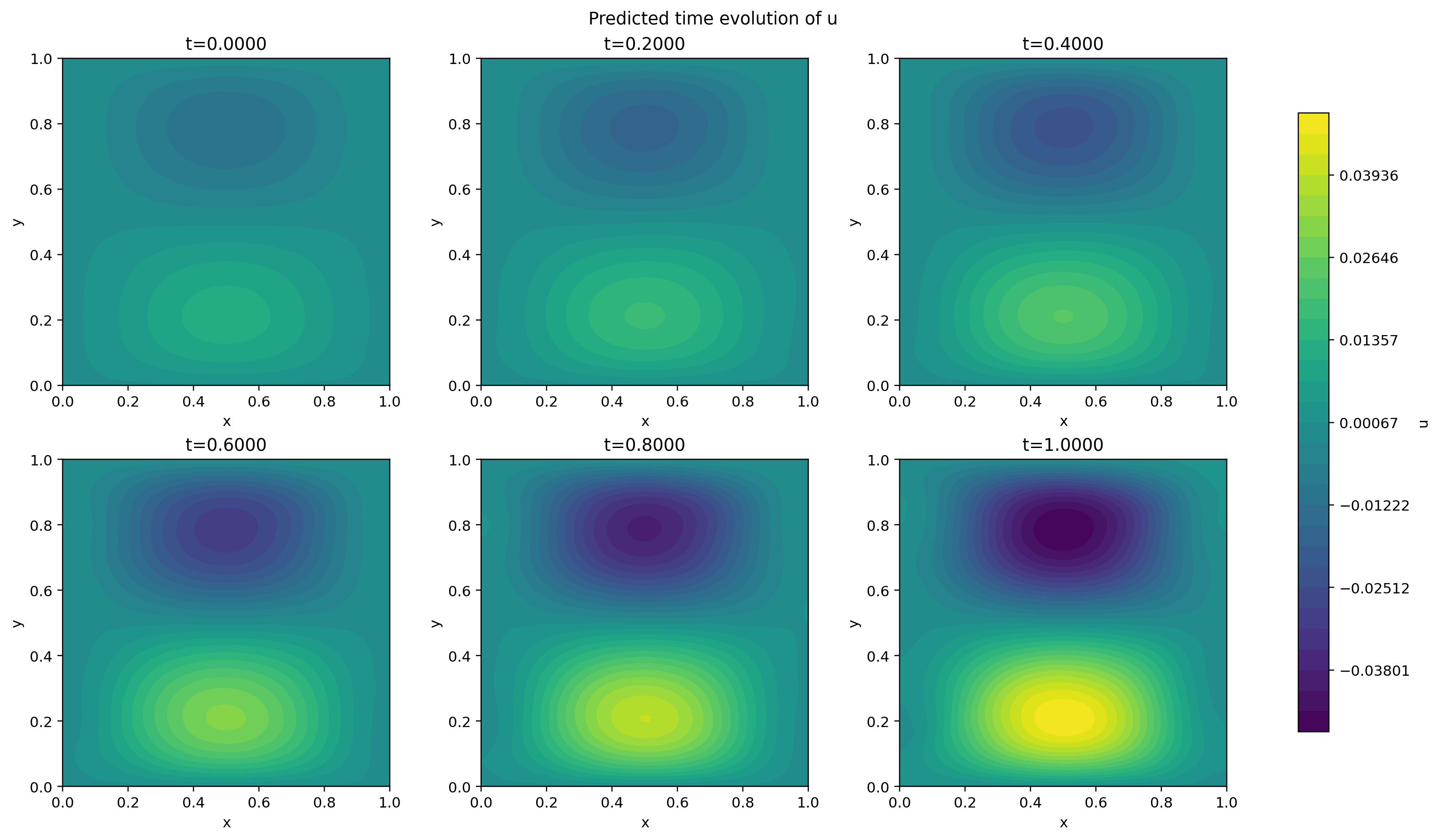}
	\caption{GeoDiv-KAN predicted temporal evolution of  the solution fields $u$ on the square domain at
		\(t=0,\,0.2,\,0.4,\,0.6,\,0.8,\) and \(1.0\).}
	\label{time_evolution_predicted_u_1}
\end{figure}
\begin{figure}[H]
	\centering	\includegraphics[width=1.0\textwidth]{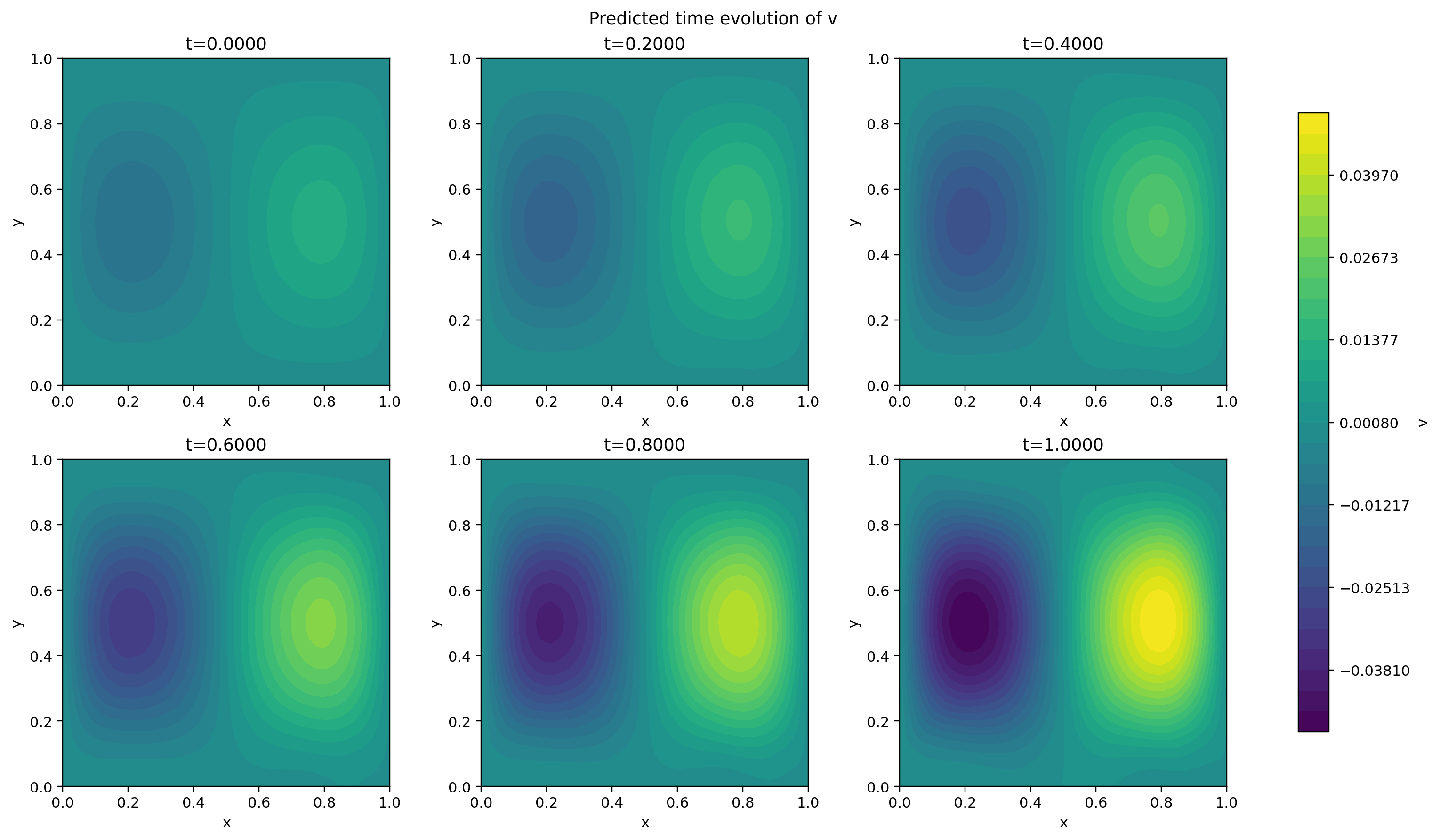}
	\caption{GeoDiv-KAN predicted temporal evolution of  the solution fields $v$ on the square domain at
		\(t=0,\,0.2,\,0.4,\,0.6,\,0.8,\) and \(1.0\).}
	\label{time_evolution_predicted_v_1}
\end{figure}
The temporal evolution of the GeoDiv-KAN predicted velocity and pressure 
fields over \(t\in[0,1]\) is presented in 
Figs.~\ref{time_evolution_predicted_u_1}, \ref{time_evolution_predicted_v_1}, and \ref{time_evolution_predicted_p_1}. The horizontal velocity \(u\) remains antisymmetric about \(y=0.5\), 
with positive and negative regions in the lower and upper halves of the 
domain, respectively. The vertical velocity \(v\) is antisymmetric about 
\(x=0.5\), exhibiting negative values on the left and positive values on 
the right. The pressure 
field retains its symmetric four-lobed structure, with positive regions 
near the upper and lower central boundaries and negative regions near the  lateral midpoints. 

\begin{figure}[H]
	\centering	\includegraphics[width=1.0\textwidth]{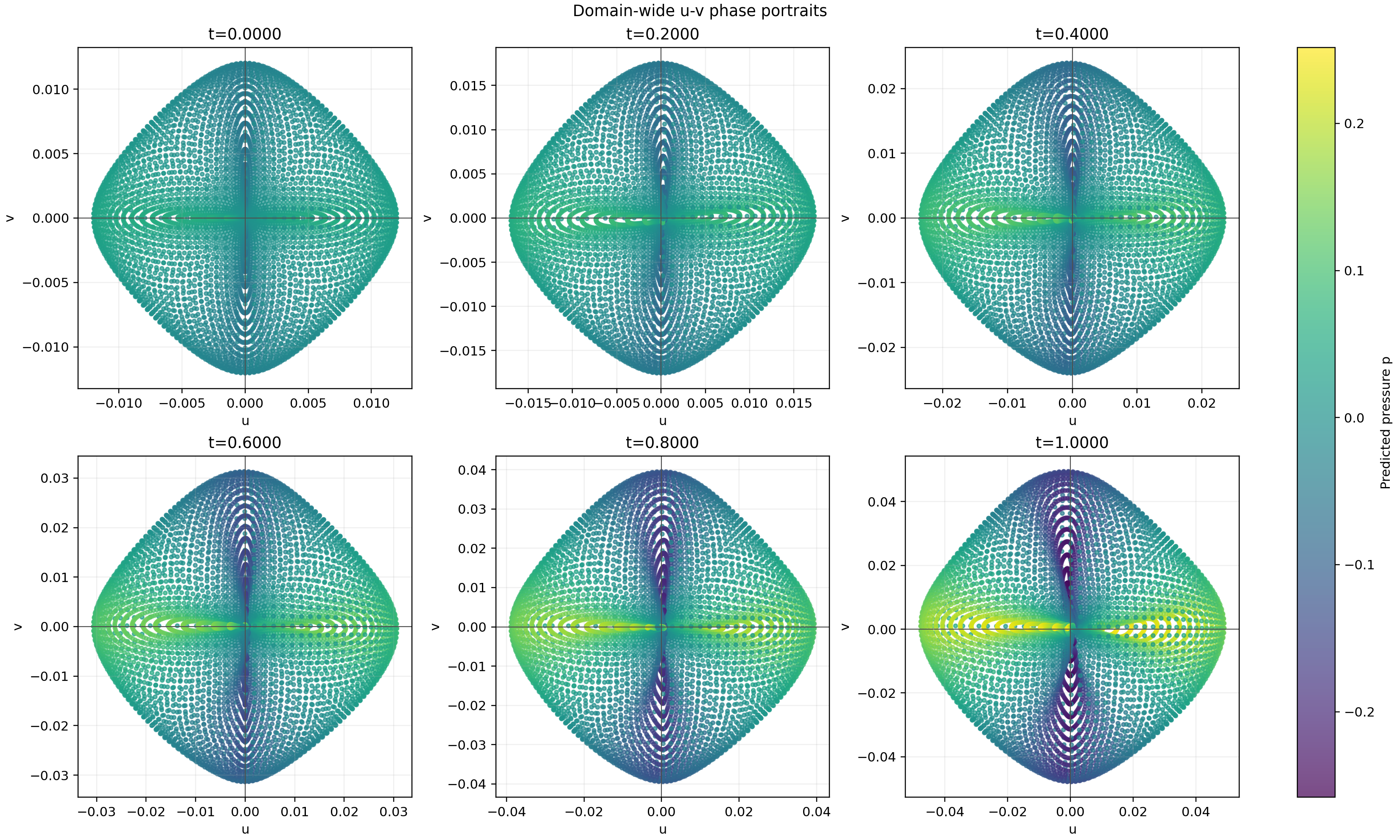}
	\caption{Domain-wide distributions of the predicted velocity pairs \((u,v)\) for square case at different t, coloured by the corresponding predicted pressure.}
	\label{uv_phase_portraits_1}
\end{figure}
The domain-wide relationship between the predicted velocity components 
is shown in Fig.~\ref{uv_phase_portraits_1} for six time levels. Each 
point represents the pair \((u(x,y,t),v(x,y,t))\) at a spatial location 
and is coloured according to the corresponding predicted pressure. The pressure colouring remains systematically organized 
across the velocity-state space. These results indicate that GeoDiv-KAN 
preserves the coupled relationships among \(u\), \(v\), and \(p\) during 
the temporal evolution.

\begin{figure}[H]
	\centering	\includegraphics[width=1.0\textwidth]{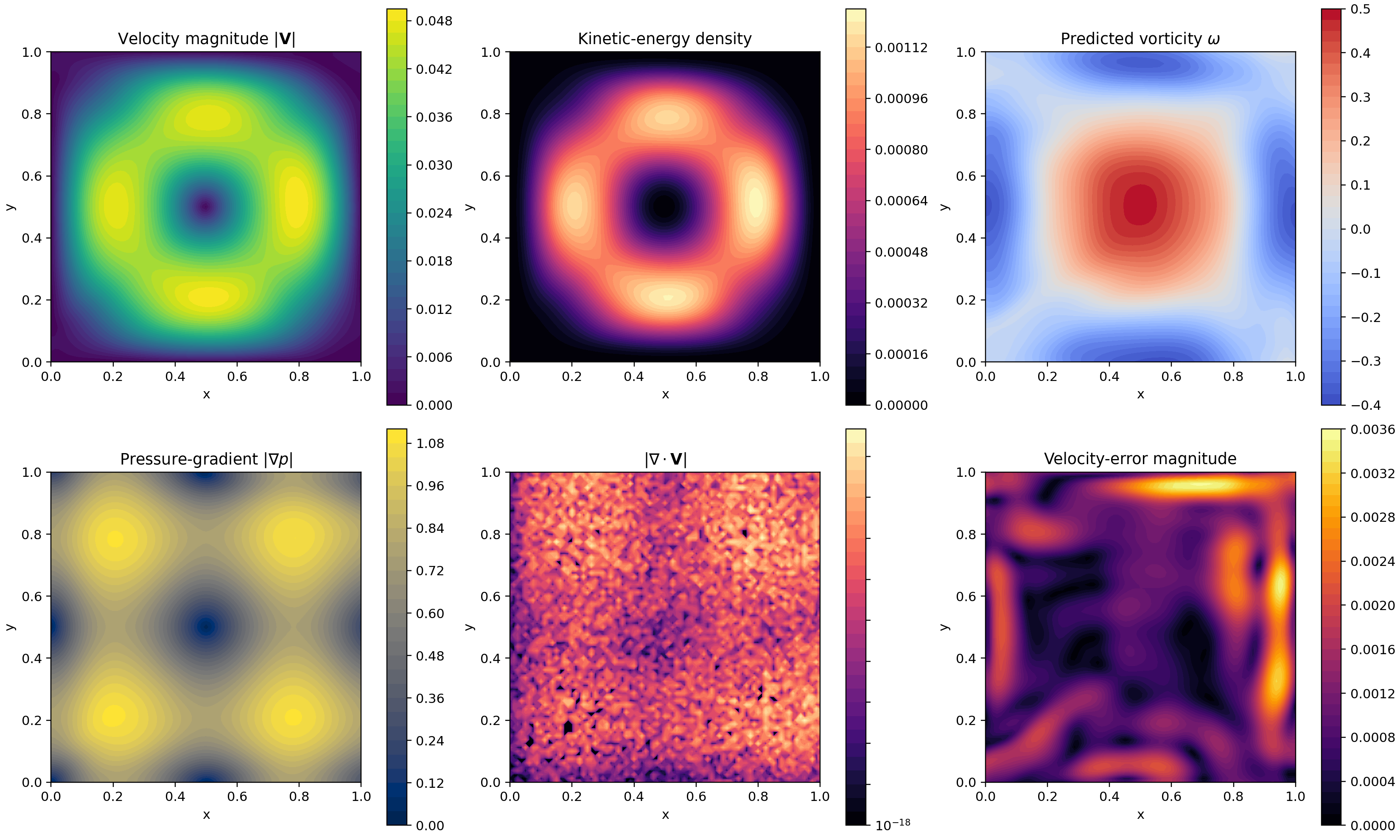}
	\caption{Physical diagnostics of the GeoDiv-KAN solution for square domain at \(t=1\): velocity magnitude,
		kinetic-energy density, vorticity, pressure-gradient magnitude,
		divergence magnitude, and velocity-error magnitude.}
	\label{physics_diagnostics_1}
\end{figure}

The physical consistency of the GeoDiv-KAN solution at \(t=1\) is 
examined in Fig.~\ref{physics_diagnostics_1}. The velocity 
magnitude,
\(
|\boldsymbol{V}|=\sqrt{u^{2}+v^{2}},
\)
and kinetic-energy density,
\(
E_k=\tfrac{1}{2}(u^{2}+v^{2}),
\)
display smooth, symmetric distributions and approach zero along the 
domain boundary. The predicted vorticity,
\(
\omega=\partial v/\partial x-\partial u/\partial y,
\)
reveals a positive central region surrounded by regions of opposite 
rotation. The pressure-gradient magnitude,
\(
|\nabla p|=\sqrt{p_x^{2}+p_y^{2}},
\)
identifies the regions in which pressure produces the strongest 
contribution to the momentum balance. Most importantly, the divergence 
magnitude remains of the order of \(10^{-18}\), indicating that the 
incompressibility condition is satisfied to machine precision through 
the streamfunction formulation. The velocity-error magnitude remains 
below approximately \(3.6\times10^{-3}\) and is primarily concentrated 
near the domain boundary, while comparatively small errors occur over 
most of the interior.
The quantitative comparison reported in
Table~\ref{tab:square_case1_results} demonstrates that GeoDiv-KAN achieves the lowest RMSE, maximum absolute error, and relative \(L^2\)
error for all three solution components.  The consistently improved performance shows that geometry-gated, Gaussian RBF edge functions, and the divergence-free
streamfunction representation provide a more accurate approximation of
the coupled TFNSE fields than the considered comparison models.
\begin{table}[H]
	\centering
	\caption{Comparison of different models for the square case at $t=1.0$.}
	\label{tab:square_case1_results}
	
	\small
	\renewcommand{\arraystretch}{1.20}
	\setlength{\tabcolsep}{4pt}
	
	\begin{tabular}{llccc}
		\toprule
		\textbf{Model} &
		\textbf{Component} &
		\textbf{RMSE} &
		\textbf{Max. Absolute Error} &
		\textbf{Relative $L_2$ Error} \\
		\midrule
		
		\multirow{3}{*}{\textbf{GeoDiv-KAN}}
		& $u$ & $\mathbf{8.9488\times10^{-4}}$ 
		& $\mathbf{3.4069\times10^{-3}}$ 
		& $\mathbf{4.1196\times10^{-2}}$ \\
		& $v$ & $\mathbf{9.8067\times10^{-4}}$ 
		& $\mathbf{3.2151\times10^{-3}}$ 
		& $\mathbf{4.5145\times10^{-2}}$ \\
		& $p$ & $\mathbf{2.9098\times10^{-3}}$ 
		& $\mathbf{2.6783\times10^{-2}}$ 
		& $\mathbf{2.3385\times10^{-2}}$ \\
		\midrule
		
		\multirow{3}{*}{DF-RBFKAN}
		& $u$ & $1.4090\times10^{-1}$ 
		& $3.7635\times10^{-1}$ 
		& $6.4863\times10^{0}$ \\
		& $v$ & $1.3969\times10^{-1}$ 
		& $6.1511\times10^{-1}$ 
		& $6.4304\times10^{0}$ \\
		& $p$ & $1.6893\times10^{-1}$ 
		& $7.6173\times10^{-1}$ 
		& $1.3576\times10^{0}$ \\
		\midrule
		
		\multirow{3}{*}{RBF-KAN}
		& $u$ & $9.4937\times10^{-2}$ 
		& $2.3327\times10^{-1}$ 
		& $4.3704\times10^{0}$ \\
		& $v$ & $2.1411\times10^{-1}$ 
		& $4.9352\times10^{-1}$ 
		& $9.8563\times10^{0}$ \\
		& $p$ & $1.5755\times10^{-1}$ 
		& $8.6372\times10^{-1}$ 
		& $1.2661\times10^{0}$ \\
		\midrule
		
		\multirow{3}{*}{Base PIKAN}
		& $u$ & $9.4999\times10^{-1}$ 
		& $2.4433\times10^{0}$ 
		& $4.3732\times10^{1}$ \\
		& $v$ & $3.6958\times10^{-1}$ 
		& $1.2028\times10^{0}$ 
		& $1.7013\times10^{1}$ \\
		& $p$ & $4.7814\times10^{-1}$ 
		& $1.0094\times10^{0}$ 
		& $3.8425\times10^{0}$ \\
		\midrule
		
		\multirow{3}{*}{PINN}
		& $u$ & $1.2758\times10^{0}$ 
		& $3.4079\times10^{0}$ 
		& $5.8729\times10^{1}$ \\
		& $v$ & $7.4382\times10^{-1}$ 
		& $1.9851\times10^{0}$ 
		& $3.4242\times10^{1}$ \\
		& $p$ & $1.1733\times10^{0}$ 
		& $4.3080\times10^{0}$ 
		& $9.4290\times10^{0}$ \\
		\bottomrule
		
	\end{tabular}
\end{table}

%%%%%%%%%%%%%%%%%%%%%%%%%%%%%%%%%%%%%%%%%%%%%%%%%%%%%%%%%%%%%%%%%%%%%%%%%%%%%%%%
\subsection{\textbf{Case 2: Circular Domain}}

\begin{figure}[H]
	\centering	\includegraphics[width=0.5\textwidth]{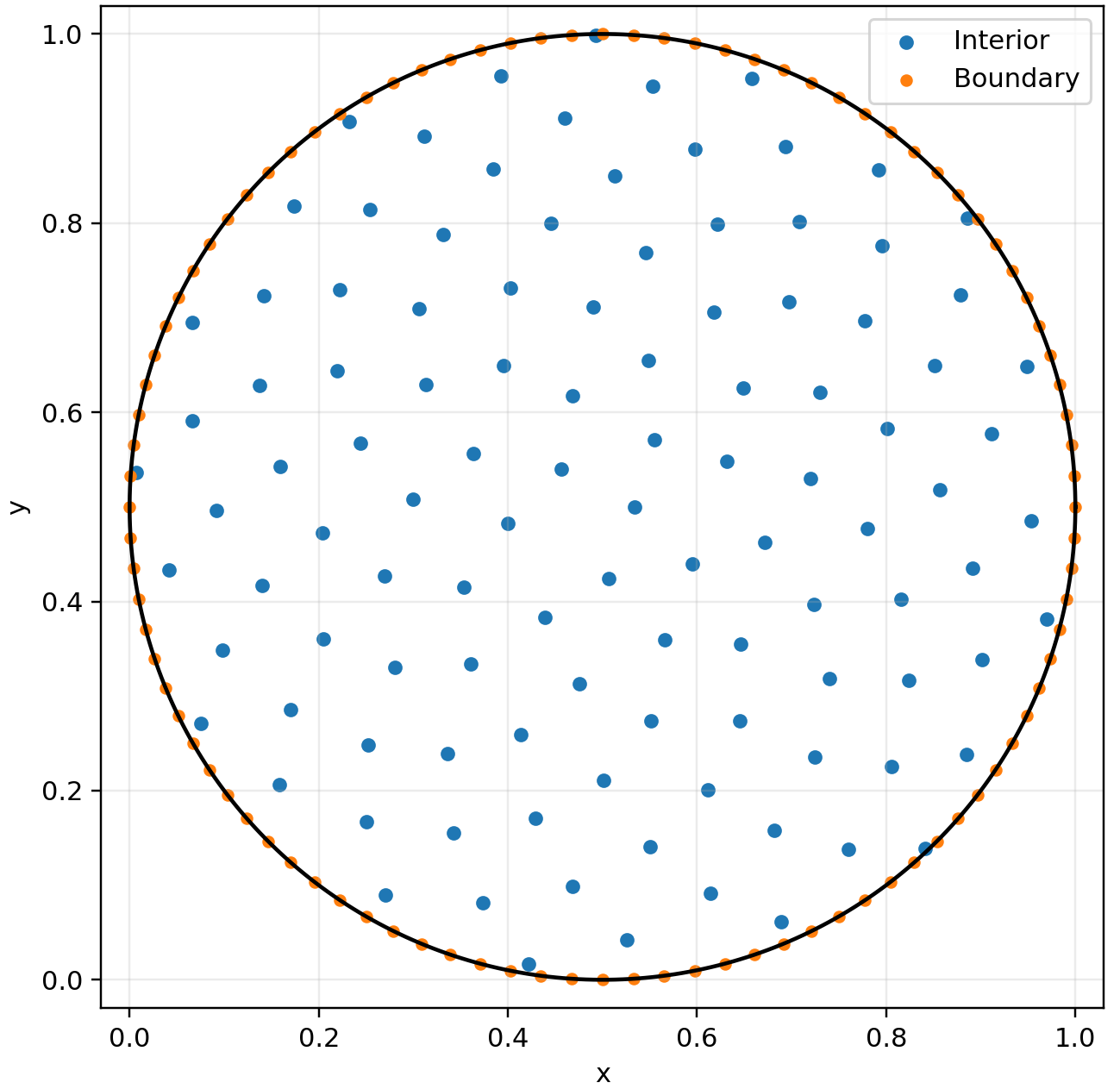}
	\caption{Distribution of interior and boundary collocation points on
		the circular domain.}
	\label{Cir_D_P_2}
\end{figure}

\begin{figure}[H]
	\centering	\includegraphics[width=1.0\textwidth]{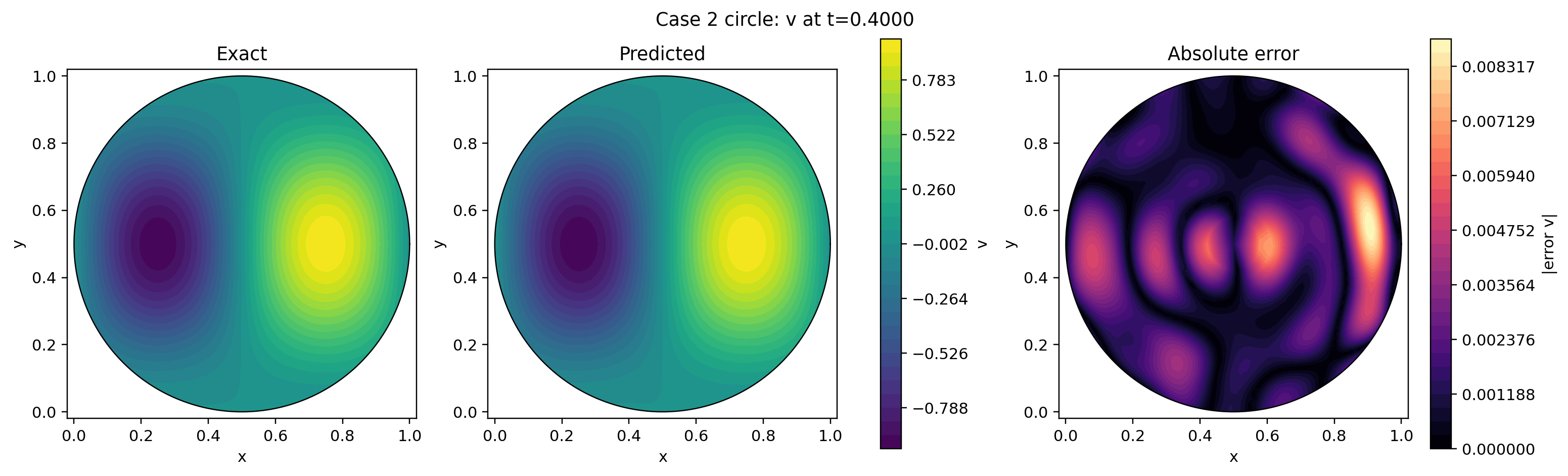}
	\caption{Comparison of the exact and GeoDiv-KAN predicted fields $v$,
		together with absolute errors, for circle domain at \(t=0.4\).}
	\label{v_exact_pred_error_t_0p4000_2}
\end{figure}

\begin{figure}[htbp]
	\centering	\includegraphics[width=1.0\textwidth]{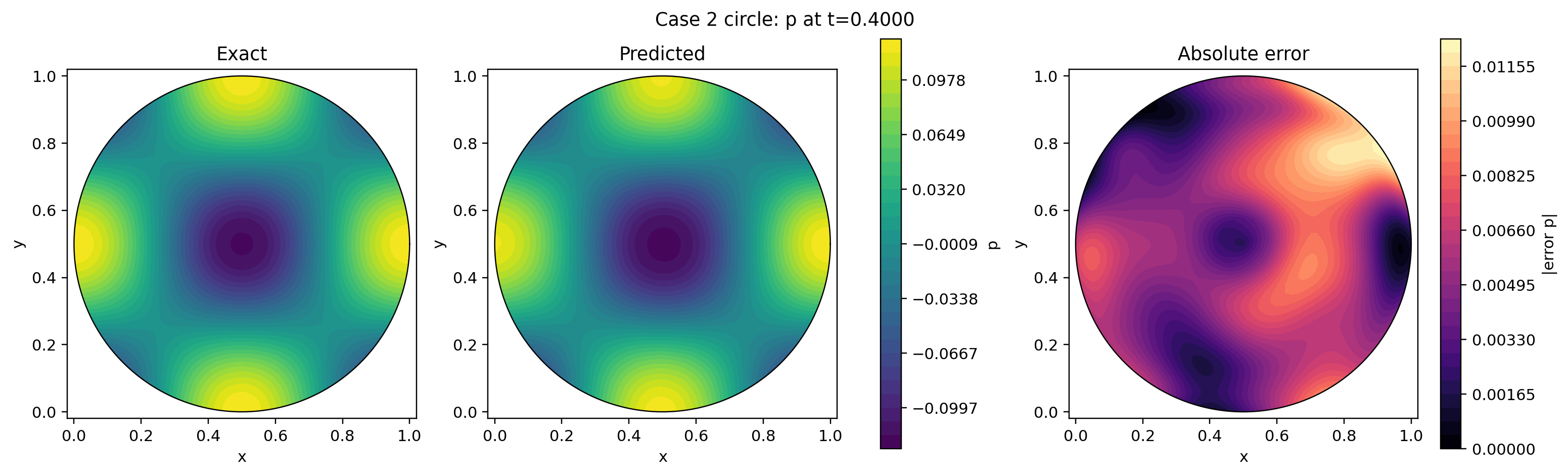}
	\caption{Comparison of the exact and GeoDiv-KAN predicted fields $p$,
		together with absolute errors, for circle domain at \(t=0.4\).}
	\label{p_exact_pred_error_t_0p4000_2}
\end{figure}

In the second case, the TFNSE system given in
Eq.~\eqref{eq:tfnse_system_M} is considered on the circular domain
shown in Fig.~\ref{Cir_D_P_2}. The domain is centred at
\((1/2,1/2)\) with radius \(1/2\) and is defined by
\begin{equation}
	\Omega_{\mathrm{cir}}
	=
	\left\{
	(x,y)\in\mathbb{R}^{2}
	\,\middle|\,
	\left(x-\frac{1}{2}\right)^{2}
	+
	\left(y-\frac{1}{2}\right)^{2}
	\leq\frac{1}{4}
	\right\}.
	\label{eq:circular_domain}
\end{equation}
Its boundary can equivalently be parameterized as
\begin{equation}
	x=\frac{1}{2}\cos\theta+\frac{1}{2},
	\qquad
	y=\frac{1}{2}\sin\theta+\frac{1}{2},
	\qquad
	0\leq\theta\leq2\pi.
\end{equation}
The corresponding source terms are given by
\begin{equation}
	\begin{aligned}
		&f_{1}(x,y,t,u,v)
		={}
		2\sin^{2}(\pi x)\sin(\pi y)\cos(\pi y)
		\left[
		\frac{t^{1-\alpha}}{\Gamma(2-\alpha)}
		+
		\frac{t^{2-\alpha}}{\Gamma(3-\alpha)}
		\right]
		\\
		&+\pi u(1+t)^{2}
		\sin(\pi y)\cos(\pi y)\sin(2\pi x)
		+\pi v(1+t)^{2}
		\sin^{2}(\pi x)\cos(2\pi y)-\frac{(1+t)^{2}}{Re}
	\\ 
	& \times \left[
		\pi^{2}\sin(2\pi y)\cos(2\pi x)
		-2\pi^{2}\sin^{2}(\pi x)\sin(2\pi y)
		\right]
		+\frac{\pi}{8}(1+t)^{2}
		\sin(2\pi x)\cos(2\pi y),
	\end{aligned}
	\label{eq:case2_f1}
\end{equation}
and
\begin{equation}
	\begin{aligned}
		&f_{2}(x,y,t,u,v)
		={}
		-2\sin^{2}(\pi y)\sin(\pi x)\cos(\pi x)
		\left[
		\frac{t^{1-\alpha}}{\Gamma(2-\alpha)}
		+
		\frac{t^{2-\alpha}}{\Gamma(3-\alpha)}
		\right]
		\\
		&-\pi u(1+t)^{2}
		\sin^{2}(\pi y)\cos(2\pi x)
		-\frac{\pi v}{2}(1+t)^{2}
		\sin(2\pi x)\sin(2\pi y)-\frac{(1+t)^{2}}{Re}
		\\
		&
		\times \left[
		2\pi^{2}\sin^{2}(\pi y)\sin(2\pi x)
		-\pi^{2}\sin(2\pi x)\cos(2\pi y)
		\right] +\frac{\pi}{8}(1+t)^{2}
		\cos(2\pi x)\sin(2\pi y).
	\end{aligned}
	\label{eq:case2_f2}
\end{equation}
The corresponding exact fields are 
\begin{equation}
	\begin{cases}
		u(x,y,t)
		=
		\displaystyle
		\frac{1}{2}
		\sin^{2}(\pi x)\sin(2\pi y)(1+t)^{2},
		\\[0.8ex]
		v(x,y,t)
		=
		\displaystyle
		-\frac{1}{2}
		\sin^{2}(\pi y)\sin(2\pi x)(1+t)^{2},
		\\[0.8ex]
		p(x,y,t)
		=
		\displaystyle
		-\frac{1}{16}
		\cos(2\pi x)\cos(2\pi y)(1+t)^{2}-c.
	\end{cases}
	\label{eq:case2_exact}
\end{equation}
To solve this case, 109 interior spatial collocation points
were distributed over the circular domain using, while 96 points were placed along the curved
boundary. The temporal interval \(t\in[0,1]\) was discretized into 41
levels with a step size of \(\Delta t=0.025\). The initial conditions
were embedded directly into the trial solution. The computations were
performed for the fractional order \(\alpha=0.8\) and Reynolds number
\(Re=400\). 
\begin{figure}[htbp]
	\centering	\includegraphics[width=1.0\textwidth]{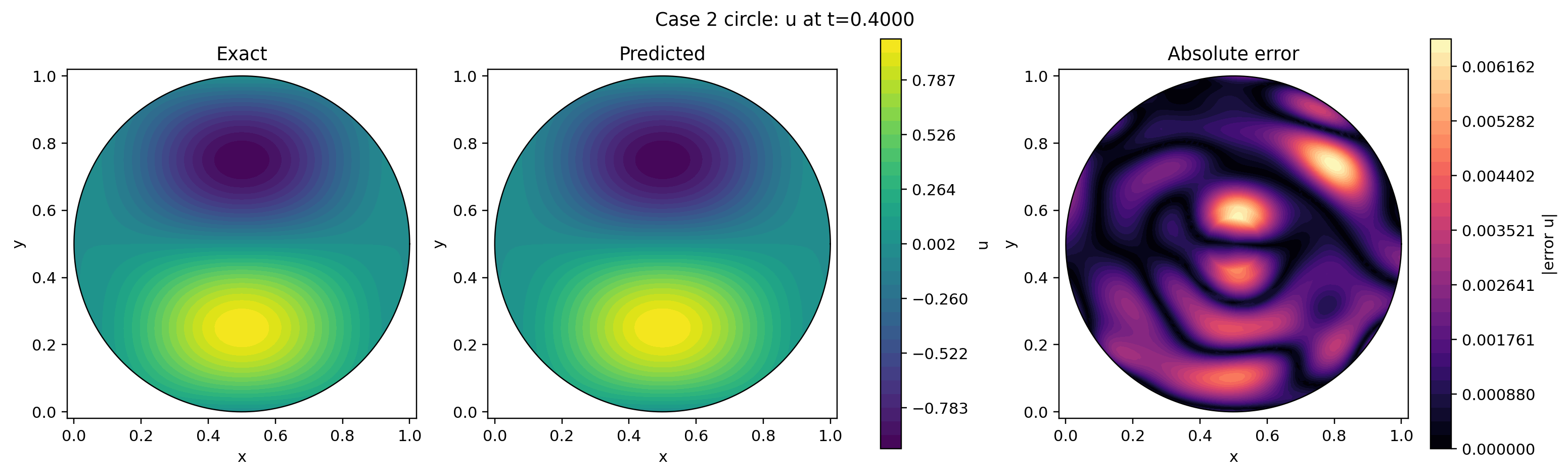}
	\caption{Comparison of the exact and GeoDiv-KAN predicted fields $u$,
		together with absolute errors, for circle domain at \(t=0.4\).}
	\label{u_exact_pred_error_t_0p4000_2}
\end{figure}
The comparison between the analytical and GeoDiv-KAN-predicted fields at
\(t=0.4\) is presented in Figs.~\ref{u_exact_pred_error_t_0p4000_2}, \ref{v_exact_pred_error_t_0p4000_2}, and \ref{p_exact_pred_error_t_0p4000_2}. The predicted velocity components $u$ and $v$ closely reproduce the analytical
fields, preserving their amplitudes, spatial symmetries, and prescribed
behaviour along the curved boundary. The predicted pressure field also closely follows the analytical
distribution, including the negative central region and the positive
regions near the upper, lower, left, and right boundaries. The maximum
absolute errors are approximately
\(6.16\times10^{-3}\), \(8.32\times10^{-3}\), and
\(1.16\times10^{-2}\) for \(u\), \(v\), and \(p\), respectively.

\begin{figure}[H]
	\centering	\includegraphics[width=1.0\textwidth]{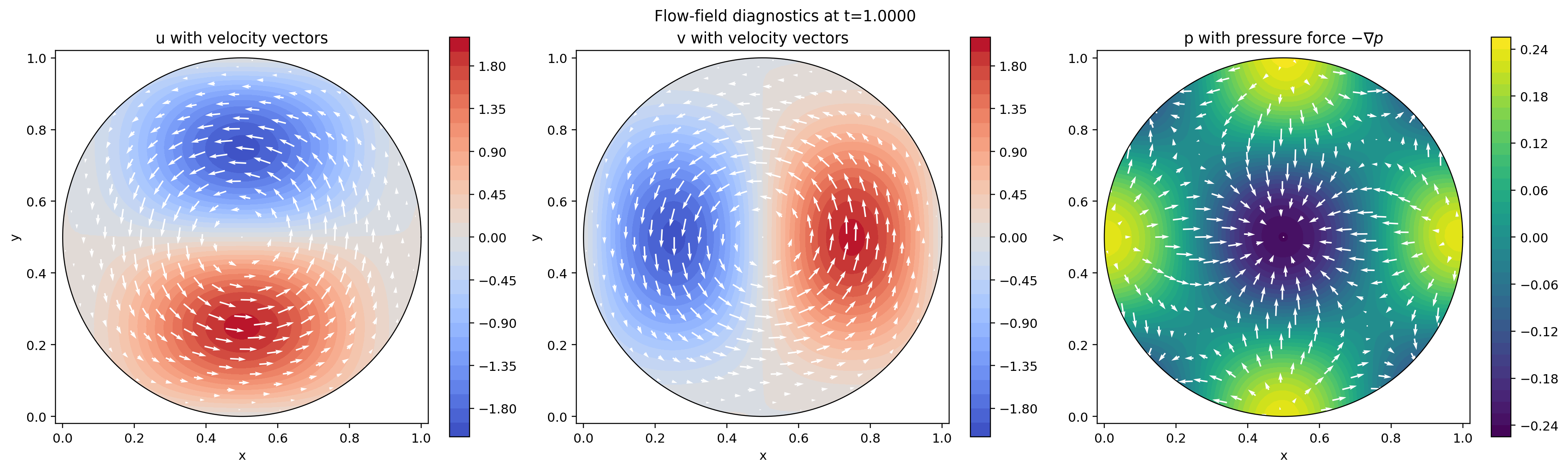}
	\caption{GeoDiv-KAN predicted flow field diagnostics for
		circle domain at \(t=1\), first two panels show \(u\) and
		\(v\) with the velocity vectors, and third presents
		\(p\) with the pressure-force vectors.}
	\label{flow_field_diagnostics_2}
\end{figure}

The predicted flow-field diagnostics at \(t=1\) are presented in
Fig.~\ref{flow_field_diagnostics_2}. The velocity components and
superimposed vectors reveal a smooth counterclockwise circulation within
the circular domain. The pressure field exhibits a central low-pressure region
surrounded by higher-pressure regions near the boundary. Accordingly,
the pressure-force vectors, \(-\nabla p\), are predominantly directed
from the high-pressure boundary regions towards the low-pressure center.

\begin{figure}[H]
	\centering	\includegraphics[width=1.0\textwidth]{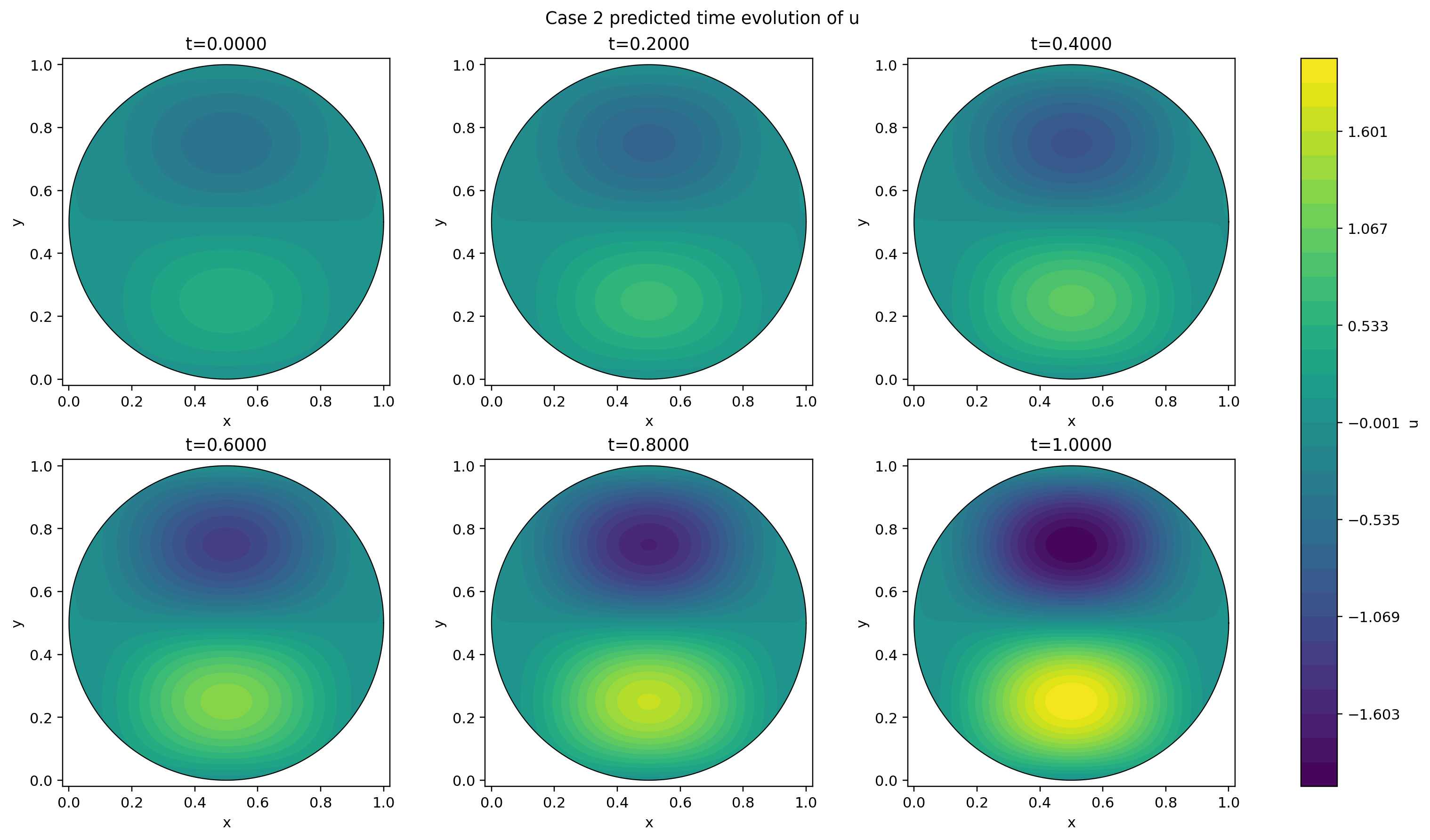}
	\caption{GeoDiv-KAN predicted temporal evolution of  the solution fields $u$ on circle domain at
		\(t=0,\,0.2,\,0.4,\,0.6,\,0.8,\) and \(1.0\).}
	\label{time_evolution_predicted_u_2}
\end{figure}

\begin{figure}[H]
	\centering	\includegraphics[width=1.0\textwidth]{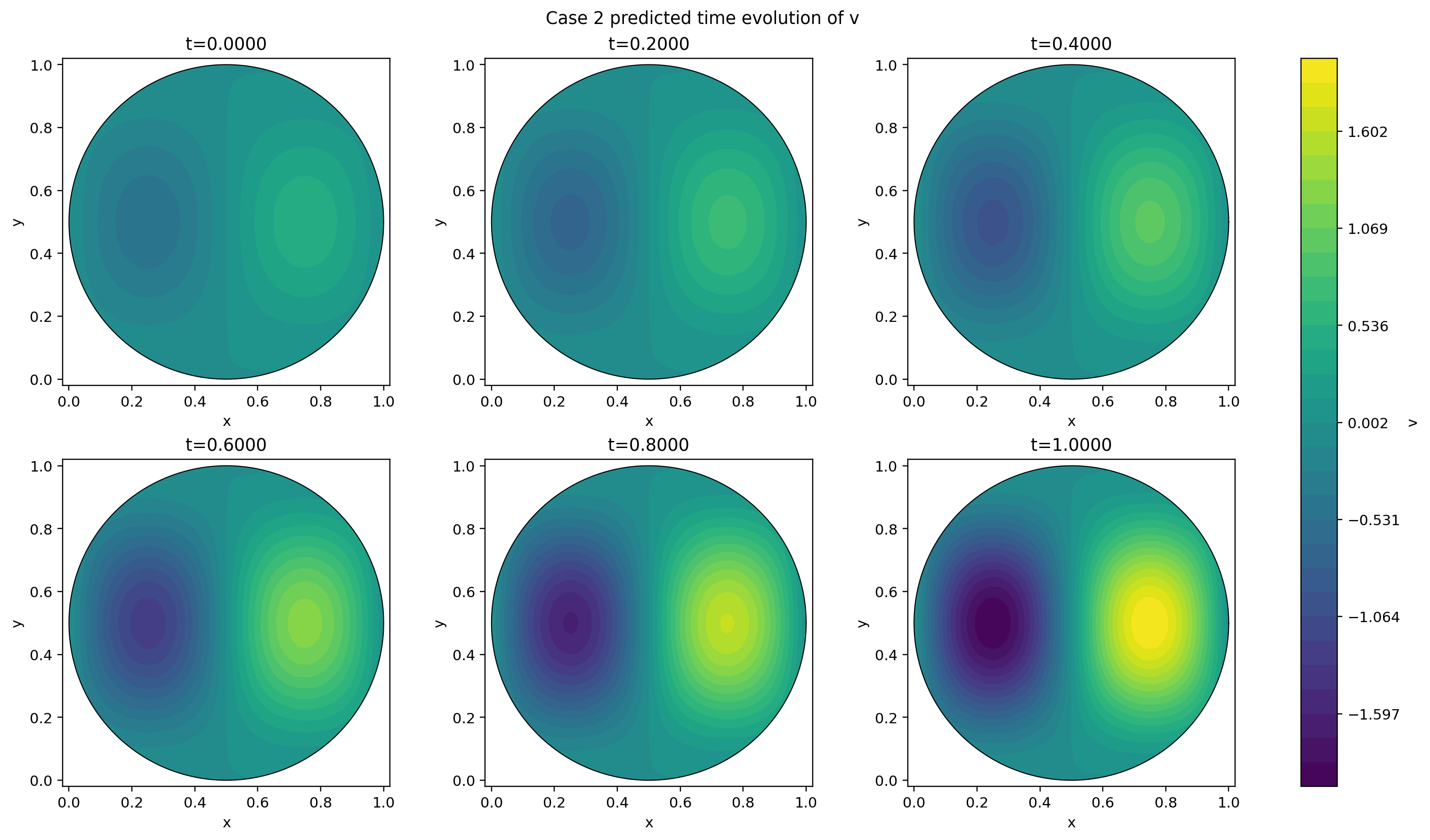}
	\caption{GeoDiv-KAN predicted temporal evolution of  the solution fields $v$ on circle domain at
		\(t=0,\,0.2,\,0.4,\,0.6,\,0.8,\) and \(1.0\).}
	\label{time_evolution_predicted_v_2}
\end{figure}

\begin{figure}[H]
	\centering	\includegraphics[width=1.0\textwidth]{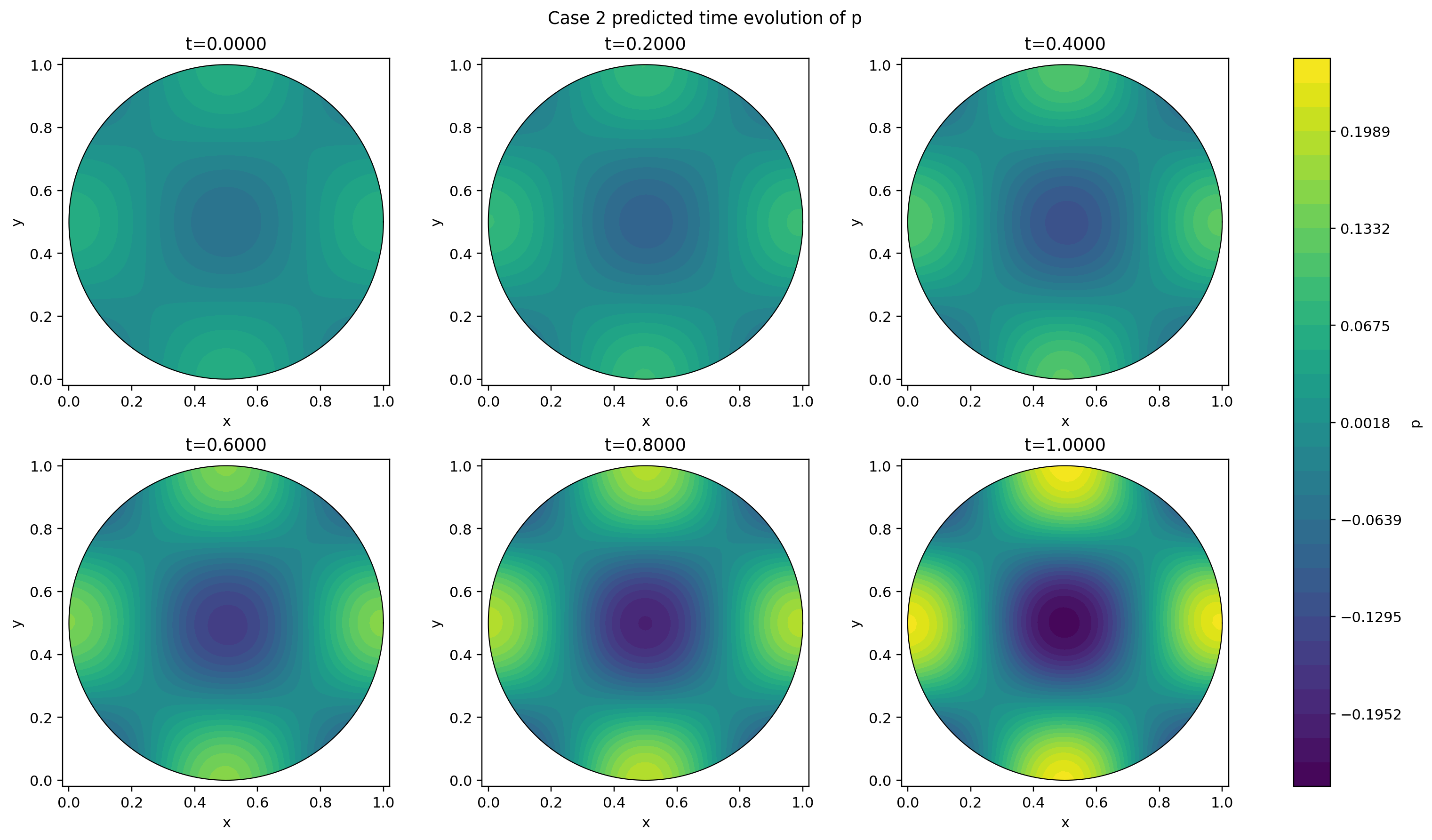}
	\caption{GeoDiv-KAN predicted temporal evolution of  the solution fields $p$ on the circle domain at
		\(t=0,\,0.2,\,0.4,\,0.6,\,0.8,\) and \(1.0\).}
	\label{time_evolution_predicted_p_2}
\end{figure}

The temporal evolution of the predicted fields over \(t\in[0,1]\) for circle case is
shown in Figs.~\ref{time_evolution_predicted_u_2}, \ref{time_evolution_predicted_v_2}, and \ref{time_evolution_predicted_p_2}. The velocity and pressure
fields preserve their characteristic spatial patterns and symmetries. The horizontal velocity \(u\) remains antisymmetric
about \(y=0.5\), whereas the vertical velocity \(v\) remains
antisymmetric about \(x=0.5\). The pressure retains a central
low-pressure region surrounded by four high-pressure regions near the circular boundary.

\begin{figure}[H]
	\centering	\includegraphics[width=1.0\textwidth]{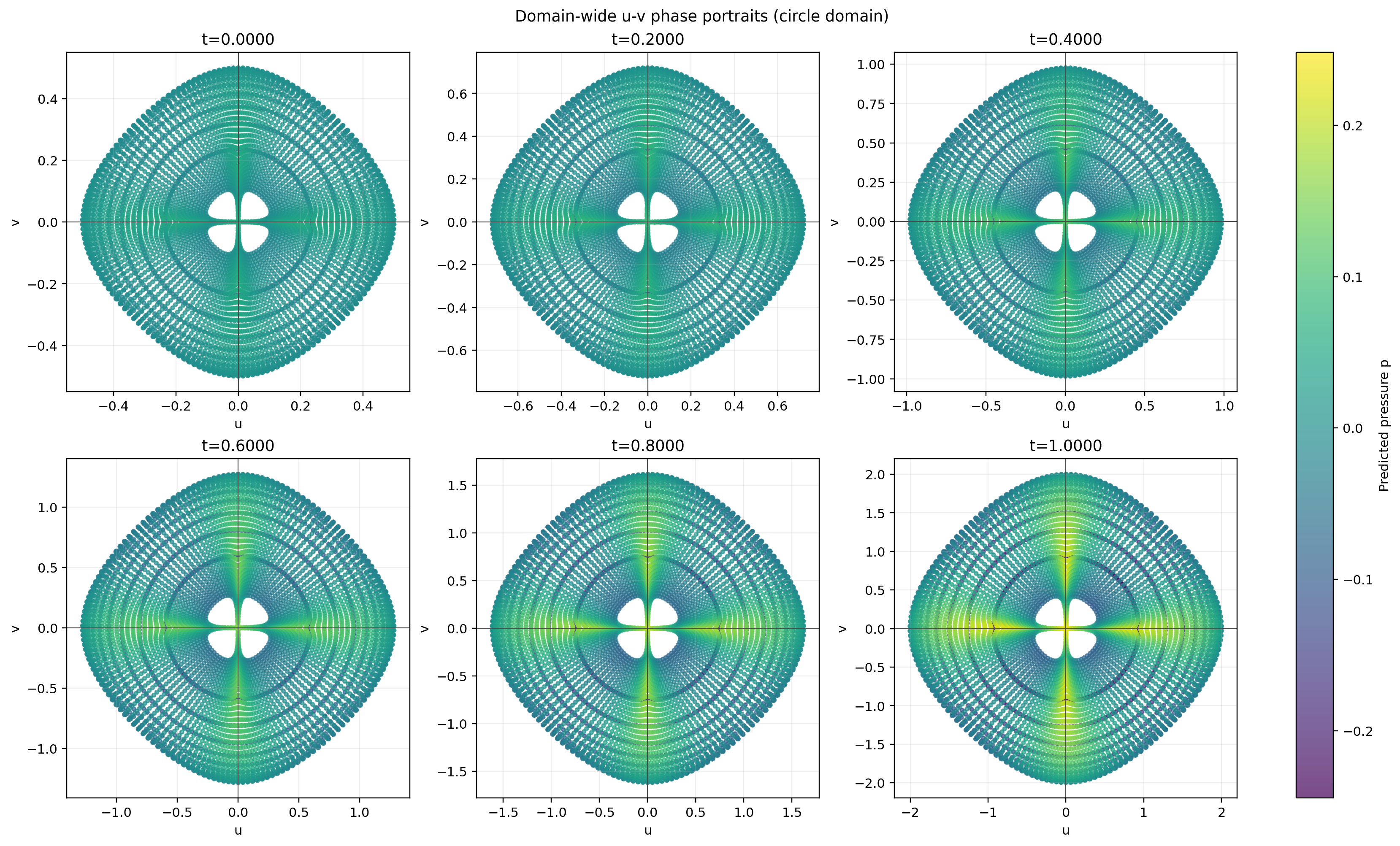}
	\caption{Domain-wide distributions of the predicted velocity
		pairs \((u,v)\) for circular case at
		different t, coloured by the
		corresponding predicted pressure.}
	\label{uv_phase_portraits_2}
\end{figure}

The domain-wide distributions of the predicted velocity pairs
\((u,v)\) for circle case are presented in Fig.~\ref{uv_phase_portraits_2} for six time
levels. Each point corresponds to a spatial location in the circular
domain and is coloured according to its predicted pressure. Their progressive expansion
reflects the temporal increase in the velocity amplitudes, while the
overall topology and organized pressure distribution remain preserved.

\begin{figure}[H]
	\centering	\includegraphics[width=1.0\textwidth]{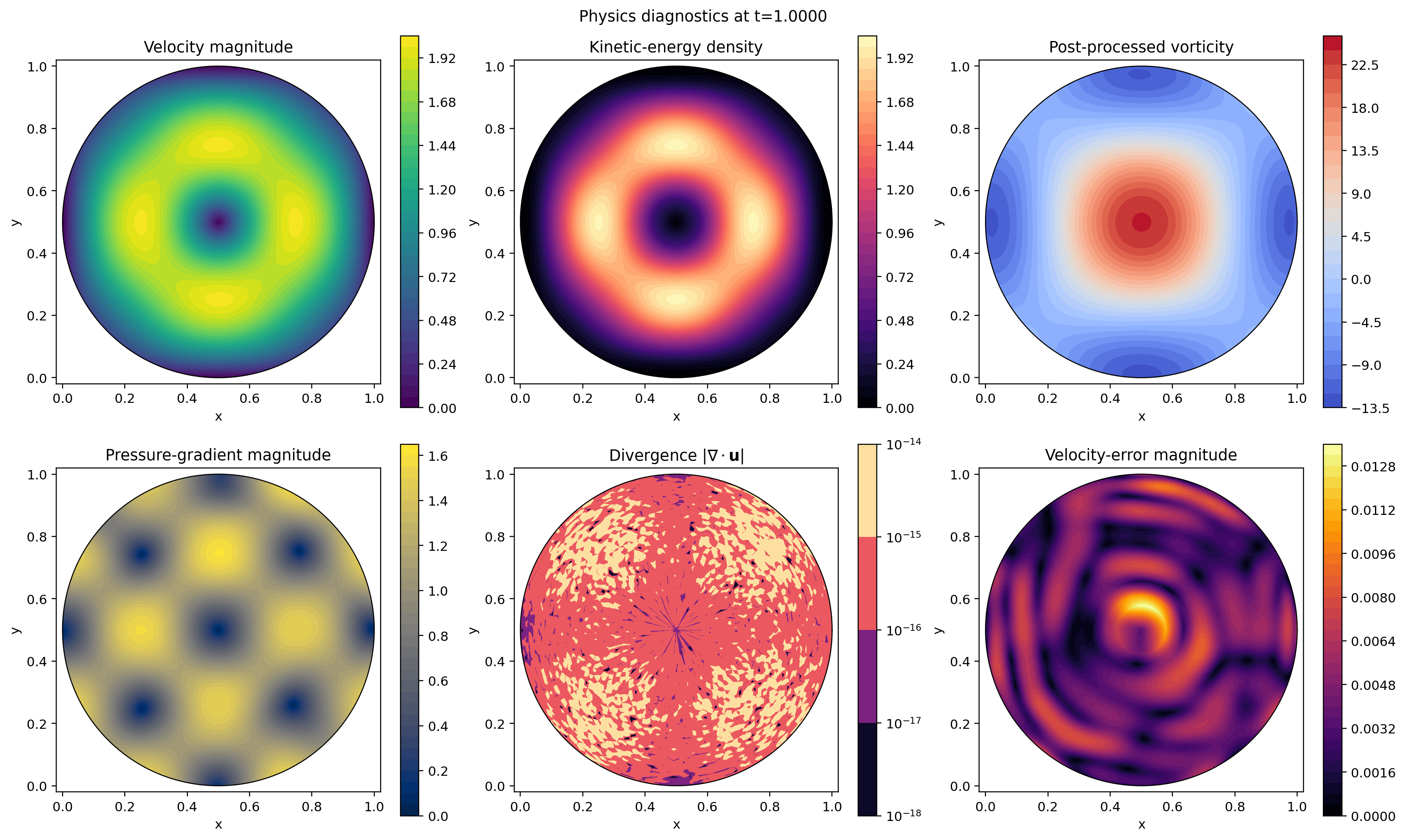}
	\caption{Physical diagnostics of the GeoDiv-KAN solution for circular  domain at \(t=1\): velocity magnitude,
		kinetic-energy density, vorticity, pressure-gradient magnitude,
		divergence magnitude, and velocity-error magnitude.}
	\label{physics_diagnostics_2}
\end{figure}

The physical diagnostics at \(t=1\), shown in
Fig.~\ref{physics_diagnostics_2}, confirm the physical
consistency of the predicted solution. The velocity magnitude and
kinetic-energy density exhibit smooth four-fold symmetric patterns, with
negligible values at the centre and circular boundary. The vorticity
field clearly resolves the dominant positive central region and the
oppositely signed structures near the boundary. The pressure-gradient
distribution also preserves the expected symmetry and remains consistent
with the predicted pressure field. The divergence
magnitude remains below \(5.33\times10^{-15}\), confirming that the
incompressibility constraint is satisfied to machine precision. The
velocity error is spatially localized and remains below approximately
\(1.3\times10^{-2}\).

The quantitative results in Table~\ref{tab:case2_results} show that
GeoDiv-KAN achieves the lowest RMSE, maximum absolute error, and relative
\(L^2\) error for all three solution components. The relative \(L^2\) errors obtained for \(u\), \(v\), and \(p\)
are \(3.4570\times10^{-3}\), \(3.3167\times10^{-3}\), and
\(6.5927\times10^{-2}\), respectively. These improvements demonstrate
the effectiveness of the geometry-dependent gating, Gaussian RBF edge
functions, and divergence-free streamfunction representation for
approximating the TFNSE solution on a circle domain.
\begin{table}[H]
	\centering
	\caption{Comparison of different models for the circular domain at $t=1.0$.}
	\label{tab:case2_results}
	
	\small
	\renewcommand{\arraystretch}{1.20}
	\setlength{\tabcolsep}{4pt}
	
	\begin{tabular}{llccc}
		\toprule
		\textbf{Model} &
		\textbf{Component} &
		\textbf{RMSE} &
		\textbf{Max. Absolute Error} &
		\textbf{Relative $L_2$ Error} \\
		\midrule
		
		\multirow{3}{*}{\textbf{GeoDiv-KAN}}
		& $u$ & $\mathbf{3.3680\times10^{-3}}$ 
		& $\mathbf{1.3169\times10^{-2}}$ 
		& $\mathbf{3.4570\times10^{-3}}$ \\
		& $v$ & $\mathbf{3.2314\times10^{-3}}$ 
		& $\mathbf{1.1034\times10^{-2}}$ 
		& $\mathbf{3.3167\times10^{-3}}$ \\
		& $p$ & $\mathbf{7.7882\times10^{-3}}$ 
		& $\mathbf{2.2770\times10^{-2}}$ 
		& $\mathbf{6.5927\times10^{-2}}$ \\
		\midrule
		
		\multirow{3}{*}{DF-RBFKAN}
		& $u$ & $3.4127\times10^{-1}$ 
		& $8.0141\times10^{-1}$ 
		& $3.4997\times10^{-1}$ \\
		& $v$ & $3.5364\times10^{-1}$ 
		& $8.8597\times10^{-1}$ 
		& $3.6265\times10^{-1}$ \\
		& $p$ & $4.5295\times10^{-1}$ 
		& $7.7488\times10^{-1}$ 
		& $3.7897\times10^{0}$ \\
		\midrule
		
		\multirow{3}{*}{RBF-KAN}
		& $u$ & $3.8925\times10^{-1}$ 
		& $9.5774\times10^{-1}$ 
		& $3.9918\times10^{-1}$ \\
		& $v$ & $3.9686\times10^{-1}$ 
		& $9.8831\times10^{-1}$ 
		& $4.0697\times10^{-1}$ \\
		& $p$ & $7.8752\times10^{-1}$ 
		& $1.1443\times10^{0}$ 
		& $6.5890\times10^{0}$ \\
		\midrule
		
		\multirow{3}{*}{Base PIKAN}
		& $u$ & $7.3374\times10^{-1}$ 
		& $1.5544\times10^{0}$ 
		& $7.5244\times10^{-1}$ \\
		& $v$ & $7.3208\times10^{-1}$ 
		& $1.5209\times10^{0}$ 
		& $7.5074\times10^{-1}$ \\
		& $p$ & $9.4623\times10^{-2}$ 
		& $2.0573\times10^{-1}$ 
		& $7.9168\times10^{-1}$ \\
		\midrule
		
		\multirow{3}{*}{PINN}
		& $u$ & $1.0866\times10^{0}$ 
		& $2.3270\times10^{0}$ 
		& $1.1143\times10^{0}$ \\
		& $v$ & $1.0401\times10^{0}$ 
		& $2.4822\times10^{0}$ 
		& $1.0666\times10^{0}$ \\
		& $p$ & $1.6286\times10^{0}$ 
		& $3.0609\times10^{0}$ 
		& $1.3626\times10^{1}$ \\
		\bottomrule
		
	\end{tabular}
\end{table}

%%%%%%%%%%%%%%%%%%%%%%%%%%%%%%%%%%%%%%%%%%%%
\subsection{\textbf{Case 3: Star-Shaped Domain}}
\begin{figure}[H]
\centering	\includegraphics[width=0.4\textwidth]{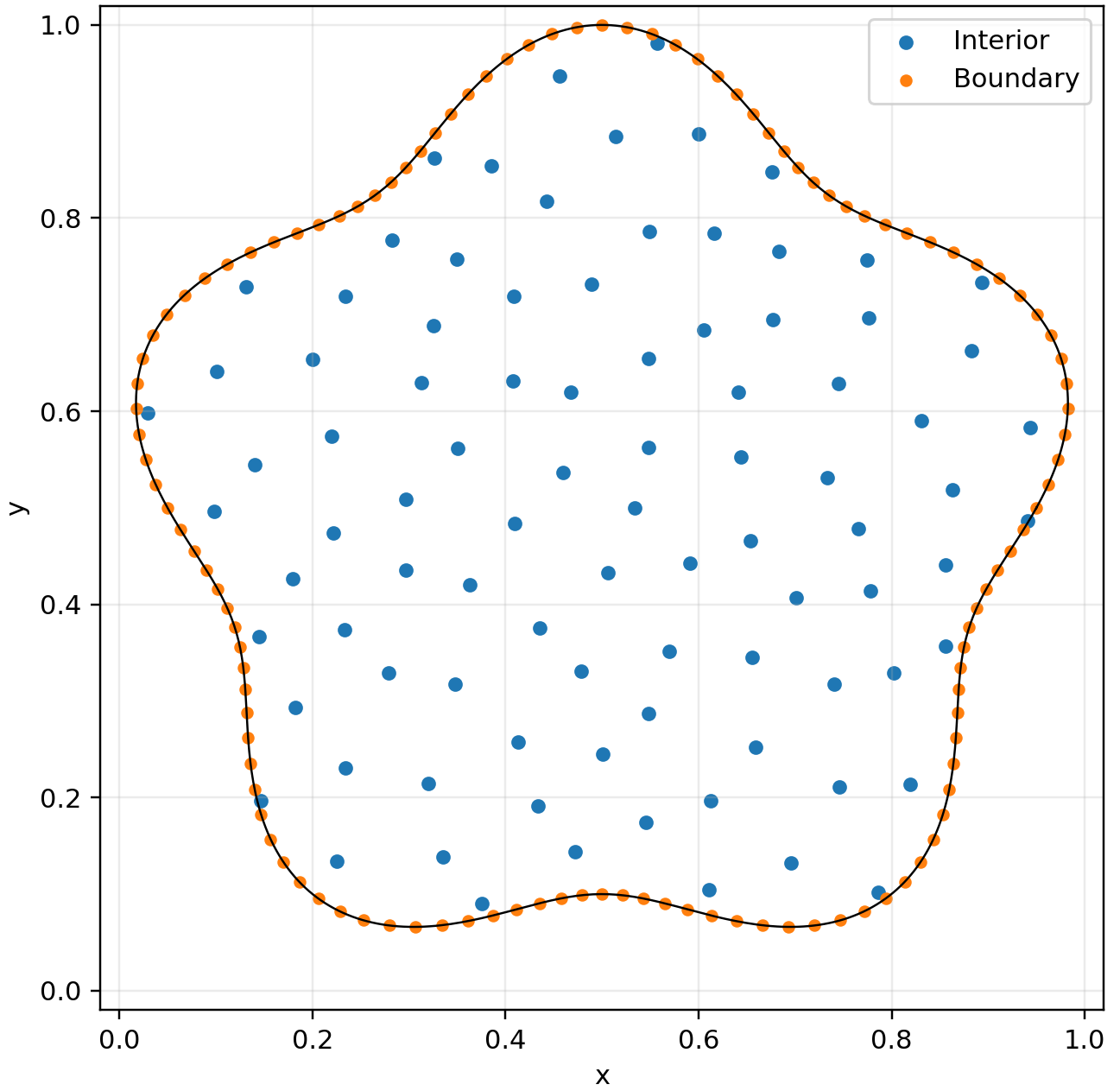}
\caption{Distribution of interior and boundary collocation points on
	the star-shaped domain.}
\label{Star_D_P_3}
\end{figure}

In the third case, we solve the TFNSE system given in
Eq.~\eqref{eq:tfnse_system_M} using the same source terms and exact
solutions as those employed in Case~2. Thus, \(f_1\) and \(f_2\) are
defined by Eqs.~\eqref{eq:case2_f1} and \eqref{eq:case2_f2},
respectively, while the exact velocity and pressure fields are given by
Eq.~\eqref{eq:case2_exact} and the
star-shaped domain shown in Fig.~\ref{Star_D_P_3},  boundary is
parameterized as
\begin{equation}
	\begin{cases}
		x(\theta)
		=
		\displaystyle
		\frac{1}{2}
		\left[0.9+0.1\sin(5\theta)\right]\cos\theta
		+\frac{1}{2},
		\\[0.8ex]
		y(\theta)
		=
		\displaystyle
		\frac{1}{2}
		\left[0.9+0.1\sin(5\theta)\right]\sin\theta
		+\frac{1}{2},
	\end{cases}
	\qquad
	0\leq\theta\leq2\pi.
	\label{eq:star_boundary}
\end{equation}
The corresponding computational domain is defined as
\begin{equation}
	\Omega_{\mathrm{star}}
	=
	\left\{
	\left(
	\frac{r}{2}
	\left[0.9+0.1\sin(5\theta)\right]\cos\theta+\frac{1}{2},
	\frac{r}{2}
	\left[0.9+0.1\sin(5\theta)\right]\sin\theta+\frac{1}{2}
	\right)
	\,\middle|\,
	0\leq r\leq1,\;
	0\leq\theta\leq2\pi
	\right\}.
	\label{eq:star_domain}
\end{equation}
The initial and boundary conditions are obtained from
Eq.~\eqref{eq:case2_exact} and imposed on
\(\Omega_{\mathrm{star}}\) and \(\partial\Omega_{\mathrm{star}}\),
respectively.\\

To solve this case, 90 interior spatial collocation points and 120
boundary points were distributed over the five-lobed star-shaped domain.
The interval \(t\in[0,1]\) was divided into 41 temporal levels with
\(\Delta t=0.025\), resulting in \(90\times40=3600\) interior
space--time residual points and \(120\times41=4920\) boundary
space--time points. The computations were performed using
\(\alpha=0.8\) and \(Re=400\).

\begin{figure}[H]
	\centering	\includegraphics[width=1.0\textwidth]{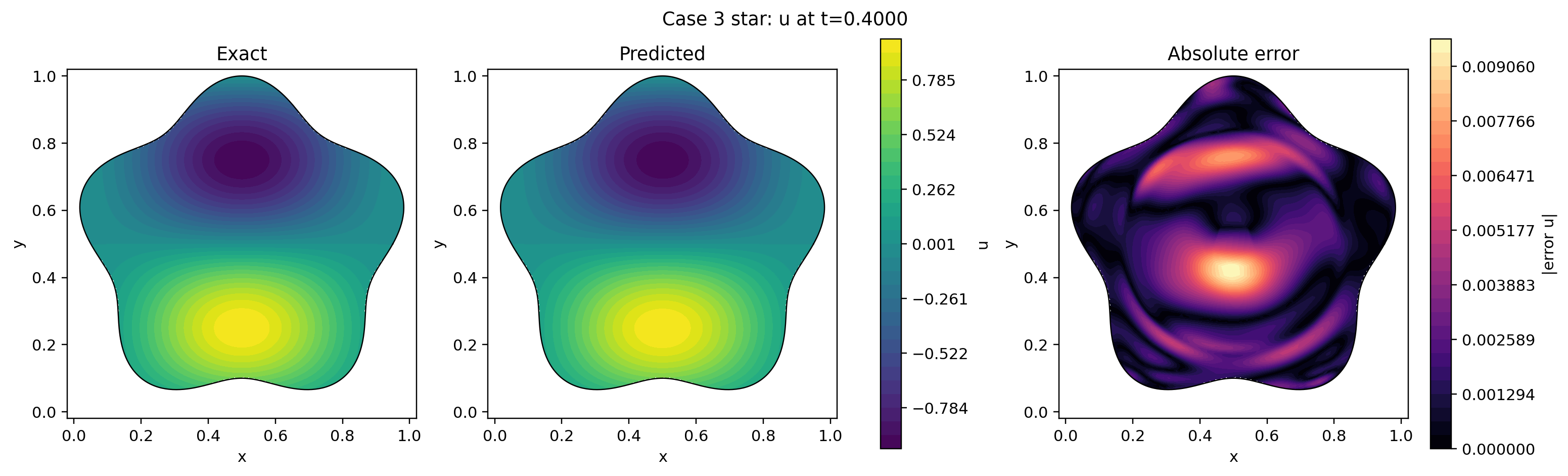}
	\caption{Comparison of the exact and GeoDiv-KAN predicted fields $u$,
		together with absolute errors, for star shape at \(t=0.4\).}
	\label{u_exact_pred_error_t_0p4000_3}
\end{figure}

\begin{figure}[H]
	\centering	\includegraphics[width=1.0\textwidth]{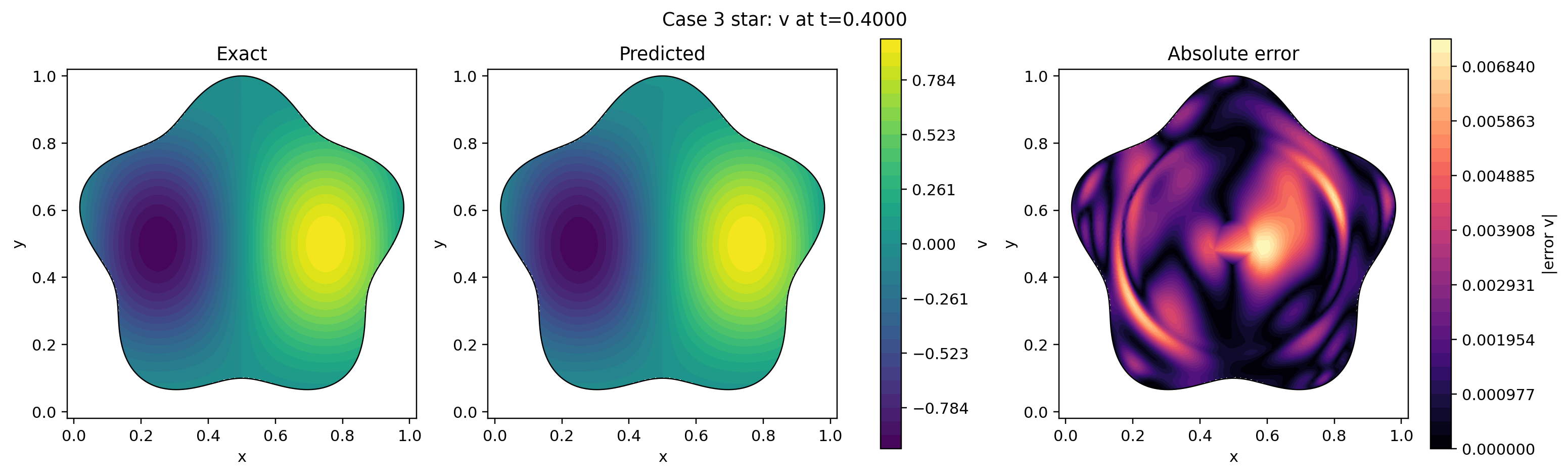}
	\caption{Comparison of the exact and GeoDiv-KAN predicted fields $v$,
		together with absolute errors, for star shape at \(t=0.4\).}
	\label{v_exact_pred_error_t_0p4000_3}
\end{figure}

\begin{figure}[H]
	\centering	\includegraphics[width=1.0\textwidth]{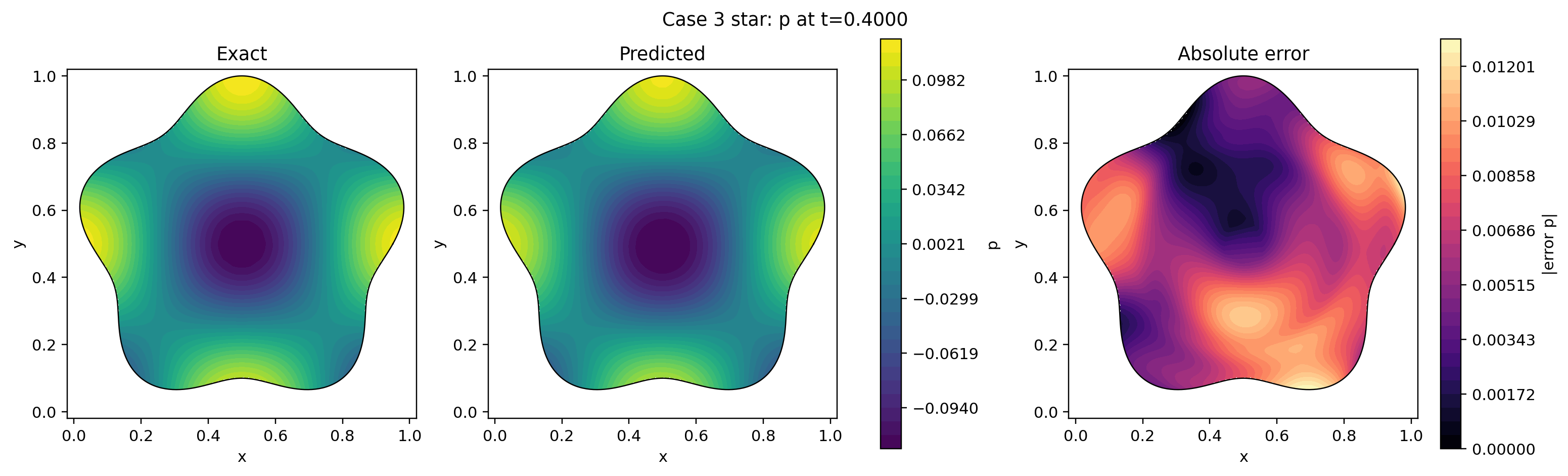}
	\caption{Comparison of the exact and GeoDiv-KAN predicted fields $p$,
		together with absolute errors, for star shape at \(t=0.4\).}
	\label{p_exact_pred_error_t_0p4000_3}
\end{figure}
Figs. \ref{u_exact_pred_error_t_0p4000_3}, \ref{v_exact_pred_error_t_0p4000_3}, and \ref{p_exact_pred_error_t_0p4000_3} compares the analytical and predicted
fields at \(t=0.4\) for star case. GeoDiv-KAN accurately reproduces the spatial
patterns, symmetries, and amplitudes of \(u\), \(v\), and \(p\) on the
star-shaped domain. The absolute errors are below
\(9.06\times10^{-3}\), \(6.84\times10^{-3}\), and
\(1.20\times10^{-2}\) for \(u\), \(v\), and \(p\), respectively. The
close agreement demonstrates the model's ability to resolve the TFNSE
solution on a complex non-convex domain.

\begin{figure}[H]
	\centering	\includegraphics[width=1.0\textwidth]{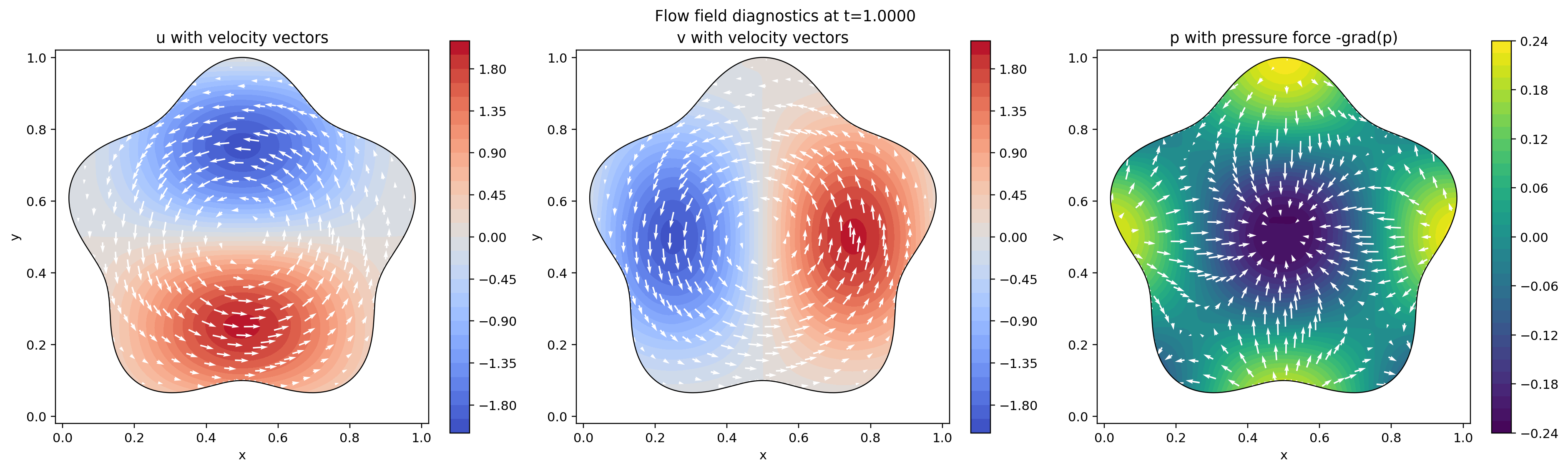}
	\caption{GeoDiv-KAN predicted flow field diagnostics for
		star-shaped domain at \(t=1\), first two panels show \(u\) and
		\(v\) with the velocity vectors, and third presents
		\(p\) with the pressure-force vectors.}
	\label{flow_field_diagnostics_3}
\end{figure}

\begin{figure}[H]
	\centering	\includegraphics[width=1.0\textwidth]{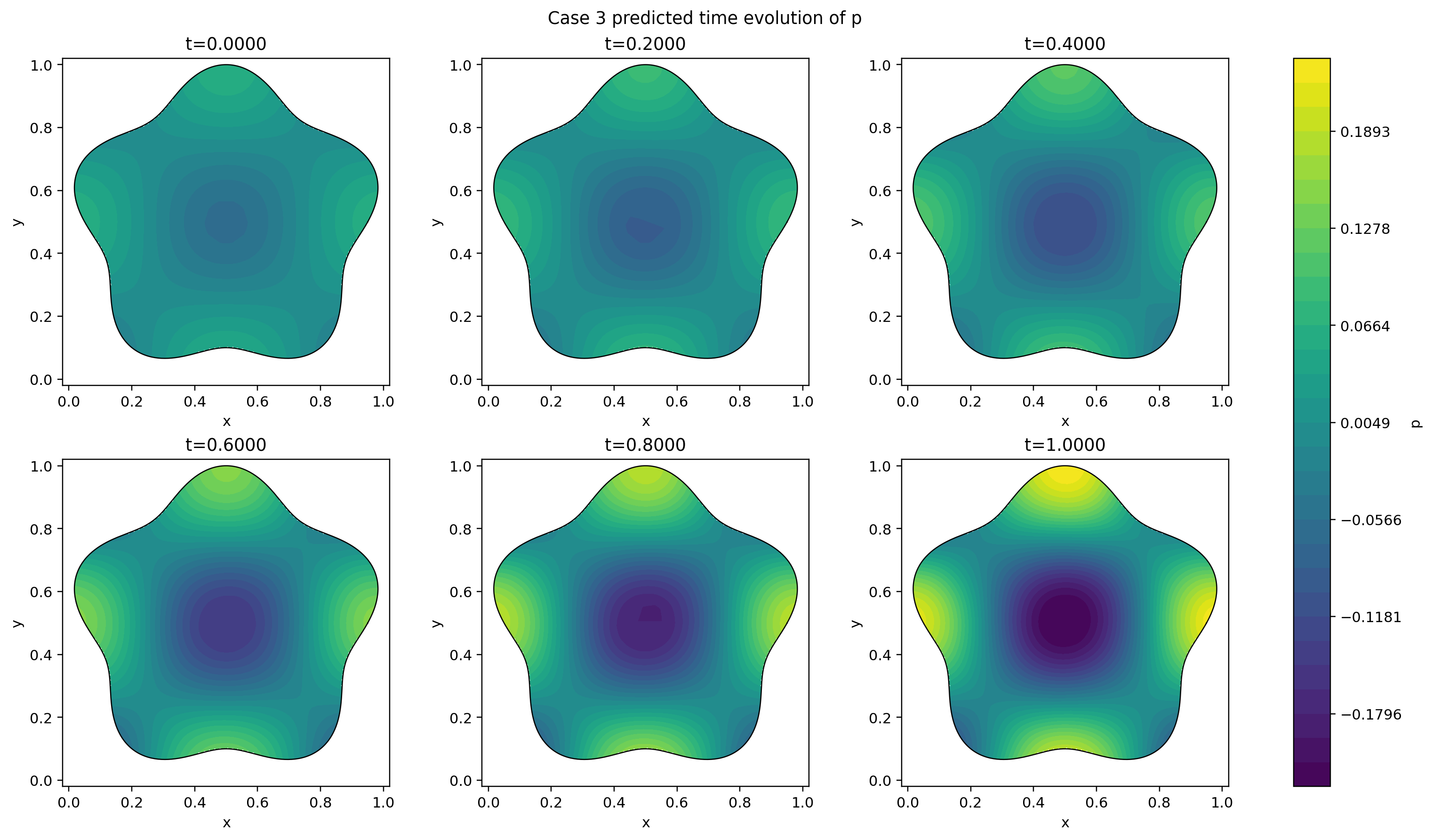}
	\caption{GeoDiv-KAN predicted temporal evolution of  the solution fields $p$ on the star-shaped domain at
		\(t=0,\,0.2,\,0.4,\,0.6,\,0.8,\) and \(1.0\).}
	\label{time_evolution_predicted_p_3}
\end{figure}

\begin{figure}[H]
	\centering	\includegraphics[width=1.0\textwidth]{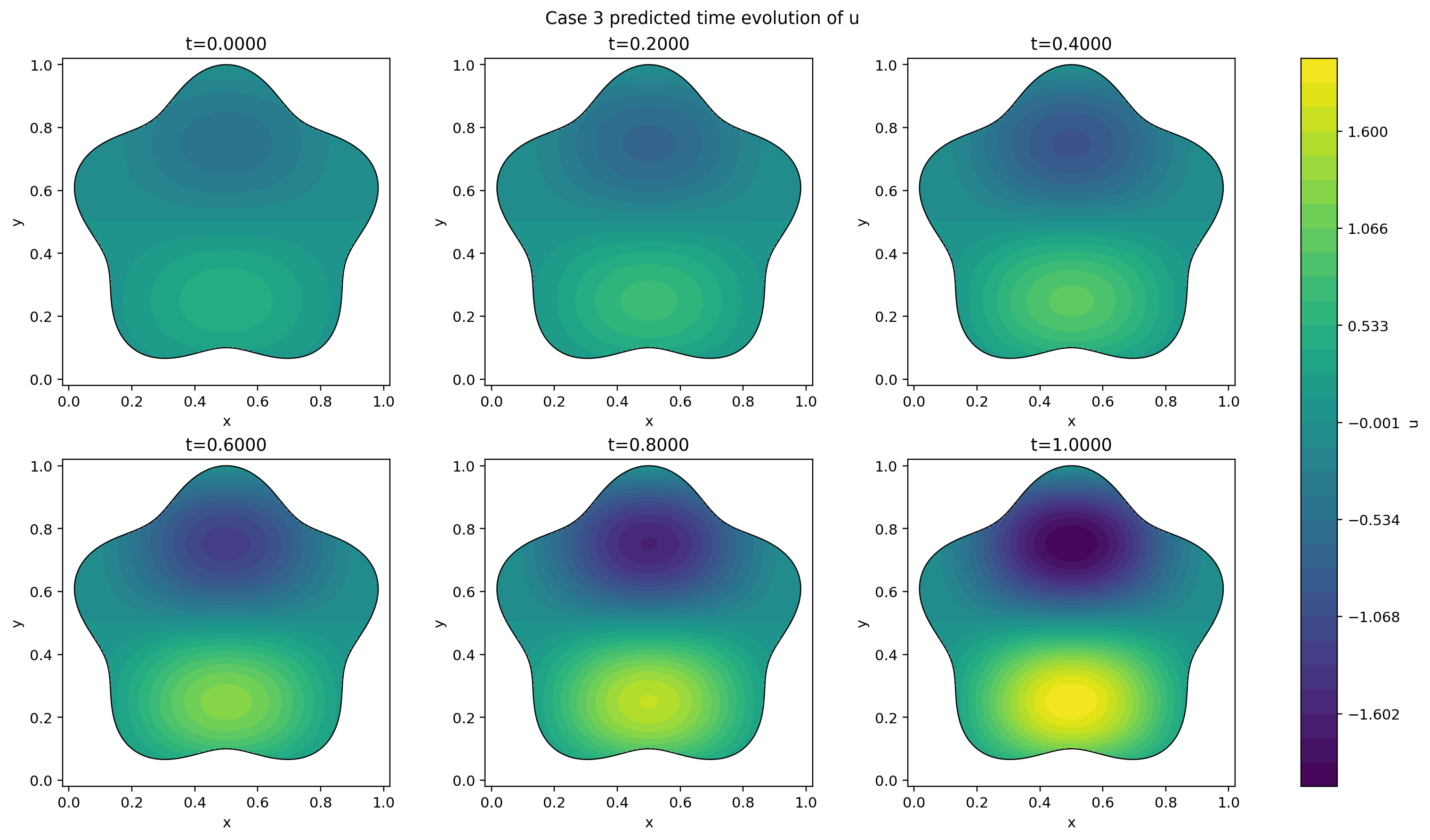}
	\caption{GeoDiv-KAN predicted temporal evolution of  the solution fields $u$ on the star-shaped domain at
		\(t=0,\,0.2,\,0.4,\,0.6,\,0.8,\) and \(1.0\).}
	\label{time_evolution_predicted_u_3}
\end{figure}

\begin{figure}[H]
	\centering	\includegraphics[width=1.0\textwidth]{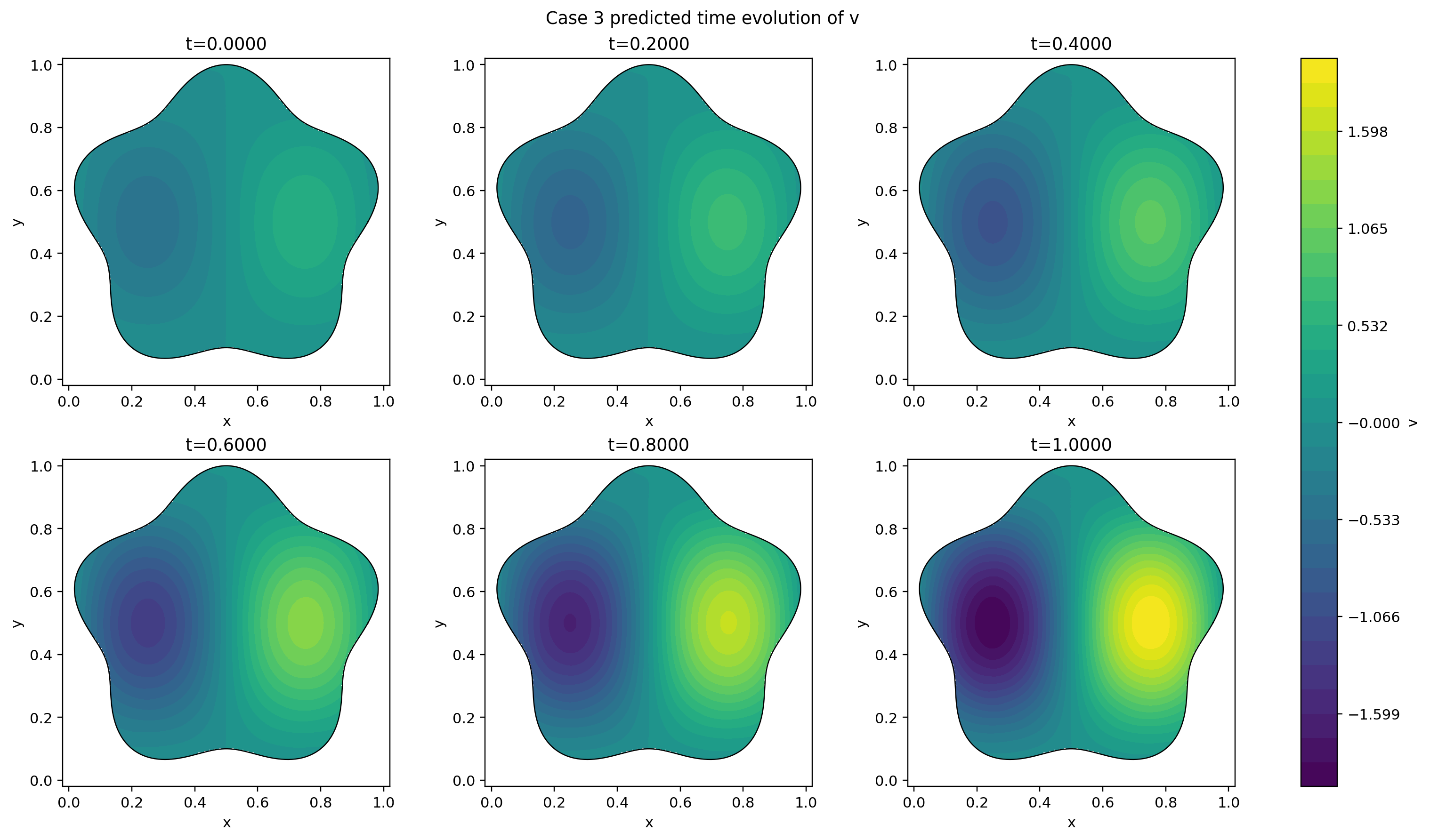}
	\caption{GeoDiv-KAN predicted temporal evolution of  the solution fields $v$ on the star-shaped domain at
		\(t=0,\,0.2,\,0.4,\,0.6,\,0.8,\) and \(1.0\).}
	\label{time_evolution_predicted_v_3}
\end{figure}

The flow-field diagnostics at \(t=1\), shown in
Fig.~\ref{flow_field_diagnostics_3}, reveal a smooth counterclockwise
circulation that conforms to the non-convex star-shaped boundary. The
predicted \(u\) and \(v\) fields preserve the expected spatial
symmetries, while the pressure field contains a central low-pressure
region and higher-pressure regions near the outward lobes. The
pressure-force vectors, \(-\nabla p\), are predominantly directed from
these high-pressure regions toward the centre. 
Figs.~\ref{time_evolution_predicted_p_3}, \ref{time_evolution_predicted_u_3},and \ref{time_evolution_predicted_v_3} illustrates the evolution of the
GeoDiv-KAN-predicted fields over \(t\in[0,1]\). Their spatial patterns
and symmetries remain stable. Smooth contours across all time levels confirm that the
model captures the TFNSE dynamics while consistently accommodating the
non-convex star-shaped boundary.

\begin{figure}[H]
	\centering	\includegraphics[width=1.0\textwidth]{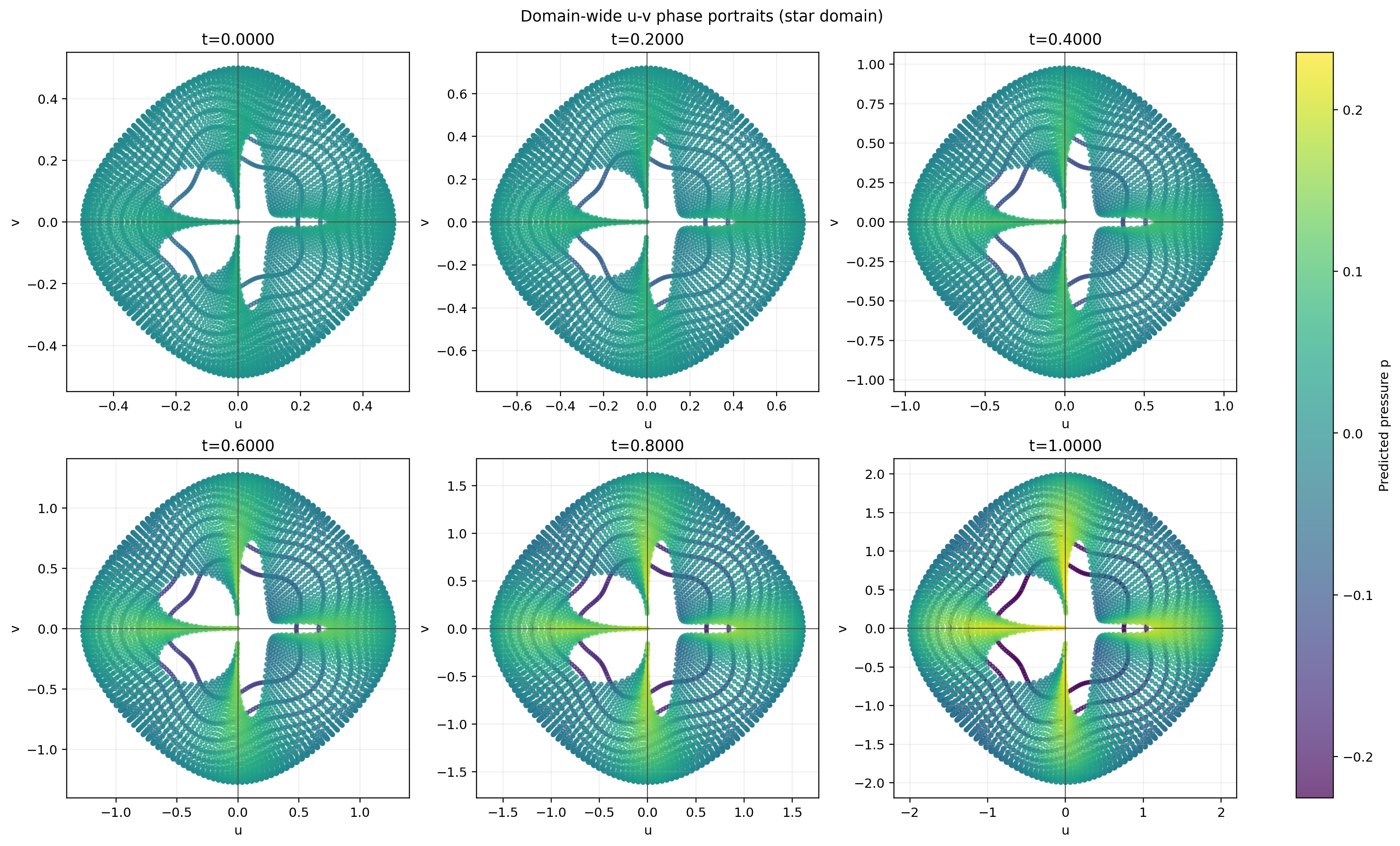}
	\caption{Domain-wide distributions of the predicted velocity
		pairs \((u,v)\) for star shape case at
		different t, coloured by the
		corresponding predicted pressure.}
	\label{uv_phase_portraits_3}
\end{figure}

The predicted $(u,v)$ distributions at various time levels are displayed in Fig.~\ref{uv_phase_portraits_3}, with color denoting the associated pressure. While the maintained pattern shows steady velocity--pressure coupling throughout time, the progressive expansion shows increasing velocity amplitudes.

\begin{figure}[H]
	\centering	\includegraphics[width=1.0\textwidth]{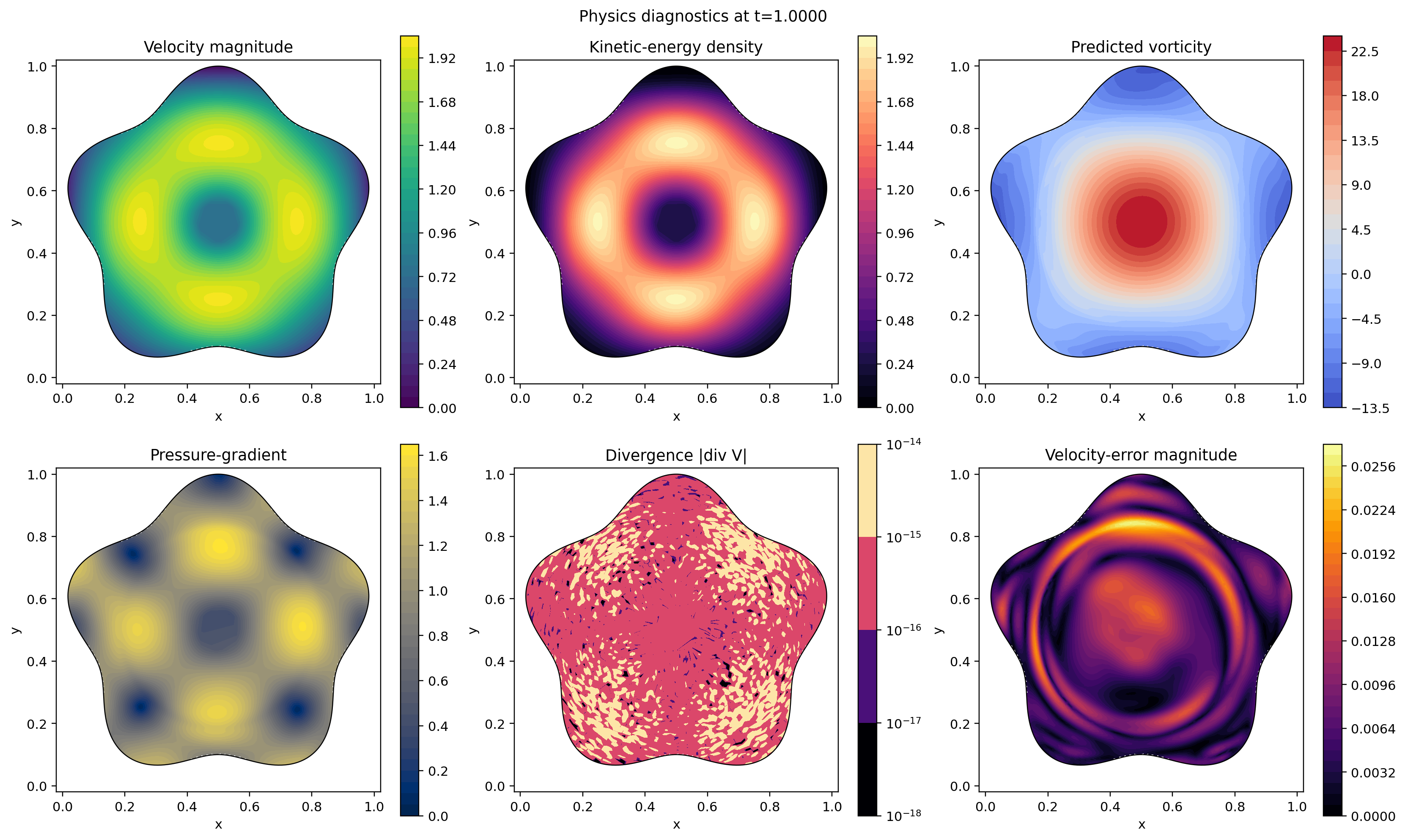}
	\caption{Physical diagnostics of the GeoDiv-KAN solution for star-shaped domain at \(t=1\): velocity magnitude,
		kinetic-energy density, vorticity, pressure-gradient magnitude,
		divergence magnitude, and velocity-error magnitude.}
	\label{physics_diagnostics_3}
\end{figure}

Fig.~\ref{physics_diagnostics_3} shows smooth and physically
consistent flow diagnostics at \(t=1\). The divergence remains below
\(4.44\times10^{-15}\), confirming incompressibility to machine
precision, while the localized velocity error remains below
\(2.6\times10^{-2}\). These results demonstrate accurate prediction on
the non-convex star-shaped domain.

Table~\ref{tab:case3_results} shows that GeoDiv-KAN achieves the lowest
errors for all solution components compared with other models. This substantial
improvement demonstrates the effectiveness of GeoDiv-KAN for solving the TFNSE on the non-convex star-shaped domain.
\begin{table}[H]
	\centering
	\caption{Comparison of different models for the star-shaped domain at $t=1.0$.}
	\label{tab:case3_results}
	
	\small
	\renewcommand{\arraystretch}{1.20}
	\setlength{\tabcolsep}{4pt}
	
	\begin{tabular}{llccc}
		\toprule
		\textbf{Model} &
		\textbf{Component} &
		\textbf{RMSE} &
		\textbf{Max. Absolute Error} &
		\textbf{Relative $L_2$ Error} \\
		\midrule
		
		\multirow{3}{*}{\textbf{GeoDiv-KAN}}
		& $u$ & $\mathbf{6.8132\times10^{-3}}$ 
		& $\mathbf{2.6302\times10^{-2}}$ 
		& $\mathbf{6.3748\times10^{-3}}$ \\
		& $v$ & $\mathbf{6.0799\times10^{-3}}$ 
		& $\mathbf{2.0223\times10^{-2}}$ 
		& $\mathbf{5.6894\times10^{-3}}$ \\
		& $p$ & $\mathbf{1.2668\times10^{-2}}$ 
		& $\mathbf{3.0300\times10^{-2}}$ 
		& $\mathbf{1.1350\times10^{-1}}$ \\
		\midrule
		
		\multirow{3}{*}{DF-RBFKAN}
		& $u$ & $3.7335\times10^{-2}$ 
		& $9.0434\times10^{-2}$ 
		& $3.5317\times10^{-2}$ \\
		& $v$ & $4.4447\times10^{-2}$ 
		& $1.0828\times10^{-1}$ 
		& $4.2060\times10^{-2}$ \\
		& $p$ & $1.7635\times10^{-1}$ 
		& $2.9138\times10^{-1}$ 
		& $1.5391\times10^{0}$ \\
		\midrule
		
		\multirow{3}{*}{RBF-KAN}
		& $u$ & $3.9613\times10^{-1}$ 
		& $8.3906\times10^{-1}$ 
		& $3.7473\times10^{-1}$ \\
		& $v$ & $4.0277\times10^{-1}$ 
		& $9.2519\times10^{-1}$ 
		& $3.8114\times10^{-1}$ \\
		& $p$ & $4.9192\times10^{-1}$ 
		& $8.1766\times10^{-1}$ 
		& $4.2932\times10^{0}$ \\
		\midrule
		
		\multirow{3}{*}{Base PIKAN}
		& $u$ & $6.2880\times10^{-1}$ 
		& $1.2938\times10^{0}$ 
		& $5.9482\times10^{-1}$ \\
		& $v$ & $5.7534\times10^{-1}$ 
		& $1.3826\times10^{0}$ 
		& $5.4444\times10^{-1}$ \\
		& $p$ & $4.0842\times10^{-1}$ 
		& $8.4697\times10^{-1}$ 
		& $3.5645\times10^{0}$ \\
		\midrule
		
		\multirow{3}{*}{PINN}
		& $u$ & $9.2025\times10^{-1}$ 
		& $1.7443\times10^{0}$ 
		& $8.7052\times10^{-1}$ \\
		& $v$ & $9.4294\times10^{-1}$ 
		& $1.8701\times10^{0}$ 
		& $8.9230\times10^{-1}$ \\
		& $p$ & $7.1896\times10^{-1}$ 
		& $1.3541\times10^{0}$ 
		& $6.2747\times10^{0}$ \\
		\bottomrule
		
	\end{tabular}
\end{table}

\section{Application}\label{sec6}
The
applicability of GeoDiv-KAN is further examined using CFD-generated flow
fields from two engineering configurations. The first application
considers two-dimensional unsteady vortex shedding behind a fixed
circular cylinder at \(Re=100\), whereas the second considers steady
turbulent RANS flow through a complex U-bend geometry at \(Re=40{,}000\).
The objective is to assess the field-reconstruction capability of the
proposed architecture.

\subsection{Unsteady Flow Past a Circular Cylinder at \(Re=100\)}
\begin{figure}[H]
	\centering	\includegraphics[width=1.0\textwidth]{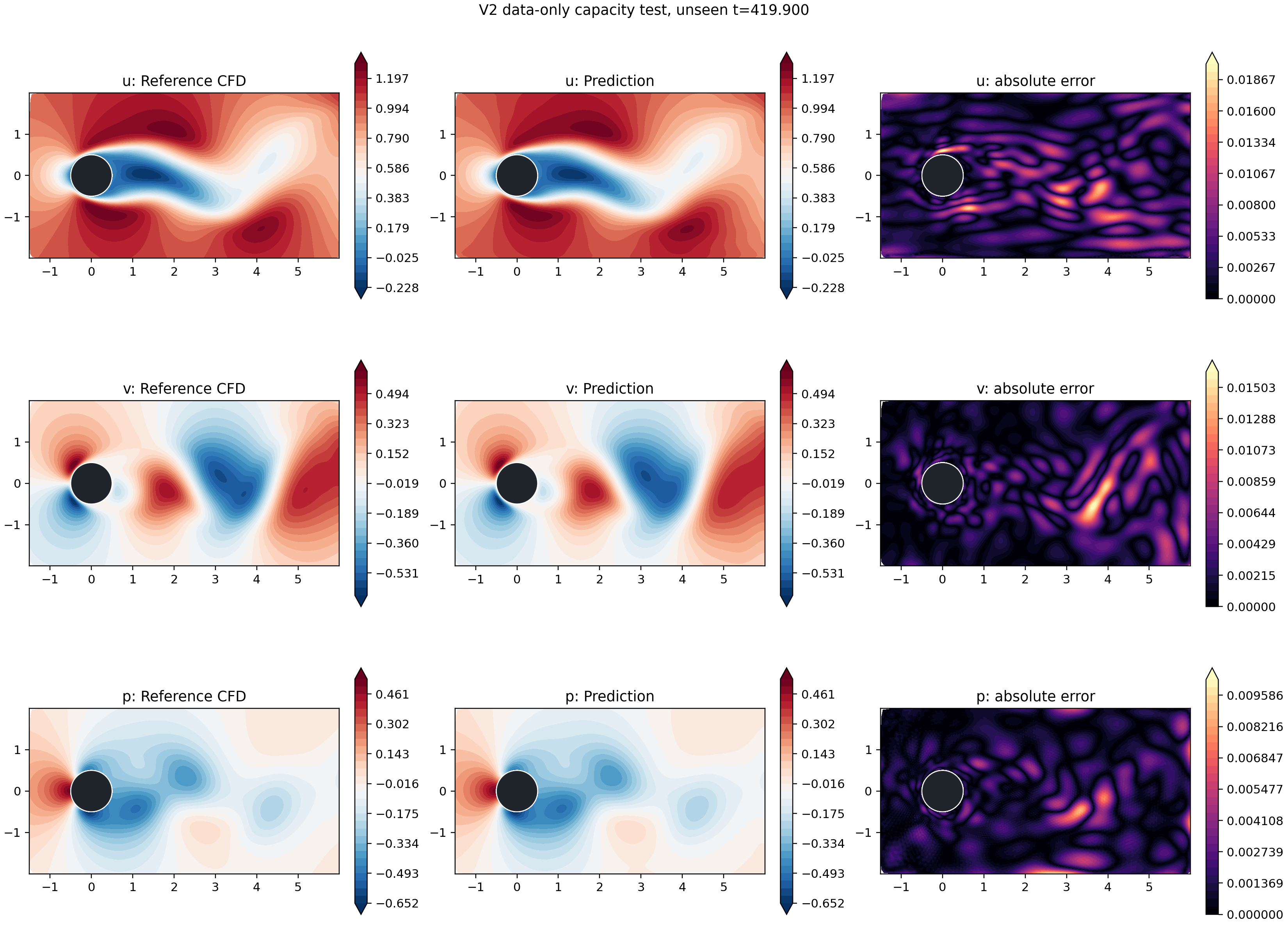}
	\caption{Comparison of the reference CFD and GeoDiv-KAN reconstructed
		fields at the unseen test time \(t=419.9\) for \(u\),
		\(v\), \(p\), and pointwise absolute error.}
	\label{Experiment_1_1}
\end{figure}

To assess GeoDiv-KAN on a realistic unsteady flow, we consider
two-dimensional incompressible flow past a fixed circular cylinder at
\(Re=100\) as shown in Fig. \ref{Experiment_1_1}. The flow exhibits periodic vortex shedding, making it a
suitable test for spatiotemporal CFD-field reconstruction. The reference
CFD data are obtained from the publicly available dataset of
Boudina~\cite{boudina2021cylinder}.
\begin{figure}[H]
	\centering	\includegraphics[width=0.8\textwidth]{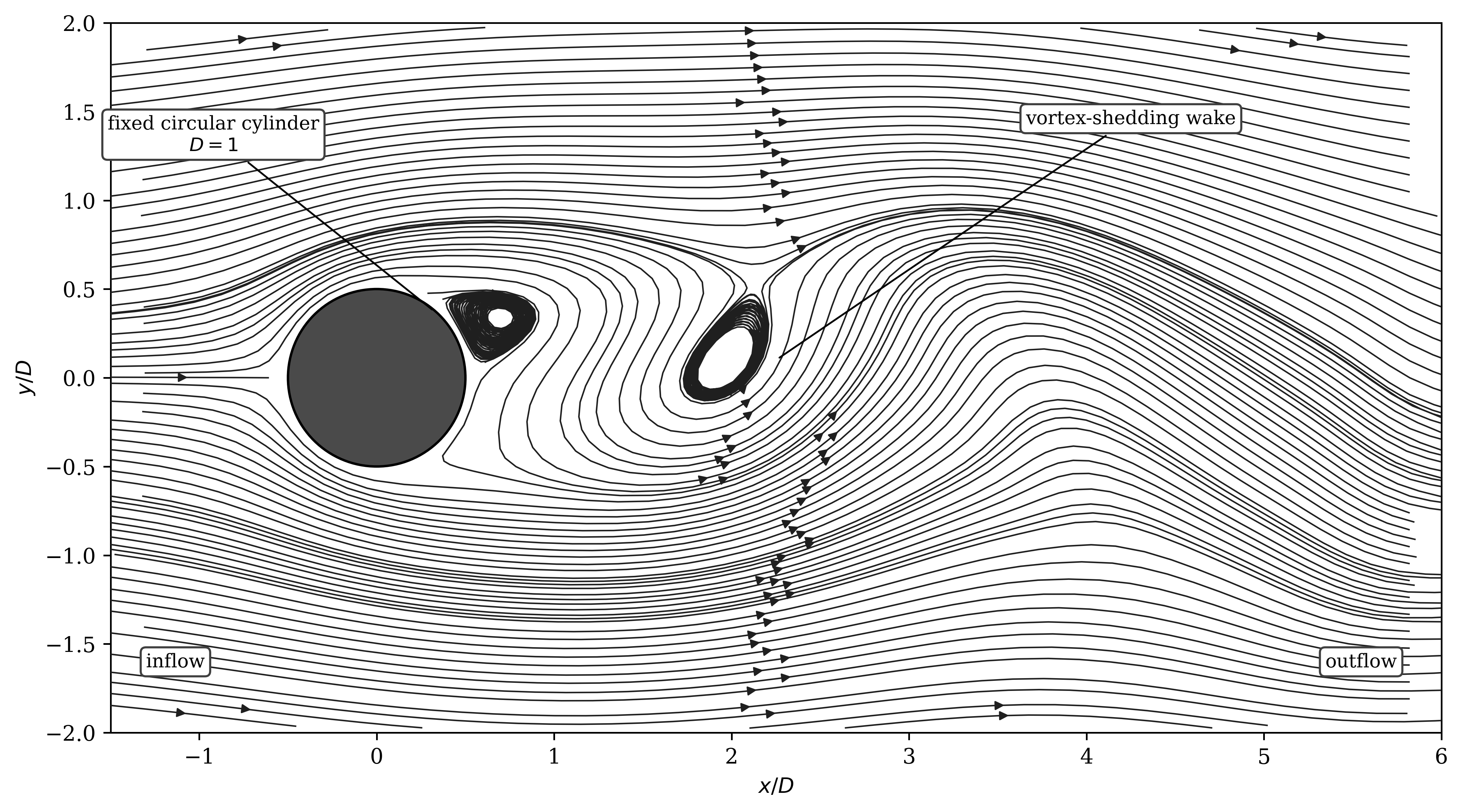}
	\caption{Streamline topology of flow past a fixed circular cylinder at \(Re=100\), showing the inflow, cylinder,
	downstream vortex-shedding wake, and outflow.}
	\label{}
\end{figure}

A compact region containing the near-cylinder flow and downstream wake
was extracted from the original computational domain. Using the cylinder
diameter \(D\) as the characteristic length, this region is defined by
\begin{equation}
	-1.5\leq x/D\leq6,
	\qquad
	-2\leq y/D\leq2.
\end{equation}
In the adopted nondimensional coordinate system, the cylinder is centred
at \((0,0)\), with \(D=1\) and radius \(R=0.5\). The selected region
retains 40,036 of the original 58,205 spatial nodes and covers the
separated shear layers and vortex-shedding wake.

To reconstruct the unsteady CFD fields while retaining the
divergence-free structure of the proposed model, GeoDiv-KAN takes the
normalized spatiotemporal coordinates \((x/D,y/D,t)\) and the
cylinder-geometry features as inputs. The network predicts the
streamfunction and pressure, while the velocity components are derived
from the streamfunction to preserve incompressibility by construction. GeoDiv-KAN employed a hidden width of 18 and 12 Gaussian basis functions
per edge, resulting in 17,388 trainable parameters. The model was trained
for 1000 accepted iterations using the L-BFGS optimizer with a
strong-Wolfe line search.

\begin{figure}[H]
	\centering	\includegraphics[width=1.0\textwidth]{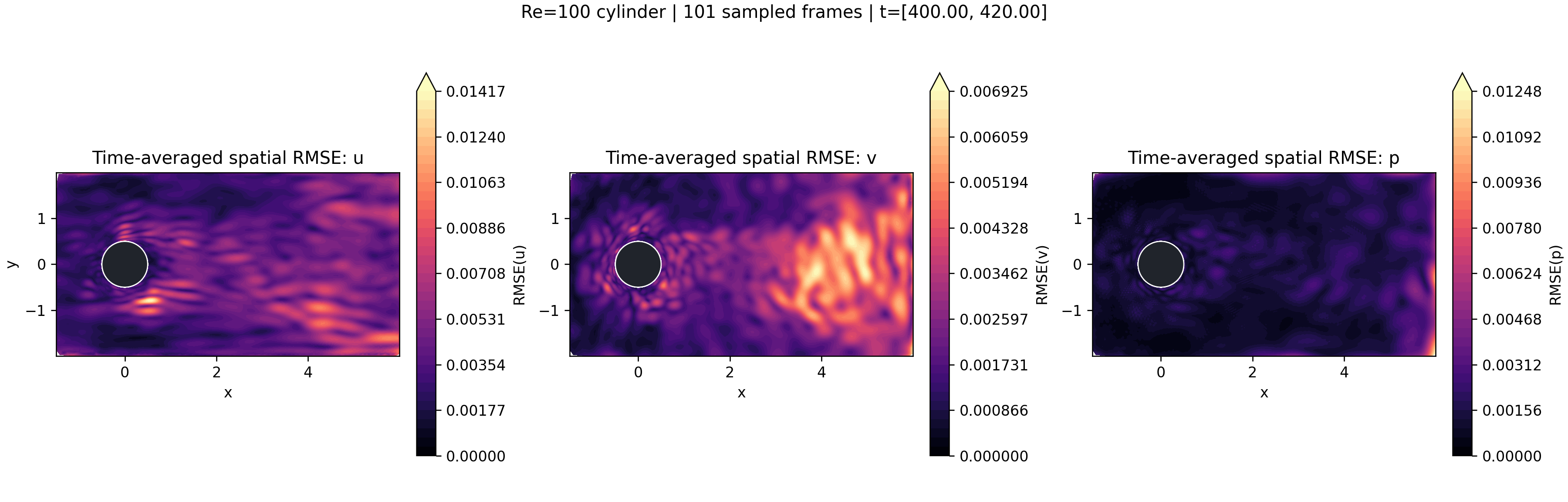}
	\caption{Time-averaged spatial RMSE distributions of \(u\), \(v\), and
		gauge-aligned \(p\)}
	\label{Experiment_2_1}
\end{figure}

At the unseen test time \(t=419.9\), GeoDiv-KAN closely reproduces the
reference velocity and pressure fields, as shown in
Fig.~\ref{Experiment_1_1}. The model captures the
streamwise velocity deficit, alternating transverse-velocity structures,
and pressure variations in the downstream wake. The errors remain
localized primarily around the separated shear layers and vortex cores.
The errors are \(6.49\times10^{-3}\),
\(1.60\times10^{-2}\), and \(9.17\times10^{-3}\) for \(u\), \(v\), and
the gauge-aligned pressure, respectively, demonstrating accurate
reconstruction of the unseen CFD snapshot. Fig. \ref{Experiment_2_1} shows low time-averaged
errors across most of the domain, with larger errors confined to the
shear layers and downstream wake. The global RMSE values are
\(3.83\times10^{-3}\), \(2.65\times10^{-3}\), and
\(1.91\times10^{-3}\) for \(u\), \(v\), and \(p\), respectively,
confirming stable reconstruction over the full time interval.

\begin{figure}[H]
	\centering	\includegraphics[width=1.0\textwidth]{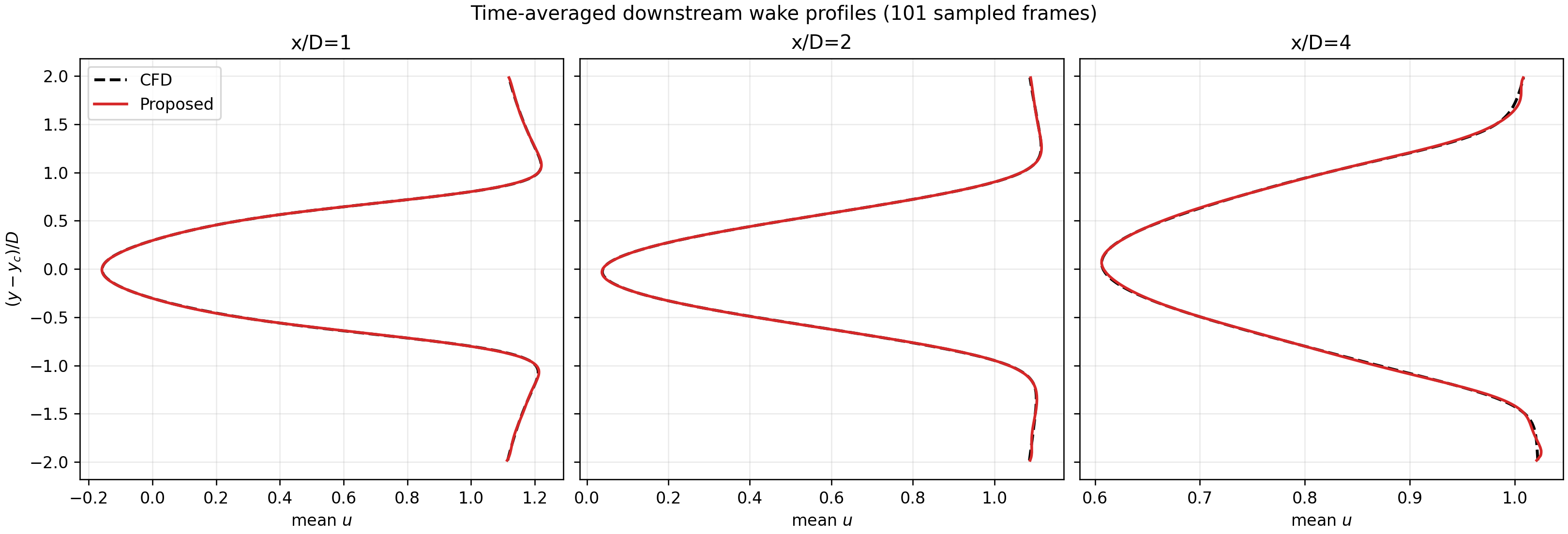}
	\caption{Comparison of the time-averaged CFD and GeoDiv-KAN
		streamwise velocity profiles at \(x/D=1,2,\) and \(4\).}
	\label{Experiment_3_1}
\end{figure}
Time-averaged streamwise-velocity profiles at \(x/D=1,2,\) and \(4\)
are compared in Fig.~\ref{Experiment_3_1}. The predicted
profiles closely follow the reference CFD curves, capturing both the
near-wake velocity deficit and its gradual downstream recovery.
\begin{figure}[H]
	\centering	\includegraphics[width=0.4\textwidth]{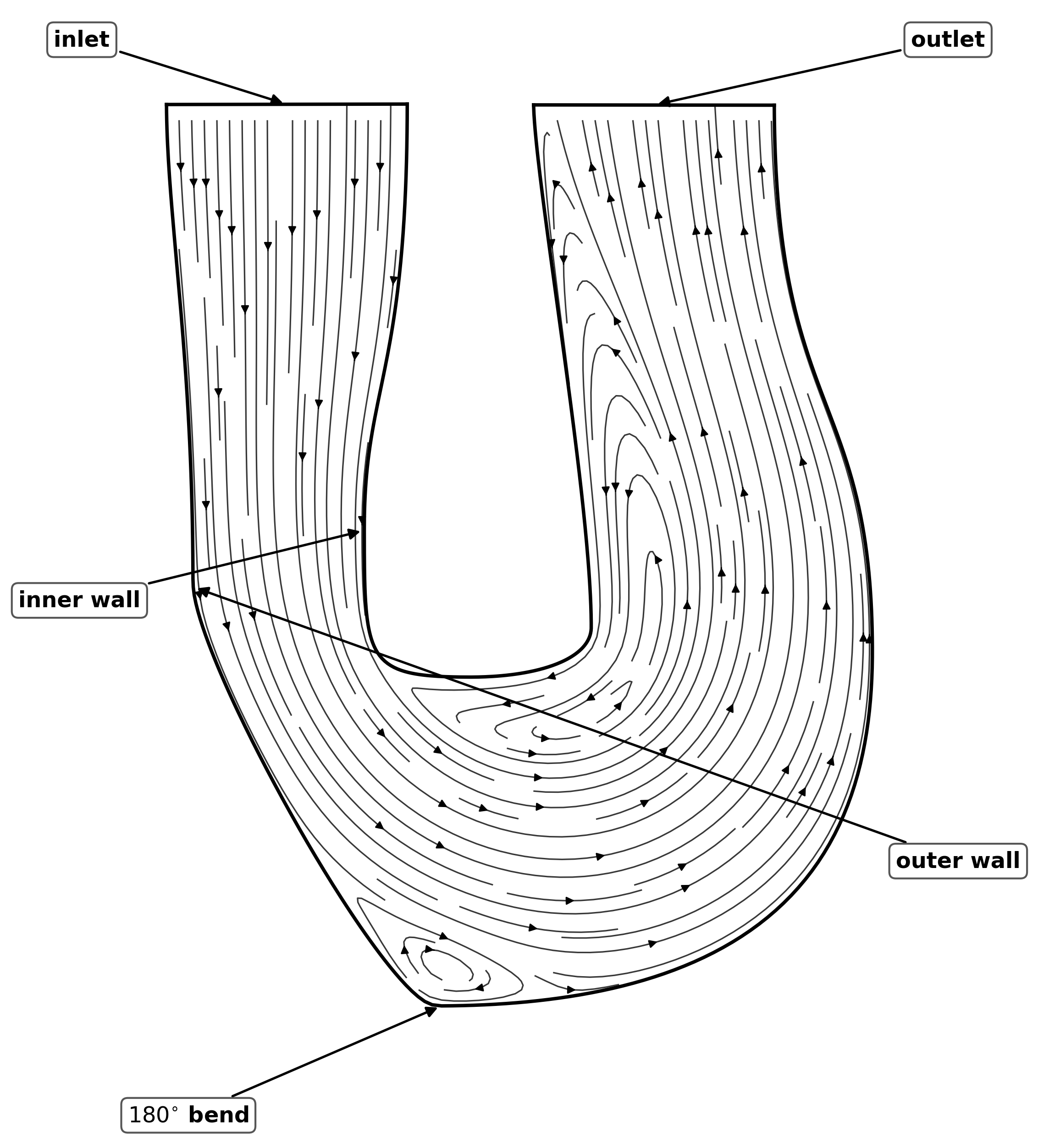}
	\caption{Physical configuration of the selected U-bend flow problem,
		showing the inlet, outlet, inner and outer walls.}
	\label{E_2_initial_problem_configuration}
\end{figure}

\begin{figure}[H]
	\centering	\includegraphics[width=0.8\textwidth]{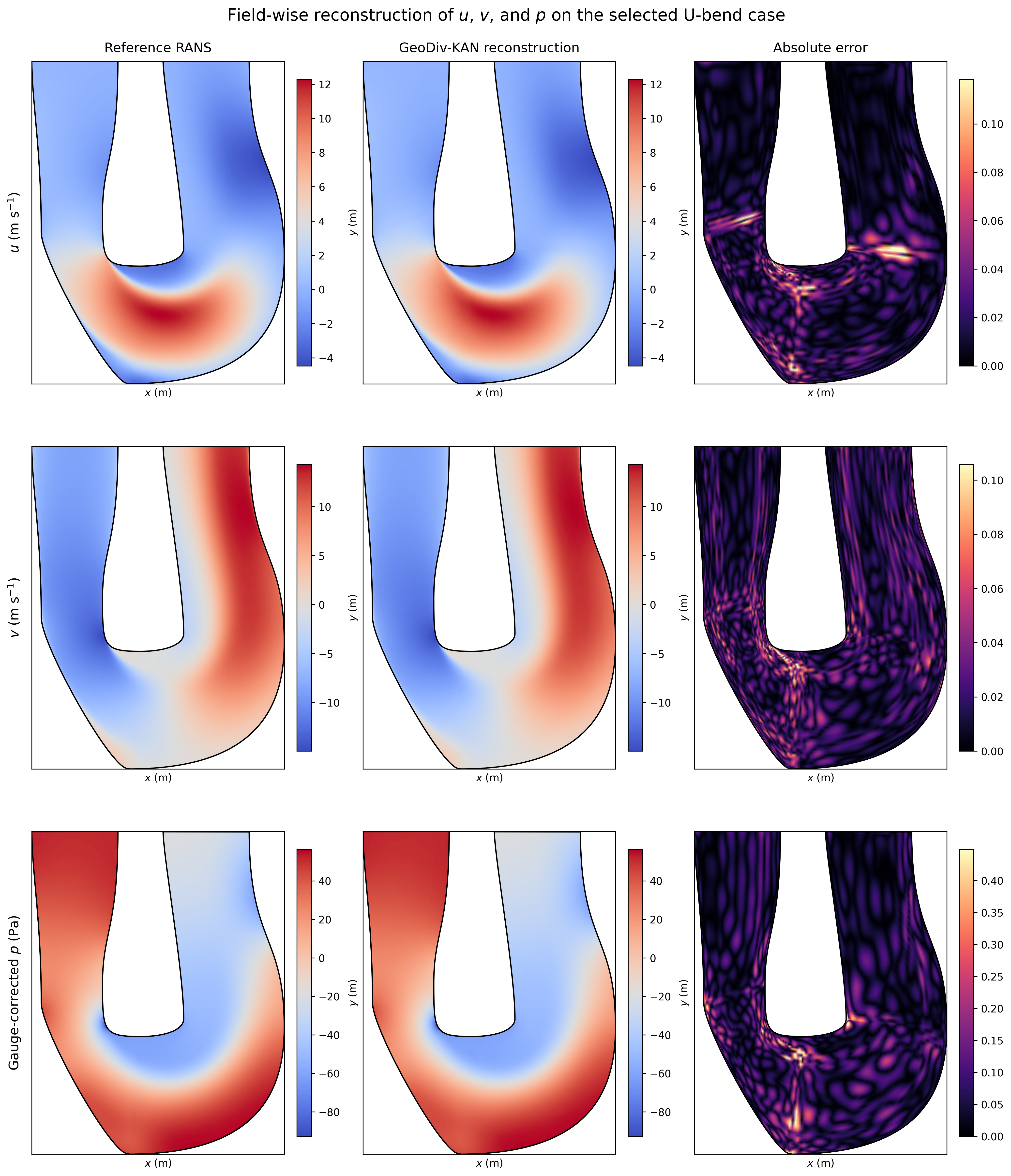}
	\caption{Reference RANS fields, GeoDiv-KAN reconstructions, and absolute
		errors for \(u\), \(v\), and gauge-corrected pressure \(p\).}
	\label{E_2_uvp_reference_reconstruction_absolute_error}
\end{figure}
\subsection{Turbulent RANS Flow Reconstruction in a U-Bend}
This application considers two-dimensional turbulent flow through a
U-bend at \(Re=40{,}000\). As shown in
Fig.~\ref{E_2_initial_problem_configuration}, the flow passes through a
\(180^{\circ}\) bend, producing strong curvature effects, flow separation,
and recirculation. 

The reference velocity and pressure fields were obtained from the
publicly available U-bend CFD dataset of Decke \textit{et al.}~\cite{decke2023ubendrepository},
distributed under the CC BY--NC 4.0 license. The selected case,
\texttt{sample\_100}, contains a \(240\times60\) fluid grid and 28
geometry parameters.

GeoDiv-KAN predicts the streamfunction and gauge-corrected pressure from
the normalized coordinates and geometry features, with velocity derived
from the streamfunction to preserve incompressibility. The model contains
\(50{,}850\) trainable parameters, with width \(32\) and ten Gaussian
basis functions. It is trained for \(4{,}000\) epochs using \(14{,}000\)
sampled grid points.
\begin{figure}[H]
	\centering	\includegraphics[width=1.0\textwidth]{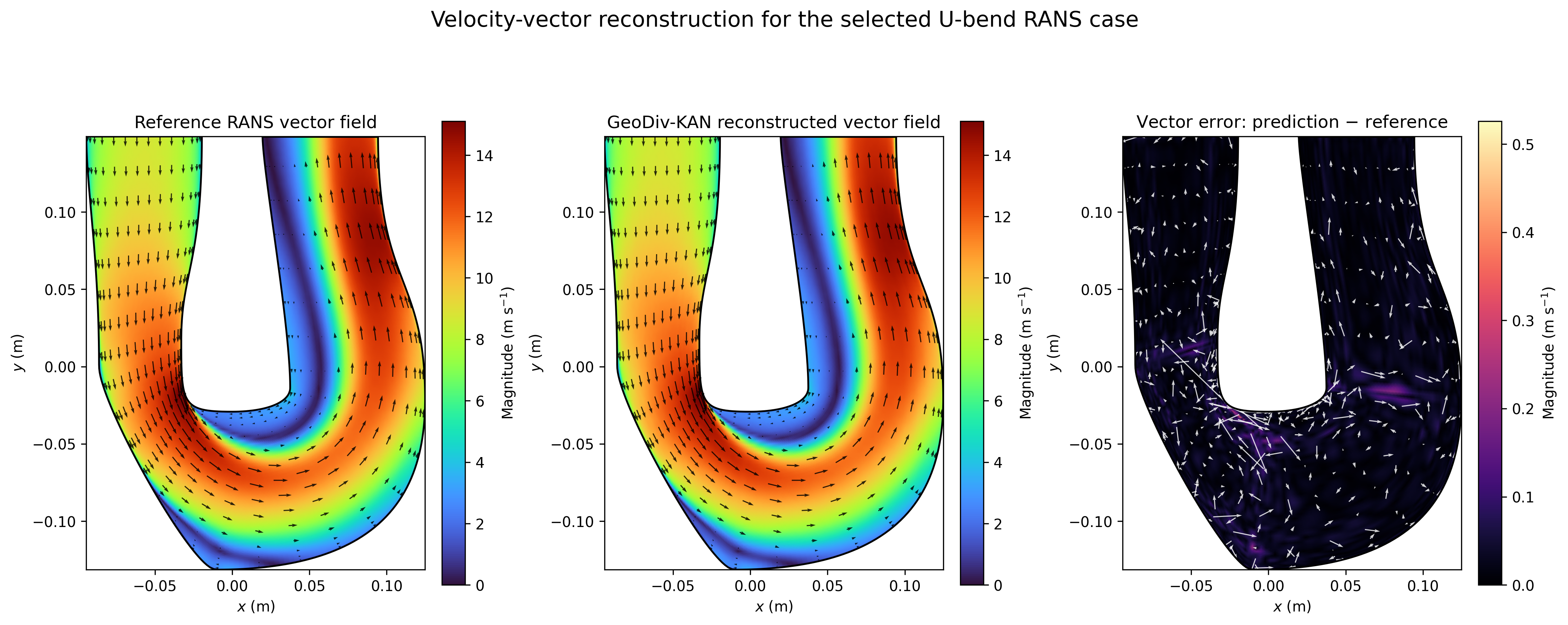}
	\caption{Reference RANS and GeoDiv-KAN reconstructed velocity-vector
		fields, together with the vector-error magnitude.}
	\label{E_2_vector_field_reference_prediction_error}
\end{figure}

\begin{figure}[H]
	\centering	\includegraphics[width=1.0\textwidth]{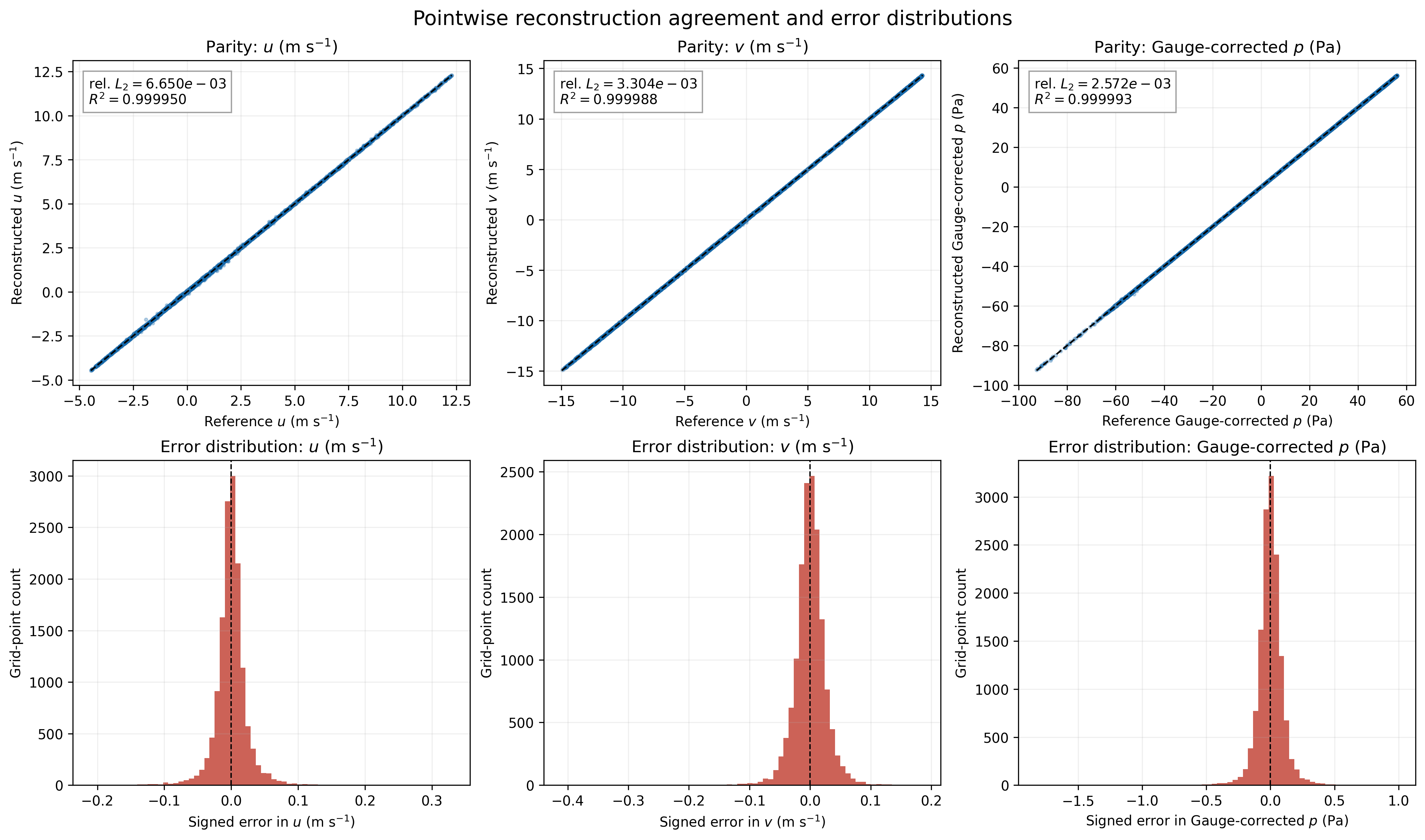}
	\caption{Pointwise parity and signed-error distributions for reconstruction of \(u\), \(v\), and gauge corrected pressure
		\(p\).}
	\label{E_2_parity_and_error_distributions}
\end{figure}
The reconstructed values closely reflect the reference values over the whole field range for \(u\), \(v\), and \(p\), respectively, as shown in Fig.~\ref{E_2_parity_and_error_distributions}. Signed errors are concentrated near zero, indicating accurate reconstruction and no consistent over or underprediction.

The reference RANS fields, GeoDiv-KAN reconstructions, and absolute errors for $u$, $v$, and gauge-corrected $p$ are compared in Fig.~\ref{E_2_uvp_reference_reconstruction_absolute_error}. The pressure distribution across the geometry and the velocity variation via the bend are precisely captured by the reconstructed fields. Errors are still mostly confined to high-gradient areas and the sharply curved inner wall.

Velocity-vector reconstruction demonstrates close agreement between
GeoDiv-KAN and the reference RANS solution in Fig.~\ref{E_2_vector_field_reference_prediction_error}. The model captures the
flow direction and acceleration through the bend, while noticeable
errors are mainly confined to the inner bend and high-gradient regions.
\section{Conclusion}\label{sec7}
In this manuscript we introduced GeoDiv-KAN, a geometry-gated, divergence-free Gaussian RBF-based KAN for solving TFNSEs on complex domains. The proposed model combines learnable Gaussian RBF edge functions with geometry-dependent feature gating and a streamfunction ased velocity representation. This structure enables the network to work with different boundary shapes while preserving incompressibility by design. The accurate approximation of velocity and pressure fields and the steady improvement over PINN, PIKAN, RBF-KAN, and DF-RBFKAN were shown by numerical tests on square, circular, and star-shaped domains. The predicted solutions preserved the expected spatial symmetries, temporal evolution, pressure variation, and flow structures across over all domains. The divergence remained close to machine precision, confirming the effectiveness of the streamfunction formulation.

Two data-driven CFD experiments were used to better investigate GeoDiv-KAN's practical applicability.  For unsteady flow past a fixed circular cylinder at \(Re=100\), the model reproduced the evolving wake, vortex shedding behaviour, and downstream mean velocity profiles. For the known U-bend case at \(Re=40{,}000\), it accurately reconstructed the mean velocity and gauge-corrected pressure fields of the reference RANS solution, including the curvature-induced flow variation through the bend. 

The current work is restricted to CFD reconstruction testing and prescribed two-dimensional geometries. Three-dimensional issues, unobserved geometries and circumstances, sparse/noisy data, and uncertainty quantification will all be covered in future study.
\section*{Conflict of Interest Statement}
The authors confirm that there are no conflicts of interest with regard to the publication of this work.
\section*{Funding Declaration}
The authors state that they did not receive any funding to carry out this investigation.
\section*{Authors’ contributions}\justifying
\textbf{Yatin Kumar:} Investigation, Methodology, Conceptualization, Visualization, Software,  Writing-original draft \& editing.
\textbf{Shivani Saini:} Investigation, Visualization,  Writing-original draft \& editing. \textbf{Ramesh Kumar Vats:} Supervision, Formal analysis, Investigation,  Validation,  Writing--review \& editing.
% \textbf{Takaharu Yaguchi:} Methodology, Formal analysis, Validation, and
%Writing--review \& editing.
\section*{Data Availability Statement}\justifying
The cylinder-flow and U-bend CFD datasets used in this study are publicly
available from \href{https://doi.org/10.5281/zenodo.5039610}{10.5281/zenodo.5039610} and
\href{https://huggingface.co/datasets/JDecke/ubend-cfd}{huggingface.co/datasets/JDecke/ubend-cfd},
under the CC BY 4.0 and CC BY-NC 4.0 licenses, respectively.
\bibliographystyle{elsarticle-num} % or "unsrt" / "ieeetr" / "apalike" depending on style
\bibliography{ref-1} % without .bib extension

\end{document}